\documentclass[twocolumn,nofootinbib,superscriptaddress]{revtex4}
\usepackage{amsmath} 
\usepackage{verbatim} 
\usepackage{graphicx} 

\makeatletter
\renewcommand*\env@cases[1][1.2]{%
  \let\@ifnextchar\new@ifnextchar
  \left\lbrace
  \def\arraystretch{#1}%
  \array{@{}l@{\quad}l@{}}%
} 
\makeatother

\begin{document}

\title{Primordial Perturbations in Einstein-Aether and BPSH Theories}

\author{Cristian Armendariz-Picon}
\affiliation{Department of Physics, Syracuse University, Syracuse, NY 13244-1130, USA}
\affiliation{Departament de F{\'\i}sica Fonamental i \\Institut de Ci{\`e}ncies
 del Cosmos, Universitat de Barcelona,\\ Mart{\'\i}\ i Franqu{\`e}s
 1, 08028 Barcelona, Spain}

\author{Noela Fari\~na Sierra}
\author{Jaume Garriga}
\affiliation{Departament de F{\'\i}sica Fonamental i \\Institut de Ci{\`e}ncies
 del Cosmos, Universitat de Barcelona,\\ Mart{\'\i}\ i Franqu{\`e}s
 1, 08028 Barcelona, Spain}

\begin{abstract}
We study the primordial perturbations generated during a stage of single-field inflation in Einstein-aether 
theories. Quantum fluctuations of the inflaton and aether fields seed long wavelength adiabatic and 
isocurvature scalar perturbations, as well as transverse vector perturbations. Geometrically, 
the isocurvature mode is the potential for the velocity field of the aether with respect to matter.
For a certain range of parameters, this mode may lead to a sizable random velocity of the aether 
within the observable universe. The adiabatic  mode corresponds to curvature perturbations of 
co-moving slices (where matter is at rest). In contrast with the standard case, it has a non-vanishing 
anisotropic stress on large scales. Scalar and vector perturbations may leave significant imprints on the 
cosmic microwave background. We calculate their primordial spectra, analyze their contributions to the temperature 
anisotropies, and  formulate some of the phenomenological constraints that follow from observations. 
These may be used to further tighten the existing limits on the parameters for this class of theories. 
The results for the scalar sector also apply to the extension of Ho\v{r}ava gravity recently proposed by Blas, 
Pujol\`as and Sibiryakov.

\end{abstract}

\maketitle

\section{Introduction}

The enigmas of dark matter and cosmic acceleration have motivated the exploration of theories where
gravity is ``modified" at large distances. On the other hand, the range of possibilities for constructing such 
theories is severely 
limited by the requirement of general covariance, and for that reason, most of the proposed alternatives to General Relativity (GR) can in fact be cast 
as GR coupled to new fields.\footnote{
A counterexample is the DGP brane-world scenario, where gravity is modified in the infrared by a continuum of 
Kaluza-Klein gravitons \cite{DGP}. Because of the continuum, DGP cannot be 
formulated as a standard four dimensional GR with additional fields.
See also \cite{galileon} and \cite{galileoncov}, 
for recent related proposals in the four dimensional context.}

Cosmic acceleration may be due to a scalar field slowly rolling down 
a potential \cite{Peebles:1987ek,mod},
or simply sitting in one of its local minima \cite{bopo}.
Alternatively, it can be driven by the non-minimal kinetic term of a k-essence scalar field  with a Lagrangian
of the form $p(X,\phi)$, where $X=\partial_\mu\phi\partial^\mu\phi$. This form is quite versatile, 
and can be used to mimic cosmic fluids with a wide range of possibilities for
the effective equation of state and speed of sound, including those which are characteristic 
of dark energy and cold dark matter \cite{kessence}.

The gradient of the k-essence field,  $\partial_{\mu} \phi$, is a time-like vector which spontaneously breaks 
Lorentz invariance, in a way that is parametrically independent of its effects on the time evolution 
of the background geometry. 
In particular, Lorentz invariance can be spontaneously
broken by $\partial_\mu \phi$ while the background spacetime remains maximally symmetric,
a situation which is known as ghost condensation \cite{ghostcondensation}. Still, the ``fluid" responds 
to the gravitational pull of ordinary matter, leading to modifications of the long range potentials. 

More generally, theories with a massive graviton can be written in a covariant form as
GR coupled to a set of ``St\"uckelberg" scalar fields $\phi^A$ with non-minimal kinetic terms,
whose gradients have non-vanishing expectation values \cite{massive,phases}. 
Depending on the interactions and the expectation values of the condensates, this
can describe different phases of massive gravity. Aside from the Lorentz preserving
Fierz-Pauli phase \cite{massive} (see also \cite{slavaali}), 
Lorentz breaking phases have been investigated in \cite{phases}. Some of these have interesting phenomenology,
such as the absence of ghosts in the linearized spectrum, a massive graviton with just two transverse polarizations,
and weak gravitational potentials which differ from those in standard GR by terms proportional to the square of the 
graviton mass \cite{lbmass,phases,pheno}. 

Additional fields of spin 2 have been considered in bi-gravity (or multi-gravity) 
theories \cite{Damour:2002ws}, where space-time is endowed with several metrics interacting
with each other non-derivatively. Due to general covariance, only one of the gravitons in the linearized 
spectrum stays massless, while the remaining ones acquire masses proportional to the non-derivative 
interaction terms. Lorentz invariance can be broken spontaneously even in cases where all metrics are flat,
provided that their light-cones have different limiting speeds.  This leads to phenomenology \cite{lbbig} similar to that of certain phases of Lorentz breaking massive gravity  referred to above \cite{lbmass,phases,pheno}, of which multigravity can be seen as a particular realization.

Finally, additional vector fields have received considerable attention in cosmology.
Effective field theories for vectors are strongly constrained by stability requirements. Typically, 
those with non-trivial 
cosmological dynamics contain a massive ghost \cite{ArmendarizPicon:2009ai}, which can be removed from the spectrum 
by sending its mass to infinity. This amounts to imposing a fixed-norm constraint on the vector, 
which in turn forces a Lorentz-breaking vacuum expectation value. 
This led Jacobson and Mattingly to dub this type of models Einstein-aether theories \cite{Jacobson:2000xp,mond}. 
Their low-energy excitations are the Goldstone bosons of the broken Lorentz symmetry,\footnote{In theories with 
spontaneously broken spacetime symmetries, the number of Goldstone bosons does not generally agree with the number of broken generators. 
However, if the order parameter that breaks the spacetime symmetry is spacetime-independent (as the constant aether field), 
then both numbers do agree \cite{Low:2001bw}.} 
which will participate in the dynamics of the long range gravitational interactions. 

An interesting recent development 
is the proposal by Ho\v{r}ava \cite{hg} that a Lorentz-breaking theory of gravity 
may be renormalizable and UV complete. The breaking of Lorentz invariance in this case is implemented by introducing
a preferred foliation of space-time, but no additional structure.
As pointed out in \cite{bps1}, any theory with a preferred 
foliation can be written in a generally covariant form 
by treating the time parameter which labels the different hypersurfaces as a St\"uckelberg scalar field ${\cal T}$. 
The foliation is considered to be physical, but not the parametrization, and therefore
the covariant theory should be invariant under field redefinitions ${\cal T}\to f({\cal T})$. In other words,
the Lagrangian can have a dependence on the unit normal to the hypersurfaces, but not 
on the magnitude of the gradient ${\cal T},_\mu$ (in contrast with the examples of k-essence and ghost 
condensation mentioned above).
From this observation, Blas, Pujol\`as and Sibiryakov  showed \cite{bps2} that Ho\v{r}ava gravity
could be extended  by including in the action all terms compatible with reparametrization symmetry, and consistent 
with power counting renormalizability. Interestingly enough, this extension also cured certain problems in the scalar sector
of the original proposal (such as instabilities and strong coupling at low energies \cite{bps1}). 
Jacobson \cite{jbpsh}, has recently 
clarified the relation between the Einstein-aether theory and this extended version of Ho\v{r}ava gravity, which he dubbed 
BPSH gravity. In particular, he pointed out that any solution of Einstein-aether where the vector field is hypersurface 
orthogonal is also a solution of the low energy limit of BPSH gravity.

Since the aether only interacts gravitationally, any signal of it must be proportional to a 
power of $(E/M_P)^2$, where $M_P$ is the reduced Planck  mass, and $E$ is an energy scale. 
Thus, even though the aether contains massless fields, its presence is hard to detect. 
In that respect, inflation provides an interesting window 
to probe the aether and its implications. During inflation, short-scale vacuum 
fluctuations of light fields are transferred to cosmological distances, 
where they may leave an observable imprint. 
It is thus natural to look for signatures of Einstein-aether on the spectrum of primordial perturbations, 
which is the subject to which we devote this article. 

Previous work on this subject
has been done in Refs. \cite{Lim:2004js,Li:2007vz}, although in a somewhat narrower region of parameter
space and with somewhat different conclusions. In the scalar sector, we find that there
is a primordial isocurvature mode, which can be interpreted as the velocity potential for the aether with respect 
to matter. Depending on the aether parameters, this mode can grow on superhorizon scales, leading to a large 
random velocity field for the aether. Similar results apply to the transverse vector sector. 
These perturbations may thus be of phenomenological interest. We also find that the isocurvature mode is strongly 
correlated with the usual adiabatic mode, which corresponds to curvature perturbations in the co-moving slicing. 

For previous work on the impact of adiabatic scalar perturbations  on the cosmic microwave background radiation 
(CMB) and large scale structure in (generalized) aether theories, see \cite{Zuntz:2008zz, Zuntz:2010jp}.

While this paper was being prepared, an interesting related paper by Kobayashi, Urakawa and Yamaguchi 
appeared \cite{yuko}, which analyzes the post-inflationary evolution of the adiabatic scalar mode in BPSH theory.  
Where we overlap, our conclusions agree.

The plan of the paper is the following. In Section \ref{sec:Theory} we review the basics of Einstein-aether theory
and the homogeneous cosmological solutions. Sections \ref{sec:tensor perturbations}, \ref{sec:scalar perturbations}
and
\ref{sec:vector perturbations} are devoted to the analysis of tensor, scalar and transverse vector perturbations
respectively. Section \ref{ncmbs} analyzes the contribution of vector modes to the CMB spectrum. 

Readers familiar with the details Einstein-aether or BPSH theory are encouraged to jump directly to the concluding
Section \ref{sumcon}, for a self contained summary of the main results. 

Appendix A summarizes the existing bounds on the parameters 
of Einstein-aether theories. Appendix B discusses the equations of motion for the scalar sector of the theory in
the longitudinal gauge. Appendix C deals with the canonical reduction of the scalar sector to the two physical 
degrees of freedom (a necessary step for the proper normalization of the vacuum fluctuations). Appendix D contains 
a derivation of the long wavelength adiabatic and isocurvature scalar modes, for generic matter content and
expansion history. Appendix E derives the CMB temperature anisotropies due to vector modes.

\section{Einstein-Aether Theories}
\label{sec:Theory}

The Einstein-aether  is described by the most general Lagrangian with two derivatives acting on a vector field of 
constrained norm \cite{Jacobson:2000xp},
\begin{multline}\label{eq:L aether}
\mathcal{L}_A=c_{1} \nabla_{\alpha}A^{\gamma} \nabla^{\alpha}A_{\gamma}+c_{2} \nabla_{\alpha}A^{\alpha} 
\nabla_{\gamma}A^{\gamma}+c_{3} \nabla_{\alpha}A^{\gamma} \nabla_{\gamma}A^{\alpha}- \\
{}-c_{4}A^{\alpha}A^{\beta}
\nabla_{\alpha}A^{\gamma} \nabla_{\beta}A_{\gamma}+ \lambda (A^{\alpha}A_{\alpha}+1).
\end{multline}
Here,  the $c_i$ are dimensionless coefficients, and $\lambda$ is a Lagrange multiplier that enforces 
the constraint 
\begin{equation} \label{eq:fixed norm}
A^{\mu}A_{\mu}=-1.
\end{equation}
The Lagrangian (\ref{eq:L aether}) can be thought of as the low-energy 
description of a theory in which boost invariance is spontaneously broken by the expectation value of $A_\mu$, 
while spatial rotations and translations remain unbroken. 
The fixed-norm constraint eliminates the ``radial" degree of freedom in field space, which is typically a ghost. 
We assume that $A_\mu$ is ``minimally  coupled" to gravity and to the rest of matter, so the total 
action is of the form,  
\begin{equation}\label{eq:action}
	S= \frac{M_P^2}{2}\int d^{4}x\sqrt{-g}\left[R+\mathcal{L}_A\right]
	+\int d^{4}x\sqrt{-g}\, \mathcal{L}_{m}.
\end{equation}
Here $M_P$ is the reduced Planck mass, and $\mathcal{L}_m$ is the Lagrangian of ordinary matter, 
which we assume does not contain couplings to the aether field.

The gravitational equations involve the energy-momentum 
tensor of the vector, ${T_{\mu\nu}=(-1/\sqrt{-g})(\delta S_A/\delta g^{\mu\nu})}$. This is given by
\begin{multline}
\label{eq:tmn}
	T_{\mu \nu}= 
\nabla_{\sigma}\left(J^{\;\;\;\sigma}_{(\mu}A_{\nu)}-J^{\sigma}_{\;(\mu}A_{\nu)}-J_{(\mu\nu)}A^{\sigma}\right)
+Y_{\mu \nu}+\\
	\shoveright{}+\frac{1}{2}g_{\mu \nu}\mathcal{L}_A+\lambda A_{\mu}A_{\nu}
	-c_{4}A^{\alpha}A^{\beta}(\nabla_{\alpha}A_{\mu})(\nabla_{\beta}A_{\nu}),
\end{multline}
where 
\begin{equation}
J^{\alpha}_{\;\;\;\sigma} =c_{1}\nabla^{\alpha}A_{\sigma}+c_{2}\delta^{\alpha}_{\sigma}\nabla_{\beta}A^{\beta}+c_{3}\nabla_{\sigma}A^{\alpha}-c_{4}A^{\alpha}A^{\beta}\nabla_{\beta}A_{\sigma},
\end{equation}
and 
\begin{equation}
Y_{\alpha \beta}= c_{1}\left[(\nabla_{\gamma}A_{\alpha})(\nabla^{\gamma}A_{\beta})-(\nabla_{\alpha}A_{\gamma})(\nabla_{\beta}A^{\gamma})\right].
\end{equation}
Variation of the Lagrangian density (\ref{eq:L aether}) with respect to $A$ leads to the field equation
\begin{equation}\label{eq:vector field motion}
	\nabla_{\alpha}(J^{\alpha}_{\;\;\;\beta})
	+c_{4}A^{\alpha}(\nabla_{\alpha}A^{\gamma})(\nabla_{\beta}A_{\gamma})
	= \lambda A_{\beta},
\end{equation}
whilst variation of the Lagrangian density with respect to the Lagrange multiplier $\lambda$ imposes the 
fixed norm constraint (\ref{eq:fixed norm}).

The coefficients $c_i$ are subject to both theoretical and phenomenological restrictions, which we collect in 
Appendix \ref{sec:other constraints} and summarize in Table \ref{tab:conditions}. Their 
magnitude, relative to the symmetry breaking scale, can be estimated from dimensional analysis. 
The field redefinition ${A_\mu=\tilde{A}_\mu/M}$ leads to the fixed norm constraint ${\tilde{A}_\mu \tilde{A}^\mu=-M^2}$, from which we may interpret $M$ as the scale at which Lorentz symmetry is spontaneously broken. In terms of the coefficients 
$\tilde{c}_i$ that would multiply the action for the rescaled  field $\tilde{A}$, the original coefficients
are given by $c_{1,2,3}=(M/M_P)^2 \tilde{c}_{1,2,3}$ and $c_4=(M/M_P)^2 M^2 \tilde{c}_4$. We expect the 
dimensionless  $\tilde{c}_{1,2,3}$ to be of order one, and the dimensionful $\tilde{c}_4$ to be of order $M^{-2}$, 
which leads to
\begin{equation}
	  c_i\sim \frac{M^2}{M_P^2}.
\end{equation}
For convenience, in what follows we use the abbreviations
\begin{subequations}\label{abbreviations}
\begin{eqnarray}
	c_{13}=c_1+c_3, \quad c_{14}&=&c_1+c_4,\\
\alpha=c_1+3c_2+c_3, \quad \beta&=&c_1+c_2+c_3.
\end{eqnarray}
\end{subequations}
Note that $\alpha=3\beta-2c_{13}$, so these abbreviations are not supposed to be an independent parametrization.
Note also that our coefficients $c_i$ and  those of other works in the aether literature may have opposite signs.

\begin{table*}
\begin{tabular}{| l | c | c |}
\hline
Condition & Constraint & \,  Equation \,  \\
\hline
Solution of Einstein's equations & $\alpha<2$  & (\ref{eq:condition alpha})  \\
Stability of Tensors & $c_{13}>-1$ & (\ref{eq:tensor no ghost})\\
Stability of Scalars & $-2 \leq c_{14}< 0,\,\ \beta < 0$ & (\ref{eq:scalar stability}) \\
Stability of Vectors &  $2 c_1 \leq c_{13}^2 (1+c_{13})$  & (\ref{eq:vector stability}) \\
PPN Limits & {\it see} Equation & (\ref{eq:PPN constraints}) \\
Big-Bang Nucleosynthesis  & $c_{14}+\alpha\lesssim 0.2$ & (\ref{eq:BBN})  \\
\hline\hline
Cherenkov radiation (assumes subluminality)& {\it see} Equation & (\ref{eq:Cherenkov constraints}) \\
\hline \hline
Superluminal Tensors & $c_{13} \leq 0$  & (\ref{eq:tensor c})\\
Superluminal Scalars & $(2+c_{14}) \beta \leq (2-\alpha) (1+c_{13}) c_{14}$ & (\ref{eq:scalar cs}) \\
Superluminal Vectors & $ 2c_4 \geq -c_{13}^2/(1+c_{13})$ & (\ref{eq:vector c}) \\
\hline\hline
Anisotropic stress of long wavelength adiabatic modes & $|c_{13}|\lesssim 1$ & (\ref{asam2})\\
Non-growing scalar isocurvature modes & $\alpha/c_{14}\geq -1$ & (\ref{kappaeq})\\
Subdominant contribution of vectors to CMB & $C^V_\ell\lesssim C^\zeta_\ell$ & (\ref{comparativa})\\
\hline
\end{tabular}
\caption{Summary of the theoretical and phenomenological conditions on the parameters of aether theories.
We use the abbreviations $\alpha,\beta,c_{13}$ and $c_{14}$, which are related to the standard aether parameters
$c_i$ through Eqs. (\ref{abbreviations})}\label{tab:conditions}
\end{table*}

\subsection*{Cosmological dynamics}

Let us consider the dynamics of a spatially flat unperturbed 
Friedman-Robertson-Walker universe in the presence of the aether. 
Homogeneity and isotropy constrains the form of the metric and of the aether. With the line element given by  
$ds^2=a^2(\eta)\left[-d\eta^2+d\vec{x}^2\right]$ we have, from Eq.~(\ref{eq:fixed norm}), 
\begin{equation} \label{eq:background A}
	A^\mu=(a^{-1},0,0,0).
\end{equation}
Substituting into the expression for the energy-momentum tensor (\ref{eq:tmn}), 
we find that the energy density and pressure of the vector field are respectively given by 
\begin{equation}
\rho_A=\frac{3\alpha}{16\pi G a^2} \mathcal{H}^2, \quad \text{\and}
p_A=-\frac{\alpha}{16\pi G a^2}\left(\mathcal{H}^2+2\mathcal{H}'\right),
\end{equation}
where $G=1/8\pi M_P^2$, $\mathcal{H}=a'/a$ and a prime denotes a derivative with respect to conformal time.
Thus, Einstein's equations read
\begin{subequations}\label{eq:Einstein}
\begin{eqnarray}
	\mathcal{H}^{2}&=&\frac{8\pi G_\mathrm{cos}}{3} a^{2}\rho,  \label{eq:Friedman}\\
	\mathcal{H}'&=&-\frac{4\pi G_\mathrm{cos}}{3} a^{2}(\rho+3p), \label{eq:acceleration}
\end{eqnarray}
\end{subequations}
where $\rho$ and $p$ are 
the energy density and pressure of the remaining matter fields (aether excluded) and 
\begin{equation}
	G_\mathrm{cos} = \left(1-{\alpha \over 2}\right)^{-1} G.
\end{equation}
A comparison with the same equations in the absence of the aether shows that the effect 
of the vector field is merely to ``renormalize" the value of Newton's gravitational constant \cite{Carroll:2004ai}; 
the energy density and pressure of the vector field mimic that of the remaining components in the universe. 

On the other hand, the gravitational field created by isolated bodies is not exactly the same as that
of General Relativity, and in that sense the aether is a bona-fide modification of gravity. 
To lowest order in a post-Newtonian expansion, the potential 
sourced by a static and spherically symmetric body satisfies the  Poisson 
equation $\Delta \phi=4\pi G_N \rho$, but with a modified gravitational constant \cite{Jacobson:2008aj}
\begin{equation}\label{eq:Newtonian}
	G_N=\left(1+ {c_{14}\over 2} \right)^{-1}G.
\end{equation}
Hence, the aether also renormalizes the gravitational constant measured in ``local" experiments, 
but by a different amount than in the cosmological case. Post-Newtonian corrections lead
to further deviations of General Relativity, which place severe constraints on the aether parameters.
A summary of these and other constraints is given in Appendix \ref{sec:other constraints}. 
Nucleosynthesis, in particular, places a bound on the relative magnitude of the two Newton constants,
of the form \cite{Carroll:2004ai}
\begin{equation}
\left|{G_\mathrm{cos}\over G_{N}} - 1\right|< 10\%. 
\end{equation}

Note that, for  positive matter energy density and positive Newton's constant  
$G$, Eq. (\ref{eq:Friedman}) can only be solved if\footnote{We could have $\alpha>2$ if we allow $G<0$. 
However, this leads to instabilities in 
the tensor modes, as we shall discuss in Section III.}
\begin{equation}\label{eq:condition alpha}
	\alpha<2.
\end{equation}
Remarkably, this condition does not follow from any of the perturbative stability arguments which we shall consider 
below, but merely from the existence of a cosmological solution with positive energy density for ordinary matter.
Note also that the Lagrange multiplier has 
a finite value along the cosmological solutions. Contracting the vector 
field equations of motion (\ref{eq:vector field motion}) with $A^\beta$ we have
\begin{equation}\label{eq:background lambda}
	\lambda=\frac{3}{a^2}\left(\beta \mathcal{H}^2-c_2 \mathcal{H}'\right).
\end{equation}

For later reference, let us consider the case where the 
matter sector consists of a scalar field with an exponential potential,
\begin{equation}\label{eq:exponential potential}
	\mathcal{L}_m=-\frac{1}{2}\partial_\mu \varphi \partial^\mu \varphi-V_0 \exp\left[-\mu \frac{\varphi}{M_P}\right].
\end{equation}
It is well-known that this potential leads to power-law inflation \cite{Lucchin:1984yf}, with a constant equation of state parameter 
$w\equiv p_\varphi/\rho_\varphi$ determined by the coefficient $\mu$ in the exponential.
With  a constant equation of state $w$ the solution of Eqs. (\ref{eq:Friedman}) and (\ref{eq:acceleration}) is then
\begin{equation}\label{eq:power law inflation}
	a\propto (-\eta)^q, \quad \text{with} \quad q=\frac{2}{1+3w}=\frac{1}{\epsilon-1},
\end{equation}
where 
\begin{equation}
\epsilon\equiv - H'/(aH^2) = {2-\alpha\over 4} \mu^2,  
\end{equation}
is the conventional slow-roll parameter.
Note that if $2-\alpha$ is sufficiently small, inflation may be de Sitter like even if $\mu$ 
is of order one. This broadens the class of ``natural" inflationary models that do not require 
particularly flat potentials, though we shall not explore this possibility here.

\subsection*{Cosmological Perturbations}

The background vector field (\ref{eq:background A}) preserves rotational invariance, 
and so it is still convenient to use the standard decomposition of perturbations in scalars, vector and tensors 
under spatial rotations:
\begin{equation}\label{eq:metric perturbations}
\begin{split}
	ds^2&=a^2(\eta)\Big[-(1+2\phi)d\eta^2 +2 (B_{,i}+S_i) d\eta dx^i +\\
	&+ \left(\delta_{ij}-2\psi \delta_{ij}+E_{,ij}+F_{i,j}+F_{j,i}+h_{ij}\right)dx^i dx^j\Big],
\end{split}
\end{equation}
and
\begin{equation} \label{eq:field perturbations}
	A^0=\frac{1}{a}+ \delta A^0 ,\quad A^i=\frac{1}{a}\left(C_{,i}+V_i-S_i\right).
\end{equation}
Since the metric and vector fields are related to the Lagrange multiplier by Eq. (\ref{eq:vector field motion}), 
we also need to perturb the Lagrange multiplier,
\begin{equation}\label{eq:lambda}
	\lambda=\lambda_0+\delta\lambda,
\end{equation}
where $\lambda_0$ is the background value, given by Eq. (\ref{eq:background lambda}). 
Variation of the second order action with respect to $\delta \lambda$ 
leads to the linearized form of the constraint (\ref{eq:fixed norm}),
\begin{equation}
\delta A^0=-{\phi\over a}.
\end{equation}
Here, $\phi, B, \psi, E, C$ are scalars, $S_i, F_i, V_i$ are transverse vectors, 
and $h_{ij}$ is a transverse and traceless tensor. Note that $h_{ij}$ and $V_i$ are gauge-invariant. 
Scalars, vectors and tensors decouple from each other in the linearized theory, so we consider each sector 
separately. In momentum space, our convention for the Fourier components is
\begin{equation}
	f_{\bf k}(\eta) \equiv f(\eta,{\bf k})=\int \frac{d^3 x}{(2\pi)^{3/2}}\,  f(\eta,{\bf x})
	\exp\left(-i{\bf k} \cdot {\bf x}\right).
\end{equation}

\section{Tensor Perturbations}\label{sec:tensor perturbations}

As discussed in \cite{Lim:2004js} the presence of the aether modifies the propagator and 
the dispersion relation of the tensor modes. Substituting Eq.
(\ref{eq:metric perturbations}) into the action (\ref{eq:action}),
with matter Lagrangian given by (\ref{eq:exponential potential}), 
expanding to quadratic order in $h_{ij}$ and using the background equations 
of motion we obtain,
\begin{equation}
\label{eq:Lt_frw}
	L^{(2)}_{t}={} \frac{M_P^2 a^2}{8}\left[(1+c_{13}) {\bf h}'\cdot  {\bf h}'
	-\partial_i {\bf h} \cdot \partial^i {\bf h}\right],
\end{equation}
where $\bf h$ stands for a matrix with components $h_{ij}$ and the dot indicates contraction of both indices (with the Euclidean
metric).
On short (subhorizon) scales, gravity waves propagate at a speed \cite{Jacobson:2004ts}
\begin{equation}\label{eq:tensor c}
	c_h^2=\frac{1}{1+c_{13}}.
\end{equation}
Classical stability of tensors thus imposes the condition 
\begin{equation}\label{eq:tensor no ghost}
	1+c_{13}>0,
\end{equation}
since, otherwise, high frequency modes grow exponentially fast.

In the previous section we noted that the background solution only exists for $\alpha<2$, 
implicitly assuming that the ``bare" Newton's constant is positive $G>0$.
Here, we note that for  $M_P^2<0$, the coefficient in front of the kinetic term of $h_{ij}$ has the 
``wrong" sign, and the two independent transverse and traceless tensor modes are ghosts.  

A theory with ghosts is quantum mechanically unstable. The vacuum can decay by emitting positive energy particle plus 
negative energy quanta while conserving 
energy. In a Lorentz-invariant theory, the phase space available for the decay of the vacuum would be infinite, and the 
lifetime of the vacuum is then infinitely short, which makes the theory unviable. In a non-Lorentz invariant 
theory, the decay rate may be finite, and the vacuum may be sufficiently long-lived (see for instance 
\cite{Cline:2003gs}). In our case, Lorentz invariance is spontaneously broken, and the effective theory we are using 
is supposed to be valid only well below the symmetry breaking scale $M$. The decay rate is UV 
sensitive, so strictly speaking it is unclear whether the theory can be made sense of in the presence
of ghosts. However, to be conservative, we shall systematically exclude from parameter space the cases 
when ghosts are present.

The primordial spectrum of tensor modes seeded during inflation is immediately obtained from (\ref{eq:Lt_frw}), 
and is inversely proportional to their propagation speed, 
\begin{equation}\label{eq:tensor power}
	\mathcal{P}_h(k)=\frac{1}{\pi^2 c_t}
		\frac{H^2}{M_P^2}\Bigg|_{c_s k=\mathcal{H}}.
\end{equation}
Hence, the amplitude of the primordial tensor modes differs from that in general relativity 
(for the same values of $H$ and $M_P$.)

\section{Scalar Perturbations}
\label{sec:scalar perturbations}

The scalar sector of Einstein-aether theories consists of the five scalars $\phi, \psi, B, E$, $C$
 defined in Eqs. (\ref{eq:metric perturbations}) and 
(\ref{eq:field perturbations}).\footnote{The perturbation in the Lagrange multiplier $\delta\lambda$ disappears from the 
Lagrangian after substituting the constraint to which it leads. To linearized order,
this constraint is $\delta A^0 =-\phi/a$, which we use to eliminate the scalar $\delta A^0$ in favour
of the potential $\phi$.} Thus, the aether enlarges the scalar sector by the aether perturbation $C$.

It is convenient to introduce a gauge-invariant description of the dynamical degrees of freedom.
To this end, following \cite{bps2,jbpsh} we  note that the scalar part of the aether field $A_{\mu}$
can be represented by means of an auxiliary scalar field ${\cal T}$ through the identification
$$
A_{\mu} \equiv {-{\cal T},_\mu \over (-{\cal T},_{\nu} {\cal T},^{\nu})^{1/2}},
$$
where it is assumed that the gradient of ${\cal T}$ is everywhere 
time-like. Surfaces of constant ${\cal T}$ define a foliation of space-like surfaces, 
and we can think of ${\cal T}$ as a time variable.
Since the background $A_\mu$ is aligned with the FRW temporal coordinate, the background field is given
by ${\cal T}={\cal T}(\eta)$. The perturbations $\delta {\cal T} (\eta,{\bf x})$ lead to the 
linearized spatial components
$A_i = - (a /{\cal T}') \delta {\cal T},_i$. From Eq.~(\ref{eq:field perturbations}) we have $A_i= a\, \partial_i(B+C)$, so it follows that
\begin{equation}
{\delta {\cal T} \over {\cal T}'}=-(B+C). \label{chi}
\end{equation}

\begin{figure}
  \begin{center}
 \includegraphics[width=9cm]{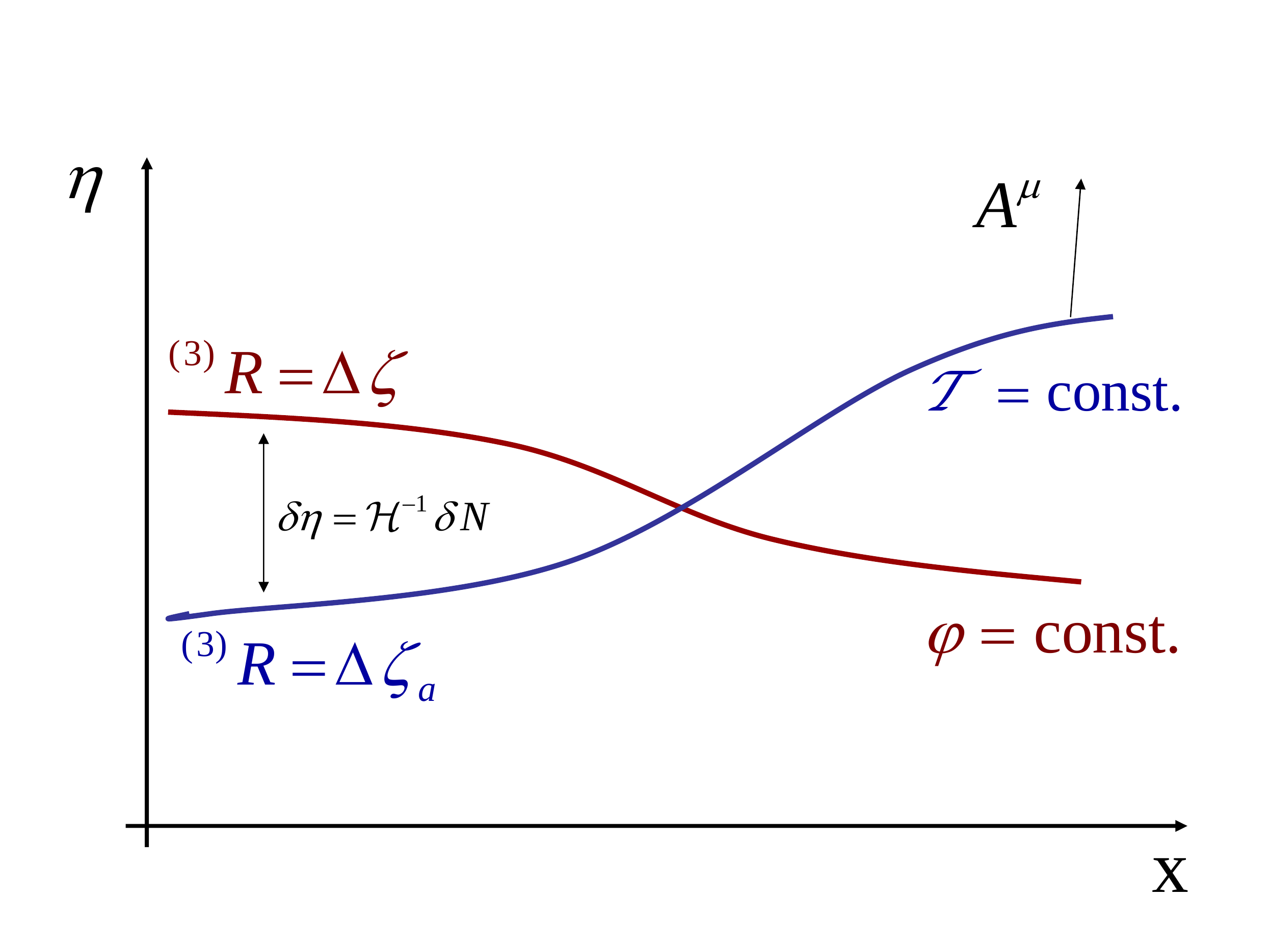}
  \end{center}
  \caption{Geometrical interpretation of different perturbation variables. 
On hypersurfaces of constant inflaton $\varphi$, the curvature perturbation is $\zeta$, while  
on hypersurfaces of constant aether $\mathcal{T}$ the perturbation of the spatial curvature is $\zeta_a$.  
In the presence of isocurvature modes, both hypersurfaces do not agree. 
Their distance in conformal time is the variable $\delta\eta$, which  measures departures from adiabaticity.}
   \label{fig:surfaces}
\end{figure}

In addition to the Einstein-aether sector, we must also include the matter sector.
When the dominant matter component is the inflaton field $\varphi$, a convenient set of gauge-invariant 
variables is given by:
\begin{subequations}
\begin{align}
	\zeta_a\equiv{}& \psi - \mathcal{H}(B+C),
	\label{eq:zetaa}\\
	\delta N\equiv{}& \frac{\mathcal{H}}{\varphi'}\delta\varphi + \mathcal{H} (B+C).
	\label{deltaN1}
\end{align}
\end{subequations}
Geometrically, these can be interpreted as follows (see Fig. \ref{fig:surfaces}.) Using (\ref{chi}), it is clear that the variable $\zeta_a$ is the curvature 
perturbation on surfaces of constant field ${\cal T}$ (i.e. on hypersurfaces orthogonal
to the aether field $A^{\mu}$). From the definition of $\zeta_a$ and $\delta N$ it also follows that 
\begin{equation}
	\zeta\equiv\zeta_a +\delta N \label{comovz}
\end{equation}
is the curvature perturbation on surfaces of constant inflaton $\varphi$. 
At the end of inflation and afterwards, $\zeta$ will describe the curvature perturbation on hypersurfaces 
comoving with matter (excluding the aether.)  On the other hand, 
\begin{equation}
\delta N = \mathcal{H}\left({\delta\varphi\over \varphi'}- 
{\delta {\cal T}\over {\cal T}'}\right) =\mathcal{H}\delta\eta, 
\label{deltaN}
\end{equation}
where $\delta\eta$ is the amount of conformal time separating the surfaces of constant $\varphi$ from the surfaces of 
constant ${\cal T}$. Hence $\delta N$ can be interpreted as the differential e-folding 
number between these two types of surfaces.
The velocity of aether with respect to the matter is given by
\begin{equation}\label{vpote}
v_i = \delta \eta,_i  = \mathcal{H}^{-1}\delta N,_i.
\end{equation}
Hence, we can also think of the isocurvature perturbation ${\cal H}^{-1} \delta N$ as a velocity potential for 
the aether with respect to matter.

In what follows, we consider the case of an exponential inflaton 
potential, Eq. (\ref{eq:exponential potential}). This somewhat simplifies the analysis because the background 
solutions have a constant equation of state parameter $p=w \rho$. 

In addition, the behaviour of long wavelength perturbations of such a scalar field can mimic those of radiation and matter dominated eras 
for $w=1/3$ and $w=0$ respectively. The ``equivalence" applies only {\em on large scales}, because scalar perturbations and fluid 
perturbations have different sound speeds. Nonetheless, in Appendix D we derive the form of the long wavelength adiabatic and isocurvature
scalar modes for generic matter content and expansion history.

\subsection{Short wavelength Lagrangian and stability.}
\label{sec:scalar stability}

In Appendix \ref{sec:gauge invariant} we discuss the Lagrangian for the scalar sector, and its reduction to a set of two 
gauge-invariant degrees of freedom $(\zeta_a, \delta N)$ given 
by Eqs. (\ref{eq:zetaa}, \ref{deltaN1}). In the short wavelength limit,
this Lagrangian is
\begin{align}
\mathcal{L} =\frac{a^2}{2Z_{N}}\left[(\delta N)'^2 - k^2 (\delta N)^2\right] + 
\frac{a^2}{2Z_{a}}(\zeta_a'^2 - c_{a}^{2} k^2 \zeta_a^2) +\ldots,\label{lsw}
\end{align}
where the ellipsis denotes terms which are subleading in the momentum expansion, and we have introduced
\begin{equation}\label{zsn}
Z_N={ 4\pi G_\mathrm{cos}\over \epsilon}, \quad  Z_a=-2\pi c_t^2\beta G_\mathrm{cos}, 
\end{equation}
and
\begin{equation}\label{eq:scalar cs}
	c_a^2=\frac{G_\mathrm{cos}}{G_N}{\beta \over c_{14}} c_t^2.
\end{equation}
Here we have also introduced the slow roll parameter ${\epsilon=(3/2)(1+w)}$ and $G_N$ as given in Eq. 
(\ref{eq:Newtonian}). For constant scale factor $a$, the residue $Z_a$ and sound speed $c_a$ 
agree with the corresponding quantities in a perturbed flat space, as discussed in 
\cite{Jacobson:2004ts}.\footnote{The above expressions are singular for $\beta=0$, but it is easy to 
show, following the derivation in Appendix \ref{sec:gauge invariant} that $\zeta_a$ is not 
dynamical in this case.}
Quantum stability requires $Z_a>0$, and classical stability 
requires 
$c_a^2\geq 0$. Recalling that stability of tensors demands $c_t^2>0$, and that (\ref{eq:condition alpha})
requires $G_\mathrm{cos}>0$, we are led to the conditions
\begin{equation}\label{eq:scalar stability}
-2\leq c_{14}< 0 \quad \text{and}\quad {\beta}< 0,
\end{equation}
which in turn guarantees $G_N>0$.
(The case $c_{14}=0$ is singular, and has to be treated separately.)
 
From Eq. (\ref{lsw}) we can read off the normalization of the positive frequency modes
associated with the ``in" vacuum in the limit $k|\eta|\to\infty$, corresponding to
wavelengths well within the horizon. The two independent mode functions are given by
\begin{subequations}\label{vacnorm}
\begin{align}
	\label{eq:mode phi}
	\zeta_a^{(\varphi)} &\to 0, 
	&\delta N^{(\varphi)} &\to
	{Z_N^{1/2}\over a}\ \frac{e^{-i k\eta}}{\sqrt{2k}},\\
	\label{eq:mode a}
	\zeta_a^{(a)} &\to {Z_{a}^{1/2}\over a}\ \frac{e^{-i c_{a}k\eta}}{\sqrt{2c_{a}k}},
	\quad 
	&\delta N^{(a)} &\to 0.
\end{align}
\end{subequations}
In the first mode,  where $\zeta_a\to 0$, the surfaces of constant aether field
coincide initially with the so-called flat slicing, and $\delta N$ is the number of 
e-folds separating the surfaces of 
constant inflaton field from the flat slicing. This mode survives in 
the limit when there is no aether field (since one
can still define the flat slicing surfaces). Hence, we may call this the inflaton perturbation. 
In the second mode, where $\delta N \to 0$, the inflaton is initially aligned with the aether, so that there is
no inflaton perturbation in the aether frame. This mode survives in the flat space limit even when there 
is no inflaton field. Hence, we call this the aether perturbation. This can be thought of as one of the 
Goldstone bosons of the spontaneous Lorentz 
symmetry breaking. 

In the previous discussion we have assumed that the action 
(\ref{eq:action}) gives an accurate description of the aether up to sufficiently high momenta, so that 
(\ref{vacnorm}) applies to scales well within the horizon. Hence, we require that
the Einstein-aether as an effective theory should be valid at least up to some spatial cut-off $\Lambda \gg H$,
where $H$ is the Hubble rate during inflation. We expect the corrections introduced by the unknown physics 
above the cut-off scale to be at most of order of $H/\Lambda$ (see for instance \cite{ArmendarizPicon:2003gd}).

It should be noted that, on large scales, the fluctuations due to the aether
will mix with those due to the inflaton. Hence, while in single field inflation 
perturbations have to be adiabatic, in Einstein-aether theories there should exist additional 
non-adiabatic modes, as we discuss next.

\subsection{Long wavelength modes}

The full Lagrangian in terms of the gauge-invariant variables $\zeta_a$ and $\delta N$ is somewhat cumbersome 
away from the short wavelength limit, because the two modes are no longer decoupled. However, for long wavelengths
the Lagrangian can be easily obtained from (\ref{eq:first order invariant})
and diagonalized, but now in terms of a new pair of gauge-invariant variables $(\zeta,\delta N)$,
where $\zeta$ is the curvature perturbation on hypersurfaces of constant inflaton field, which we shall also refer to as comoving  hypersurfaces, defined in  Eq.~(\ref{comovz}). The long wavelength Lagrangian is given by
\begin{equation}\label{lwlzn}
\mathcal{L}_{k\eta\ll 1} = \mathcal{L}_\zeta + \mathcal{L}_{\delta N} + \cdots .
\end{equation}
Here the ellipsis denotes subdominant terms and the dominant ones are given by
\begin{equation}
\mathcal{L}_{\zeta} = \left(\frac{4}{3(1+w)}-\beta c_t^2\right)^{-1}{a^{2}\over 4\pi G_\mathrm{cos}}\ \zeta'^2,
\end{equation}
and 
\begin{equation}
\mathcal{L}_{\delta N}= -c_{14}(1+3w)^2\ \frac{a^{2}(k\eta)^2}{64\pi G}\left[(\delta N)'^2+{\kappa\over \eta^2} 
(\delta N)^2\right],
\end{equation}
where we have introduced 
\begin{equation}\label{kappaeq}
\kappa = - 6\left(1+{\alpha\over c_{14}}\right) {1+w \over (1+3w)^2}.
\end{equation}
As we shall see, there are a total of four independent long wavelength modes, which we derive in Appendix \ref{sec:long wavelength}.
Two of them have the property that $\delta N=0$. For these, matter and aether are mutually at rest, and so we
call these modes adiabatic. The other two have $\delta N\neq 0$ and $\zeta=0$, so we call them isocurvature,
since there is no curvature perturbation on co-moving hypersurfaces.

\subsubsection{Adiabatic modes ($\delta N=0$)}

In standard single field inflation, the non-decaying solution for the ``adiabatic" perturbation 
$\zeta$, which we denote by $\zeta_1$, stays constant on superhorizon scales. 
In Appendix \ref{sec:adiabatic modes} we show that the same is true in the presence of the aether:
\begin{equation}
\zeta_1 = const. 
\end{equation} 
for any expansion history (including the case where the equation of state changes abruptly in time).
The corresponding gravitational potentials in the longitudinal gauge are given by Eqs. (\ref{eq:R1}):
\begin{subequations}
\begin{eqnarray}
\phi_1&=&\frac{3(1+w)}{(5+3w)}\ c_t^2\ \mathcal{\zeta}_1,  \label{asam} \\
	\psi_1&=&\phi_1 \label{asamu}
	+{c_{13}}\ c_t^2\ \mathcal{\zeta}_1.	 
\end{eqnarray}
\end{subequations}
The form of the two adiabatic 
modes (non-decaying and decaying) for an arbitrary expansion history and matter content is derived in 
Appendix \ref{sec:adiabatic modes}. The decaying adiabatic mode is given by (\ref{2ad}), and it is characterized by
\begin{subequations}
\begin{eqnarray}
\phi_2&=&\psi_2\propto \mathcal{H} a^{-2}, \\
\zeta_2&=&0.
\end{eqnarray}
\end{subequations}
It is worth mentioning that, although these  adiabatic modes have the properties described in \cite{Weinberg:2003sw}, 
they do not share the properties postulated in \cite{Weinberg:2004kr, Weinberg:2004kf, Weinberg:2008zzc}. 
In particular, for the first adiabatic mode $\zeta_1$, the anisotropic
stress is non-vanishing ($\phi_1\neq \psi_1$).

\subsubsection{Isocurvature modes ($\delta N\neq 0,\zeta=0$)}

As shown in Appendix \ref{sec:long wavelength}, for the case where the background equation of state parameter $w$
is constant, the two isocurvature modes behave as powers of conformal time:
\begin{equation}\label{longisoc}
\delta N \propto (-\eta)^{t},
\end{equation}
where the exponents $t$ are given by
\begin{eqnarray}\label{exponents}	
t_{\pm} &=& -{1\over 2}\left({5+3w\over 1+3w}\right) \pm \sqrt{{1\over 4}\left({5+3w\over 1+3w}\right)^2 + \kappa},
\end{eqnarray}
and $\kappa$ is given in Eq. (\ref{kappaeq}).
Note that for $\kappa > 0$, there is always a growing isocurvature mode.
If we don't want this mode to grow out of control, then $\kappa$ should not be too large and positive,
\begin{equation}\label{kllo}
-\infty < \kappa \ll 1. 
\end{equation}
In the following subsection we shall be more 
precise about the upper limit of this range (after discussing the overall normalization of the corresponding
power spectrum).
Note that for $\alpha=-c_{14}$, we have $\kappa=0$ and the dominant isocurvature 
mode stays constant on large 
scales, just like the adiabatic one. Hence, from the point of view of 
observability of isocurvature modes, the interesting range of parameters is around $\alpha \approx -c_{14}$.

From Eq. (\ref{isocmodes3}), the gravitational potentials for the isocurvature modes are given in terms of $\delta N$ by
\begin{equation}
\psi = -c_{13}\ c_t^2\ \delta N. \label{gdn}
\end{equation}
and 
\begin{equation}\label{anisocst}
{\psi-\phi \over \phi} \sim 1.
\end{equation}
Hence, isocurvature modes have sizable anisotropic stress. It is also straightforward to check from the relations
in the Appendix \ref{sec:long wavelength} that for the long wavelength isocurvature mode,
the velocity of the aether with respect to matter is given by
\begin{equation}\label{vandc}
v_i={\cal H}^{-1}\delta N,_i = c_t^{-2} C,_i,
\end{equation}
where in the first equality, we use Eq. (\ref{vpote}) and $C$ is the scalar aether perturbation in the 
longitudinal gauge. From Eq.~(\ref{efc}) in Appendix \ref{sec:adiabatic modes}, 
it is clear that at the time of a sudden transition in the equation of state parameter, the variable $C$ and its 
derivative remain continuous for the long wavelength isocurvature mode. This means that the velocity field  
matches trivially:
\begin{equation}
[v_i] = [v_i'] = 0,
\end{equation}
where the square brackets indicate the discontinuity at the time of the transition.
On the other hand, since the pressure changes abruptly at the transition, so does ${\cal H}'$, and therefore
the matching conditions for $\delta N$ are
${[\delta N] = 0}$, ${[\delta N'] = (3/2) [w]\ {\cal H} \delta N}$.

\subsection{Power spectra}

As shown in Subsection \ref{sec:scalar stability}, the variables $\zeta_a$ and $\delta N$ are
uncorrelated on subhorizon scales. Hence, from Eqs.~(\ref{comovz}) and (\ref{vacnorm}), 
it is clear that, at {\em short wavelengths}, the power spectra associated to  $\delta N$ and $\zeta$
are given by
\begin{subequations}
\begin{eqnarray}
{\cal P}_{\delta N}&=& {Z_N\over (2\pi)^2} \left({k\over a}\right)^2, \label{swps} \\
{\cal P}_{\zeta} &=& {\cal P}_{\zeta_a} + {\cal P}_{\delta N} = 
{Z_a +Z_N \over (2\pi)^2} \left({k\over a}\right)^2.
\end{eqnarray}
\end{subequations}
These spectra are valid for  $k\eta\gg 1$.\footnote{Here, and for the rest of this section, we shall
assume that the speed of propagation of aether is larger than or comparable to 1. This is
convenient so we do not have to introduce the scale of sound horizon crossing in the discussion of 
the adiabatic mode. Also, this assumption avoids the need of imposing the constraints due to Cherenkov 
radiation discussed in the Appendix.}


\subsubsection{Adiabatic modes}

For wavelengths comparable to the cosmological horizon,
$\delta N$ and $\zeta$ are coupled to each other, and their evolution will not have a 
simple form. 
Nonetheless, the evolution of $\delta N$ and $\zeta$ is again simple in the long wavelength limit,
as we saw in the previous subsection. In particular, $\zeta$ stays constant at long wavelengths. 
The power spectrum for $\zeta$ will be approximately equal to its value
at the time of horizon crossing, which we can estimate from (\ref{swps}) by setting $k/a = H$,
\begin{equation}\label{powzet}
{\cal P}_{\zeta} \sim 
{Z_a +Z_N \over (2\pi)^2} H^2,
\end{equation}
where $H^2$ is evaluated at the time of horizon crossing.

From  Eqs. (\ref{zsn}) and (\ref{eq:scalar cs})
we have 
\begin{equation}
Z_N+Z_a = \left(1-{\beta \epsilon c_t^2\over 2}\right) Z_N.
\end{equation}
Note that $Z_a$ is parametrically suppressed with respect to 
$Z_N$ by one power of aether parameters ${c_i\sim (M/M_p)^2 \ll 1}$ and by one power of the slow roll parameter 
$\epsilon$. Hence
\begin{equation}\label{psz}
{\cal P}_{\zeta} (k|\eta|\ll 1) \approx \left.{8 G_\mathrm{cos}^2\rho\over  3\epsilon}\right|_{\eta_k}
\left[1+O(\beta \epsilon c_t^2)\right],
\end{equation}
where $\rho$ the energy density and $\eta_k$ is the time of horizon exit during inflation. 
Up to the small corrections introduced by the fluctuation of the aether, which are controlled by $Z_a$,
this expression is the same as the one in Einstein gravity, with Newton's constant $G$ replaced with the effective 
Newton's constant $G_\mathrm{cos}$ which appears
in the Friedmann equation (\ref{eq:Friedman}). 

In summary, due to the smallness of the aether parameters $c_i$, 
the spectrum of primordial adiabatic modes does not change significantly with respect
to the case of standard Einstein gravity. As we saw in 
the previous subsection, 
in the presence of the aether the adiabatic modes do not have the properties generally 
attributed to adiabatic perturbations. In particular, from Eq. (\ref{asamu}), 
on super-horizon scales the non-decaying adiabatic mode has a non-vanishing anisotropic stress 
\begin{equation}
	\frac{\psi_\mathrm{adiab}-\phi_\mathrm{adiab}}{\phi_\mathrm{adiab}} \sim c_{13} \label{asam2}
\end{equation}
both in the matter and radiation dominated era.
It is easy to see from Eq. (\ref{eq:final C}) that for $\alpha+c_{14}=0$, the aether perturbation $C$ behaves exactly 
like a massless field which propagates at the speed $c_a$. Hence $C$ oscillates while its amplitude decays as the
inverse of the scale factor, $C \propto (1/a) e^{-i c_a k\eta}$. It then follows from  (\ref{eq:scalar stress})
that $\phi-\psi$ also decays in inverse proportion to the scale factor, and hence it is suppressed by a factor
of $a(t_k)/a(t_0)$, where $t_k$ is the time of horizon crossing. For modes which cross the horizon during the
matter era, this means that the effect is suppressed with distance as\footnote{Assuming that the aether parameters are small, these conclusions are easily extended to the case $\alpha\neq -c_{14}$.
In this case, Eq.~(\ref{eq:final C}) can be solved as the sum of the ``homogeneous" equation which is obtained by
ignoring terms proportional to $\phi$, plus the contribution of a particular ``inhomogeneous" solution.
The first one takes the form
$C_h \propto a^{-(1+d/2)} e^{i c_a k \eta}$,
where
\begin{equation}
d=c_{13}\ c_t^2\left(1+{\alpha\over c_{14}}\right).
\end{equation}
This leads to 
\begin{equation}\label{decay1}
	\frac{\psi-\phi}{\phi} \sim  c_{13} \left[(k t_0)^{-(2+d)}+ O(k t_0)^{-2}\right]
\quad   (t_{eq}  \ll k^{-1} \ll t_0).
\end{equation}
This applies to modes that entered the horizon during the matter era. For those
which crossed the horizon during the radiation era, the scaling is with one less power of $k$ in 
the denominator.}
\begin{equation}\label{decay2}
	\frac{\psi-\phi}{\phi} \sim  c_{13}\ (k t_0)^{-2} \quad   (t_{eq}  \ll k^{-1} \ll t_0, 
\quad \alpha\approx -c_{14}),
\end{equation}
where we adopt the standard convention $a(t_0)=1$.
On the other hand, for modes that crossed the horizon during the radiation era, the scaling is 
with $k^{-1}$. This is in agreement with the result that at small scales the post-Newtonian parameter 
$\psi/\phi$ equals one, as in general relativity \cite{Eling:2003rd}. Nonetheless, as a matter of principle, 
there could still be a distinct phenomenological signatures in the adiabatic sector imprinted on large scales.

Constraints on the ratio $\psi/\phi$ on cosmological scales have been derived 
under several different assumptions, using  combinations of different large scale structure probes  
\cite{Koivisto:2005mm, Daniel:2009kr,Giannantonio:2009gi,Bean:2009wj}. At present, however, the constraints are quite weak, 
and it appears that values of $\psi/\phi$ of order one are still consistent with the data. 

\subsubsection{Isocurvature modes}

Next, let us consider the spectrum of long wavelength isocurvature modes ${\cal P}_{\delta N}$. 
The phenomenological situation depends on whether 
$\alpha+c_{14}$ is positive or negative. 
If $\alpha < -c_{14}$, then $\delta N$ decays on superhorizon scales,
during and after inflation. Hence, these modes will remain insignificant with respect to the 
adiabatic ones. If $\alpha=-c_{14}$, then there is a constant non-decaying isocurvature mode, and 
$\delta N$ stays constant on superhorizon scales. Finally, for $\alpha > -c_{14}$ there is a growing 
mode and $\delta N$ can be very large at the 
time of re-entry even if it was small at the time of horizon exit.

Phenomenologically, the most interesting case seems to be the limit $|\alpha+c_{14}|\ll |c_{14}|$,
in which the supercurvature mode $\delta N$ stays approximately constant on large scales. Otherwise, 
either the mode is too suppressed to be of any significance, or it grows 
too fast to be compatible with observations.
In this case, the exponent $t_{\pm}$ for the dominant mode can be approximated by 
\begin{equation}
\hat t \approx {1+3w \over 5+3w}\ \kappa,
\end{equation}
where $\kappa$ is given in (\ref{kappaeq}),
and we have $|\hat t| \ll 1$ both during inflation and afterwards.
At the time of horizon crossing, the adiabatic and isocurvature
modes have comparable amplitudes, ${\cal P} \sim (2\pi)^{-2} Z_N H^2$, and these will remain roughly 
comparable throughout cosmic history up to the present time provided that $\hat t$ is sufficiently close to zero. 
In order to assess how small it would have to be, we can make a rough estimate of the evolution of the amplitude 
of $\delta N$ from the time of horizon crossing during inflation to the time of equality:
\begin{equation}
(\delta N)_\mathrm{eq} \sim Z_{N}^{1/2} H e^{-\hat t_i N} \left({\eta_\mathrm{eq}\over \eta_\mathrm{rh}}\right)^{\hat t_r} 
\sim Z_{N}^{1/2} H e^{(\hat t_r-\hat t_i)N}\lesssim 1.
\end{equation}
Here, the subindices $i$ and $r$ refer to inflation and radiation era respectively. Assuming 
$ Z_{N}^{1/2} H\sim 10^{-5}$, as follows from the normalization of the adiabatic modes, we find that for
\begin{equation}\label{smallt}
\hat t_r-\hat t_i \approx {1\over 3}\kappa_r \lesssim \ln(10^5)/N
\end{equation}
the perturbation $\delta N$ remains within the linear regime up to the time of equality of matter and radiation.
Here, we have neglected $\kappa_i$, which is suppressed with respect to $\kappa_r$ by a slow roll factor
(we are assuming that the aether parameters are the same today than they are during inflation), and
\begin{equation}
	N\sim 60
\end{equation}
is the number of e-foldings of inflation after the mode with co-moving wavenumber $k\sim \eta_\mathrm{eq}$ first
crossed the horizon. Note also that, according to (\ref{gdn}), the contribution of the isocurvature mode
to the gravitational potential is suppressed by $c_{13}$,
$$
\psi_\mathrm{isoc} = -c_{13}\ c_t^2\ \delta N \lesssim 10^{-5},
$$
where the last inequality is the observational bound on the gravitational potentials.
For $c_{13}\ c_t^2 \lesssim 10^{-5}$, $\psi_\mathrm{isoc}$ can remain small enough even if 
the inequality (\ref{smallt}) is saturated, so that $\delta N \sim 1$.
Also $\psi_\mathrm{isoc}$ becomes comparable to the contribution of the adiabatic mode $\psi_\mathrm{adiab}$, 
when the inequality $\kappa \leq -{3\ln|c_{13}\ c_t^2|/ 4N}$ is saturated, and combining with (\ref{smallt}), we require 
\begin{equation}
\kappa \lesssim {3\over N} \min\{\ln(10^5), -\ln|c_{13}\ c_t^2|\}.
\end{equation}

Let us estimate what the physical implications of the isocurvature perturbations might be. On one hand, they would
induce maximal anisotropic stress on large scales, as can be seen from Eq. (\ref{anisocst}), 
$$
{\psi_\mathrm{isoc}-\phi_\mathrm{isoc}\over \psi_\mathrm{isoc}}\sim 1,
$$
which means that there would be a sizable difference between the two gravitational potentials $\phi$ and $\psi$ provided 
that the contribution of the isocurvature mode is comparable to that of the adiabatic mode. For $\kappa=0$ we have $\psi_\mathrm{isoc}\sim c_{13}c_t^2 \delta N \sim c_{13}c_t^2 \psi_\mathrm{adiab}$, and so from (\ref{asam2})
both adiabatic and isocurvature modes contribute
to the anisotropic stress in a similar amount (unless $c_t^2$ is very large). However, if $\kappa$ is small and positive, then the isocurvature
mode grows on superhorizon scales, and will contribute more to the anisotropic stress than the adiabatic one.  
As argued in the previous subsection, the difference $1-(\phi/\psi)$
decays after horizon crossing, and so does its magnitude as a function of co-moving scale, which  roughly goes as $k^{-2}$ for modes 
which crossed the horizon during the matter era, and as $k^{-1}$ for modes that crossed before the time of equality
[see Eqs. (\ref{decay1}), (\ref{decay2})].

Another possible signature might be due to preferred-frame effects due to the motion of matter with respect 
to the aether \cite{Damour:1993ev,Graesser:2005bg}, such as a dipole anisotropy in the gravitational potential of 
massive bodies. The primordial perturbations cause the aether to point 
in different directions at different 
places in the observable universe. Hence, the velocity of matter with respect to the aether 
(and the corresponding gravitational dipole, for instance) would have a random distribution.  
From (\ref{vpote}) an isocurvature perturbation with wave-number $k$, induces a relative speed 
of the aether with respect to matter given by
$$
v = {k\over \mathcal{H}} \delta N.
$$
When the mode reenters the horizon during the radiation or matter era, at time $t_k\sim a/k$, we have 
$$
v \sim \delta N(t_k).
$$
This has to be compared to the peculiar velocities in bound objects at the same scale, which
is of order ${\zeta^{1/2} \sim 10^{-3}}$. Hence, the effect of the peculiar velocity of the aether will be
subdominant unless $\delta N$ has grown from the time of horizon exit, in such a way that at the time
of reentry it is at least of the order $\zeta^{1/2}$. This possibility exists, since we have seen that
$\delta N$ has a growing mode for $\alpha > -c_{14}$. Because of that, the velocity of the aether at
the time of horizon crossing could even approach moderately relativistic speeds without compromising 
the validity of the linear approximation and without contradicting current observations. 
[Note from Eqs. (\ref{gdn}) and (\ref{anisocst}), that the gravitational potentials along the 
isocurvature mode, and hence their effects on the CMB, are suppressed by a factor $c_{13}$, which can be very 
small]. 

The velocity field of the aether is strongly correlated with the amplitude of adiabatic modes,
since both have a common origin in the amplitude of the short wavelength  mode $\delta N$
when it first crosses the horizon during inflation. Should the velocity field of the aether be 
detected, such correlation would indicate that the velocity field has a primordial inflationary origin.

We may define a power spectrum ${\cal P}_v$ for the longitudinal velocity field of the aether through the
equation
\begin{equation}\label{longvel correlation}
	\langle v_i (\eta,{\bf k}) v_j (\eta,{\bf k}') \rangle\equiv \frac{2\pi^2}{k^3}
	{\cal P}_v(\eta,k) \,{k_i k_j \over k^2} \, \delta({\bf k}-{\bf k}').
\end{equation}
Note that at the time of horizon exit during inflation, we have
\begin{equation}
{\cal P}_v\sim {\cal P}_{\delta N}\sim {\cal P}_{\zeta}\sim 10^{-5}, 
\end{equation}
where the last estimate follows
from observations. 
However, since the perturbation $\delta N$ grows on large scales for
$0< \kappa \ll 1$, we can have 
${\cal P}_v\sim {\cal P}_{\delta N} \gg {\cal P}_{\zeta}$ at the time of horizon reentry.
As we shall see in the next section, vector perturbations can give an additional contribution to the velocity
field (which can of course be disentangled from the scalar isocurvature contribution from the fact 
that the corresponding velocity field is transverse). It turns out that the scalar 
component and transverse vector component of the velocity field obey the same equation of motion on large scales. 
Hence, we defer the discussion of the spectral properties of ${\cal P}_v$ on currently observable scales to the next section.

\section{Vector Perturbations}\label{sec:vector perturbations}

In a universe dominated by a scalar field there are no vector perturbations.  
Perfect fluids do support vector perturbations, but they decay as the universe expands. 
By contrast, the aether contains a massless vector field (under spatial rotations), 
which renders vector metric perturbations dynamical. 
For a certain range of parameters, these modes can grow on large scales, 
leading to potentially interesting signals, or to constraints on the parameters of Einstein-aether theories.

\subsection{Short wavelength stability}

As in the case of the tensor modes,  we can read off from the action for the vector perturbations whether 
the vector sector in Einstein-aether theories is both quantum and classically stable on short scales. 
Inserting the expansions 
(\ref{eq:metric perturbations}) and (\ref{eq:field perturbations}) into the action (\ref{eq:action}), 
expanding to quadratic order in the vectors ${\bf S}$ and ${\bf V}$, and using the background equations
of motion, we obtain the following Lagrangian 
for the vector perturbations 
\begin{equation}\label{eq:vector L(Q,V)}
\begin{split}
	L^{(2)}_{v}=\frac{M_{P}^{2}a^{2}}{2}
	\Big[&-c_{14}{\bf V}'^{2}
	 +\frac{1}{2}(1+c_{13})\partial_i{\bf Q}\cdot \partial^i{\bf Q}+\\
	&+c_{1}\partial_i{\bf V}\cdot \partial^i{\bf V}
	+c_{13} \partial_i{\bf Q} \cdot \partial^i  {\bf V}+  \\
	&+ \alpha \left(\mathcal{H}^{2}-\mathcal{H}'\right){\bf V}^{2}
	+c_{14}(\mathcal{H}^2+\mathcal{H}'){\bf V}^{2}\Big],
\end{split}
\end{equation}
where we have introduced the gauge-invariant combination
\begin{equation}\label{qdef}
{\bf Q}\equiv {\bf F}'-{\bf S}
\end{equation} 
(the vector perturbation $\bf V$ is also gauge-invariant).
Note that $\bf Q$ is not a bona-fide Lagrangian variable, 
since its definition (\ref{qdef}) relates it to the time derivative of $\bf F$. Hence, we 
shall merely think of it as shorthand for the right hand side of (\ref{qdef}).
Variation of Eq. (\ref{eq:vector L(Q,V)})  with respect to ${\bf S}$ 
gives the response of the metric to a given perturbation of the aether field,
\begin{equation}\label{eq:Q(V)}
	{\bf Q}=-{c_{13}}\ c_t^2{\bf V}.
\end{equation}
In the canonical (first order) formalism, this equation corresponds to the vanishing of the canonical momentum conjugate 
to $\bf F$, $\Pi_{\bf F}=0.$
Upon substitution of this constraint back into the first order Lagrangian, one is left with a 
Lagrangian for the single canonical pair formed by $\bf V$ and its conjugate momentum $\Pi_{\bf V}$. Rewriting this 
reduced Lagrangian back in 
second order form gives
\begin{equation}\label{eq:vector L}
\begin{split}
	\mathcal{L}^{(2)}_{v}=\frac{M_{P}^{2}}{2}
	\Big[&-c_{14}{\bf \xi}'^{2}
	+\alpha(\mathcal{H}^{2}-\mathcal{H}')\xi^{2}+\\
	&+c_{1}\left(1-\frac{c_{13}^{2}c_t^2}{2c_{1}}\right)
	\partial_i\xi \cdot  \partial^i\xi\Big],
\end{split}
\end{equation}
where, for convenience, we have introduced the rescaled variable
\begin{equation}\label{eq:xi}
	{\xi}_i \equiv a V_i.
\end{equation}

As in the tensor case, the absence of ghosts requires the coefficient in front of $\xi'^2$  in the Lagrangian 
(\ref{eq:vector L}) to be positive, 
and classical stability requires that the squared speed \cite{Jacobson:2004ts}
\begin{equation}\label{eq:vector c}
	c_v^2=\frac{c_{1}}{c_{14}}\left(1-\frac{c_{13}^{2}c_t^2}{2c_{1}}\right)
\end{equation}
be non-negative. Therefore, stability in the vector sector demands both
\begin{equation}\label{eq:vector stability}
c_{14}\leq 0 \quad \text{and} \quad  c_1\leq \frac{c_{13}^2}{2(1+c_{13})}.
\end{equation}

In a Minkowski background, the two modes in the vector sector are massless fields, which 
we may interpret as two of the Goldstone modes of the broken boost invariance. The broken 
generators transform as a spatial vector under the unbroken group of spatial rotations, so the 
corresponding Goldstone bosons transform as a vector. This can be decomposed into a transverse part 
and a longitudinal part. The longitudinal Goldstone (with helicity zero) is of course part
of the scalar sector, which we discussed in the previous section. 
It should be noted that Lorentz invariance is generically broken in any curved spacetime. For instance,
if the spacetime curvature is non-constant, the gradient $\nabla_\mu R$ defines a
non-zero vector field 
which is not invariant under Lorentz-transformations. What is particular about the Einstein-aether is that 
the breaking of Lorentz-invariance has physical consequences, namely, the existence of Nambu-Goldstone bosons, 
whose dispersion relations approach non-relativistic expressions in the high-momentum limit, 
and whose masses vanish in flat space.\footnote{In an arbitrary spacetime, 
the ``mass" of these bosons is non-zero, as illustrated in the case of a FRW universe 
by the contribution to the effective mass of the last term in the Lagrangian (\ref{eq:vector L}).}

\subsection{Solutions during Power-Law Inflation}

Variation of (\ref{eq:vector L}) with respect to $\xi$ leads 
to the equation of motion for the vector perturbations,
\begin{equation}\label{eq:xi motion}
	\xi_i''+c_v^2\,  k^{2}\, \xi_i
	+\frac{\alpha}{c_{14}}\left(\mathcal{H}^{2}-\mathcal{H}'\right)\xi_i=0.
\end{equation}
In terms of the original variable $V_i =\xi_i/a$, we have
\begin{multline}\label{efcv}
	V_i''+2\mathcal{H}V_i'+c_v^2\,  k^{2}\, V_i+\\
	\shoveright{}+\left[\left(1+\frac{\alpha}{c_{14}}\right)\mathcal{H}^2+
\left(1-\frac{\alpha}{c_{14}}\right)\mathcal{H}'\right]V_i=0.
\end{multline}
In a universe that undergoes power-law inflation (\ref{eq:power law inflation}) this can be solved in terms
of Bessel functions and we have
\begin{equation}\label{eq:solution V}
	{V}_i=
\frac{1}{2M_P}\sqrt{\frac{\pi}{-c_{14}}}\frac{(-\eta)^{1/2}}{a} H_\nu^{(1)}(-c_v k\, \eta)\, {e_i}.
\end{equation}
Here,
\begin{equation}\label{eq:nu}
\nu={t_+-t_- \over 2} = \sqrt{{1\over 4} -{\alpha q(q+1)\over c_{14} } },
\end{equation}
where $t_{\pm}$ is given in (\ref{exponents}).
The  parameter $q$ is defined in Eq. (\ref{eq:power law inflation}) and ${e}_i$ is 
a normalized transverse polarization vector, ${\bf e}\cdot {\bf k}=0$ and ${\bf e}^2=1$. For a given wave number ${\bf k}$, there are two such linearly independent polarizations, 
orthogonal to ${\bf k}$. We have chosen the amplitude of ${V}_i$ in Eq. (\ref{eq:solution V}) 
so that the solution has the appropriate normalization of a positive frequency mode 
in the limit $\eta\to -\infty$.  Note that the factor $q(q+1)$ is non negative if the null energy condition is satisfied ($w\geq -1$).

The long wavelength power spectrum of vector 
perturbations created during inflation is defined by
\begin{equation}\label{eq:vector correlation}
	\langle V_i(\eta,{\bf k}) V_j (\eta,{\bf k}') \rangle\equiv \frac{2\pi^2}{k^3} \mathcal
	{\cal P}_V(k,\eta) \,{\Pi}_{ij} \, \delta({\bf k}-{\bf k}'),
\end{equation}
where
$\Pi_{ij}=\delta_{ij}-k_i k_j/k^2$ projects onto the subspace orthogonal to ${\bf k}$.
At the time of reheating $\eta_{\rm rh}$, we have
\begin{subequations}\label{eq:vector power}
\begin{eqnarray}
        \mathcal {P}_{V}^\mathrm{rh}(k)&=& \mathcal{A}_V^2 \times \left(\frac{k}{k_N}\right)^{n_v},\\
	\quad n_v&\equiv& 3-2\nu, \label{eq:vector spectral} \\
	\mathcal{A}_V^2&\equiv& \frac{H_\mathrm{rh}^2}{M_P^2}
	\left(\frac{-1}{c_{14}q^2}\right)\frac{\Gamma^2(\nu)}{(2\pi)^3}
	\left(\frac{c_v}{2}\right)^{-2\nu}
	\exp\left(\frac{N n_v}{q}\right). \nonumber\\
\end{eqnarray}
\end{subequations}
In  Eqs. (\ref{eq:vector power})  $H_\mathrm{rh}$ is the value of the Hubble constant 
at the end of inflation, and $k_N$ is the mode that crossed the cosmic horizon $N$ e-folds before the end of 
inflation ($|k_N \eta_N|=1$). 
For the mode that is entering the horizon today, the value of $N$ depends  logarithmically on the unknown 
reheating temperature, 
and typically equals $50$ to $70$ e-folds (see for instance \cite{Dodelson:2003vq}.) 
It is important to realize that the time at  which the spectrum is evaluated matters, 
since the vector modes do not freeze out at horizon crossing. The superscript 
``$\mathrm{rh}$" is meant to imply that the power spectrum describes the amplitude of the modes just before 
the end of inflation. 
Likewise, we may define the spectrum of the corresponding metric perturbation, which according to 
Eq. (\ref{eq:Q(V)}) is given by
\begin{subequations}\label{eq:power Q}
\begin{eqnarray}
	\mathcal{P}_Q^\mathrm{rh}(k)&=&\mathcal{A}_Q^2 \times \left(\frac{k}{k_N}\right)^{n_v},\\
	\mathcal{A}_Q^2&=& c_{13}^2c_t^2 \mathcal{A}_V^2.
\end{eqnarray}
\end{subequations}
In the limit in which de Sitter inflation is approached ($q\to -1$) the index $\nu$ tends to  $1/2$, 
so the amplitude of long-wavelength perturbations is proportional to $\exp(-2N)$. Hence,  velocity
perturbations on observationally accessible scales are very small in this limit
\cite{Lim:2004js,Li:2007vz}. 
On the other hand, in typical inflationary models $q$ differs from $-1$ and the rate of
decay can be smaller. In fact, if
\begin{equation}
	\alpha>\frac{-2c_{14}}{q(q+1)}=\frac{-2c_{14}}{\epsilon}(1-\epsilon)^2\
\end{equation}
the combination  $3-2\nu$ would be negative, and long-wavelength perturbations would be amplified 
exponentially with $N$, in stark contrast with the de Sitter case. 
It turns out, however, that we do not need to deviate much from $\nu=1/2$ in order to have an observable signal.
As we shall see, even if the long wavelength velocity field is very tiny at the end of inflation, it may resurface from 
obscurity during the radiation and matter era, so that it can be quite sizable at the moment of horizon reentry.

Indeed, the behaviour of long wavelength vector perturbations is completely analogous to that of the 
scalar component of the velocity field $v_i$ which we studied in the previous section. To see this,
we note that in the long wavelength limit, Eq. (\ref{efcv}) is the same as Eq. (\ref{efc}) for the 
isocurvature perturbation. The latter is written in terms of the variable $C$ in the longitudinal gauge, which according
to Eq. (\ref{vandc}) is proportional to the longitudinal velocity field of the aether $v_i$.
Hence, on superhorizon scales, longitudinal and transverse velocity fields satisfy the same equation of motion
Eq. (\ref{efcv}). In particular
\begin{equation}
V \propto v \propto (k/a){\cal H}^{-1} \delta N \propto \eta^{1-q+t_\pm}\quad\quad (k\eta\ll 1),
\end{equation}
where $t_\pm$ is given in (\ref{exponents}).
For $\alpha = -c_{14}$, the velocity field decays exponentially during inflation. Nonetheless, as we saw in the 
previous section, the isocurvature perturbation $\delta N$ stays frozen on superhorizon scales (except at 
the transitions where the equation of state changes, where the dominant mode changes also by factors of order one).
This can lead to a sizable velocity field $v$ of order $\delta N$ at horizon reentry. The overall normalization
of $v$ and $V$ is different, but it is clear that the relative size of $V$ and $v$ at horizon reentry is determined
by their relative size at the time when they exit  the horizon during inflation.
In other words, the spectra of long wavelength modes are related by
\begin{equation}\label{comparison}
{{\cal P}_v(\eta,k)\over  {\cal P}_V(\eta,k)} = {{\cal P}_v(\eta_k,k)\over  {\cal P}_V(\eta_k,k)} \sim
Z_N {c_{14}\over M_{P}^2} \sim {c_{14} \over \epsilon},
\end{equation}
where the relative normalization can be read off from the corresponding short wavelength actions,
and $\epsilon$ is the slow roll parameter during inflation. 
Note that $\epsilon$ is of order of a few percent, while the Einstein aether parameters
such as $c_{14}\sim (M/M_P)^4$ are suppressed by the square of the symmetry breaking scale over the Planck 
scale. Unless $M$ and $M_P$ are very close, we expect $c_{14}\ll \epsilon$. Therefore, parametrically,
we expect that the transverse vectors may give a much bigger contribution than the scalar isocurvature modes.
For that reason, it is very important to assess their impact on observables such as the CMB, as we do in 
Subsection \ref{cmbvectors}.

One may worry that if the primordial amplitude of vectors due to quantum fluctuations generated at horizon
crossing decays during inflation, then it may be insignificant compared with the contribution of non-linear 
effects which source the vector modes 
at later times. However, in order to construct a vector from a quadratic expression involving scalars and tensors,
it is necessary to use at least one derivative. Because of that, the terms which may source the vectors from
the scalar and tensor sector are momentum suppressed, and hence they also decay in inverse proportion to the
scale factor. We conclude that if the initial amplitude of the vectors at horizon crossing is sizable,
compared to that of scalars and tensors, then we can safely use linear evolution in order to determine
its amplitude at the end of inflation (even if that amplitude is exceedingly small).
Vector modes can still grow during radiation and matter domination from that initial tiny amplitude, 
so that their effect on cosmological observables may be important.

\section{Vector contribution to CMB temperature anisotropies}\label{ncmbs}

As we discussed above, the power spectrum of the transverse velocity field may easily dominate over that
of  the longitudinal component. It is therefore of interest to determine the imprint that this
power spectrum may have on CMB observations, to which this section is devoted.

\subsection{Solutions During Radiation and Matter Domination}
\label{sec:vectors deceleration}

In order to find the power spectrum of the vector modes at the time of recombination, we must first evolve it
from the time of thermalization through the radiation and matter dominated epochs.
For a set of perfect fluids which do not interact with each other, 
the conservation equation $\nabla_\mu T^{\mu\ (k)}_i=0$ holds for each fluid component, which we label by $(k)$.
This leads to a homogeneous equation for the gauge-invariant 
velocity perturbation $\delta u_i^{(k)}$ that does not contain metric perturbations,
\begin{equation}
	\frac{\partial}{\partial\eta}\left[a^3(\rho_k+p_k)\delta u^{(k)}_i\right]=0\label{velpert},
\end{equation} 
from where it follows that $\delta u_i/a \propto 1/[a^4 (\rho+p)]$. Eq.~(\ref{velpert}) tells us that 
if $\delta u_i=0$ initially, 
then it will not be generated as long as the perfect fluid description is valid. Furthermore, $\delta u_i/a$ 
decreases during cosmological 
evolution, except in the radiation dominated stage, where it stays constant.  Hence, in what follows, we assume that 
$\delta u^{(k)}_i=0$ for matter and radiation. Then, the only contribution to the metric perturbations stems from the 
aether, 
and it can be shown that the equations of motion in the vector sector are still given by Eqs. (\ref{eq:Q(V)}) 
and (\ref{eq:xi motion}).

The general solution of Eq. (\ref{eq:xi motion}) during 
a stage of cosmic expansion in  which $a\propto \eta^q$ is proportional to a linear 
combination of Bessel functions. For our present purposes, it will suffice to work 
with the long-wavelength approximations
\begin{equation}\label{eq:vector long}
	\xi=\mathcal{A}_+ \, \left(\frac{\eta}{\eta_*}\right)^{{1\over 2} +\nu}
	+\mathcal{A}_-\, \left(\frac{\eta}{\eta_*}\right)^{{1\over 2} -\nu},
\end{equation} 
where $\nu$ is given by Eq. (\ref{eq:nu}) and $\xi_i\equiv \xi \cdot e_i$. For real values of $\nu$, $\mathcal{A}_+$ is the amplitude of the dominant mode, 
and $\mathcal{A}_-$ is the amplitude of the subdominant mode.

In order to   determine the mode amplitudes $\mathcal{A}_+^r$ and $\mathcal{A}_-^r$ during radiation domination, we 
simply demand that $\xi$ and its time derivative be continuous at a sudden transition 
from inflation to radiation domination. We expect this approximation to be valid for scales much longer than the duration of reheating.  
Proceeding in this manner, and dropping the contribution from the subdominant mode we find that the amplitude of the dominant mode changes  
during reheating by a factor
\begin{equation}\label{eq:vector short}
	\frac{\mathcal{A}^r_+}{\mathcal{A}^i_+}=\frac{\nu^i+\nu^r}{2 \nu^r},
\end{equation}
where the superscripts label the expansion epoch ($i$ for inflation and $r$ for radiation domination), and the subscripts label the different modes ($+$ for the dominant mode and $-$ for the subdominant one.) 

Eq. (\ref{eq:xi motion}) has an exact solution at long-wavelengths during radiation 
and matter domination, which we can use to determine the change in the mode amplitudes 
during the transition from radiation to matter domination. Since this change is typically 
of order one, we shall neglect it, and assume that the amplitude of the growing mode at the 
transition remains unchanged. 

Once a mode enters the ``sound horizon", $c_v k \eta=1$, the field starts oscillating. 
In the limit $c_v k\eta\gg1$, the solution of Eq. (\ref{eq:xi motion}) that approaches the 
growing mode at early times is 
\begin{subequations}
\begin{eqnarray}
	\xi(\eta)&=&\mathcal{A}_+\, \mathcal{C}_\mathrm{osc} \cos\left[c_v k \eta-\varphi\right], \quad
	\text{where} \\
	\mathcal{C}_\mathrm{osc}&=&\frac{c_{13}(\nu+1)}{\sqrt{\pi}}
	\left(\frac{2}{c_v k \eta_*}\right)^{{1\over 2}+\nu},
\end{eqnarray}
\end{subequations}
and $\varphi$ is  $k$-independent phase. Note that the amplitude of the oscillations $\mathcal{A_+}\mathcal{C}_\mathrm{osc}$ is roughly the value of $\xi$ at horizon entry. 

Collecting then the results from Eqs. (\ref{eq:xi}), (\ref{eq:vector long}) and (\ref{eq:vector short}) we find that during matter domination the transfer function for the vector perturbations $\mathcal{T}_k$, which we implicitly define by the relation   ${{\bf Q}(\eta)=\mathcal{T}_k(\eta) {\bf Q}(\eta_\mathrm{rh})}$ is 
\begin{subequations}\label{eq:total vector transfer}
\begin{equation}\label{eq:vector transfer}
	\mathcal{T}_k\approx T\times
	\begin{cases}[1.6]
	\left(\dfrac{a_\mathrm{eq}}{a_\mathrm{rh}}\right)^{{1\over 2}+\nu^r}
	\!\!\left(\dfrac{\eta}{\eta_\mathrm{eq}}\right)^{{1\over 2}+\nu^m},
	 &\!\!{ k\eta_\mathrm{eq}} \ll k\eta\ll  c_v^{-1}, \\
	 \left(\dfrac{a_\mathrm{eq}}{a_\mathrm{rh}}\right)^{{1\over 2}+\nu^r}
	\!\! \!\mathcal{C}_\mathrm{osc}^m \cos(c_v k \eta),
	&\!\! k\eta_\mathrm{eq}\ll c_v^{-1} \!\ll k \eta, \\
	\mathcal{C}_\mathrm{osc}^r \cos(c_v k \eta),
	&\!\!  c_v^{-1}\ll k\eta_\mathrm{eq} \ll k\eta,
\end{cases}
\end{equation}
where 
\begin{equation}
	T=\frac{a_\mathrm{rh}}{a}\frac{\nu^i+\nu^r}{2\nu^r}
\end{equation}
\end{subequations}
and  in  $\mathcal{C}^{p}_\mathrm{osc}$ the transition time $\eta_*$ equals $\eta_\mathrm{rh}$ for 
$p=r$ and $\eta_\mathrm{eq}$ for $p=m$.  The first line  in Eq. (\ref{eq:vector transfer}) holds 
for those modes that have not entered the sound horizon at time $\eta$.  The second applies to those which 
enter between the time $\eta_\mathrm{eq}$ of
equality between matter and
radiation densities and time $\eta$, and the third one to the ones which enter between 
reheating and the time of equality.  
The power spectrum of $Q$  at any time after reheating is given by
\begin{equation}\label{eq:transfer power}
 	\mathcal{P}_Q(\eta)=|\mathcal{T}_k(\eta)|^2 \mathcal{P}_Q^\mathrm{rh}.
\end{equation}

\subsection{Impact on the CMB}\label{cmbvectors}

We derive in Appendix \ref{sec:vector anisotropies} the contribution of vector perturbations to the 
temperature anisotropies in the cosmic microwave background radiation. In order to determine the 
angular power spectrum and relate it to the primordial spectrum, it is convenient to rewrite 
Eq.~(\ref{full2}) in Fourier space,
\begin{multline}\label{eq:vector delta F}
\left(\frac{\delta T}{T_0}\right)_0 = 
	e^{-i {\bf k}\cdot {\bf l} \, \eta_0}
	 \int d^3 k
	\Bigg[{\bf l}\cdot {\bf Q}(\eta_\mathrm{dec},{\bf k})
	\exp\left(i {\bf k}\cdot {\bf l} \, \eta_\mathrm{dec}\right)+\\
	{}+ \int\limits_{\eta_\mathrm{dec}}^{\eta_0} d\eta\, {\bf l} \cdot  {\bf Q}'(\eta,{\bf k})
	\exp\left(i {\bf k}\cdot {\bf l}  \, \eta\right) \Bigg].
\end{multline}
The contribution of the two terms on the right-hand-side 
of (\ref{eq:vector delta F}) is similar to that of scalar perturbations. The first term is the analogue of the Sachs-Wolfe effect, which relates the temperature anisotropies to the state of the perturbations at last scattering. The second term is the analogue of the integrated Sachs-Wolfe effect, which takes into account the change of the metric potentials along the line of sight, and vanishes if the latter are constant. 

The angular power spectrum $C_\ell$ is defined by the relation
\begin{equation}\label{eq:definition Cell}
\left\langle \frac{\delta T}{T_0} (\hat{n}) \frac{\delta T}{T_0}(\hat{m}) \right\rangle
\equiv \sum_\ell C_\ell \frac{2\ell+1}{4\pi}P_\ell (\hat{n}\cdot \hat{m}),
\end{equation} 
where $\hat{n}$ and $\hat{m}$ are two directions on the sky, and the $P_\ell$ are Legendre polynomials.   
Because scalar, vector and tensor perturbations are uncorrelated, their contributions to the temperature 
anisotropies add in quadrature,  $C_\ell=C_\ell^s+C_\ell^V+C_\ell^h$.  
Inserting Eq. (\ref{eq:vector delta F}) into the left-hand-side of Eq. (\ref{eq:definition Cell}), 
using Eq. (\ref{eq:vector correlation}), and comparing to the right-hand-side of Eq. 
(\ref{eq:definition Cell})  we find after some work that the contribution of vector perturbations to 
the angular power spectrum is given by
\begin{subequations}\label{eq:Cell vector}
\begin{eqnarray}
	C_\ell^V&=&4\pi \ell (\ell+1) \int \frac{dk}{k} \, \mathcal{P}_Q^\mathrm{rh} \,
	\left|\mathcal{N}-\mathcal{I}\right|^2, \\
	\label{eq:non-integrated}
	\mathcal{N}&\equiv&   
	\mathcal{T}_k\left(\eta_\mathrm{dec}\right)
	\frac{j_\ell(x_\mathrm{dec})}{x_\mathrm{dec}}, \\
	\label{eq:integrated}
	\mathcal{I}&\equiv&
	\int\limits_{0}^{x_\mathrm{dec}} dx \, \frac{d\mathcal{T}_k}{dx}
	\frac{j_\ell(x)}{x}.	
\end{eqnarray}
\end{subequations}
In these equations $x\equiv k(\eta_0-\eta)$, the $j_\ell$ are the spherical Bessel functions of 
the first kind, the primordial spectrum is given by Eqs. (\ref{eq:power Q}) and the transfer 
function $\mathcal{T}_k$ by Eqs. (\ref{eq:total vector transfer}). After an integration by parts, Eqs. (\ref{eq:Cell vector}) 
agree with the expression derived in \cite{Hu:1997hp} by somewhat different methods.
Note that $x$ is the ratio of the comoving 
distance to time $\eta$ divided by the wavelength of the perturbation $1/k$. Thus, $x$ is the inverse of the  
angle that an object of comoving size $1/k$ at (comoving) distance $\eta_0-\eta$  would subtend on the sky 
at time $\eta_0$.  It will be useful to consider separately those modes that enter well before 
and well after decoupling. These contribute to the temperature anisotropies, respectively, on small and 
large angular scales.

The structure of expressions (\ref{eq:Cell vector}) still reflects the two contributions to the 
temperature anisotropies we mentioned above. The value of $\mathcal{N}$ captures the analogue of the 
Sachs-Wolfe effect, while the value of the integrated term $\mathcal{I}$ captures the analogue of the 
integrated Sachs-Wolfe effect. For scalar perturbations, the Sachs-Wolfe effect is dominant except on the 
largest scales, because the gravitational potential remains constant until relatively recently. For vector 
perturbations however, this is not always the case.  To see this, it is useful to realize that we can employ 
the same approximations developed to study the contribution of tensor modes to the temperature anisotropies 
\cite{Mukhanov:2005sc}.  We begin by noting that, for $\ell\gg 1$, the Bessel function can be 
approximated by \cite{bessel}
\begin{equation}
	j_\ell(x)\approx
	\begin{cases}
	0, & \!\!\! x<\ell+\frac{1}{2} \\
	\dfrac{\cos \left[\sqrt{y}
	-(\ell+\frac{1}{2})\arccos\left(\frac{\ell+\frac{1}{2}}{x}\right)
	-\frac{\pi}{4}\right]}{x^{1/2} y^{1/4}}, \, & \!\!\! x>\ell+\frac{1}{2}
	\end{cases}
\end{equation}
where $y\equiv x^2-(\ell+1/2)^2$. Because the integrand is negligible for $x<\ell+1/2$, 
only modes that have entered the horizon by today, $x_\mathrm{dec}\approx k\eta_0\gtrsim\ell+1/2$ 
can contribute to the temperature anisotropies on angular scales $\ell\gg 1$. 

\subsubsection*{Large Angular Scales}

Large angular scales correspond to modes that cross the sound horizon after decoupling. Let us estimate the contribution of the integrated term $\mathcal{I}$ on those scales first. From Eq. (\ref{eq:vector transfer}), the derivative of the transfer function $d\mathcal{T}_k/dx$ is 
dominant during the interval $x_k \lesssim x \lesssim x_\mathrm{deq}$, where  $x_k\equiv k(\eta_0-\eta_k)$
corresponds to the time of sound horizon crossing $\eta_k\equiv 1/(c_v k)$. After horizon crossing, 
this function oscillates with period $2\pi/c_v$ and a slowly varying amplitude. On the other hand, the ratio
$j(x)/x$ changes slowly in the interval $\Delta x \lesssim 1 \ll l \lesssim x$. Assuming that $c_v$ is not 
much smaller than $1$, we have ${\Delta x = x_\mathrm{deq}-x_k \approx k\eta_k = 1/c_v \lesssim 1}$, and we
can pull the factor $j(x)/x$ out of the integral. 
What remains is a boundary term that can be readily evaluated,
\begin{equation}\label{eq:integrated approximation}
\mathcal{I}\equiv \int\limits_{0}^{x_\mathrm{dec}} dx \, \frac{d\mathcal{T}_k}{dx}
	\frac{j_\ell(x)}{x}\approx 
	\frac{j_\ell(x_\mathrm{dec})}{x_\mathrm{dec}}
	\left[\mathcal{T}_k(\eta_\mathrm{dec})-
		\mathcal{T}_k(\eta_*)\right],
\end{equation}
where the effective lower limit of integration $\eta_*$ is of order $\eta_k$.

The dominant term in the right hand side of Eq.~(\ref{eq:integrated approximation}) depends on whether the long wavelength
modes are decaying (which happens for $\nu^m < 3/2$), or growing (${\nu^m> 3/2}$) 
before horizon crossing. In the first case 
we have
\begin{equation}\label{eq:approximation decaying}
	\mathcal{I}\approx 
	\frac{j_\ell(x_\mathrm{dec})}{x_\mathrm{dec}}
	\mathcal{T}_k(\eta_\mathrm{dec}),
\end{equation}
while in the second case we have instead
\begin{equation}\label{eq:approximation growing}
	\mathcal{I} \sim -
	\frac{j_\ell(x_\mathrm{dec})}{x_\mathrm{dec}}
	\mathcal{T}_k(\eta_k). 
\end{equation}
Therefore, comparison of Eqs. (\ref{eq:approximation decaying}) and (\ref{eq:approximation growing}) with (\ref{eq:non-integrated}) 
shows that  $\mathcal{N}$ and $\mathcal{I}$ are of the same order if there is no growing mode during matter domination 
(${1\over 2}+\nu^m \leq 2$), 
and that $\mathcal{I}\gg \mathcal{N}$ otherwise ($\nu^m> 3/2$).

We are ready now to calculate the angular power spectrum on large angular scales,  
which are dominated by modes that crossed the vector sound horizon after decoupling 
($c_v k \eta_\mathrm{dec}<1$).  The relevant expression for the transfer function is 
given by the first line of Eq. (\ref{eq:vector transfer}) at $\eta=\eta_\mathrm{dec}$. 
If there is no growing mode during matter domination, the contributions from the integrated and non-integrated terms in (\ref{eq:Cell vector}) are roughly equal, and substituting 
Eqs.  (\ref{eq:power Q}) and  (\ref{eq:transfer power}) into (\ref{eq:Cell vector}) we get
\begin{equation}\label{eq:Cell intermediate short}
	C^V_\ell\approx 4\pi \mathcal{A}_Q^2  \,
	\mathcal{T}^2_k(\eta_\mathrm{dec})
		\int \frac{dx_0}{x_0}
		\frac{\ell(\ell+1)}{x_0^2} x_0^{n_v}
		 j_\ell^2(x_0),
\end{equation}
where we have chosen $k_N$ in Eq. (\ref{eq:vector power}) to be the mode that is crossing 
the horizon today, $k_N \eta_0=1$, and used that $x_\mathrm{dec}\approx x_0\equiv k\eta_0$. 

From reheating to the time of decoupling, the amplitude of the vector modes changes
by $\mathcal{T}_k(\eta_\mathrm{dec})$. The spectrum is thus proportional to the square 
of the transfer function times the primordial amplitude $\mathcal{A}_Q^2$. This factor is independent 
of angular scale, since we are taking the long wavelength limit. The angular dependence in the last 
equation can be estimated as follows.  The Bessel function is negligible for $x_0\lesssim \ell$,  
and rapidly decays at $x_0>\ell$, so the anisotropies are dominated by $x_0\sim \ell$. In the integrand, 
the maximum of the Bessel function is of order $1/x_0$, and the two remaining factors of $x_0$ in 
the denominator ``cancel" the enhancement proportional to $\ell(\ell+1)$ one 
would otherwise have. In summary, we have $C^V_\ell \propto \ell^{n_v-2}$.

More precisely, since the period of oscillations of the Bessel function is much shorter that any 
other characteristic scale in the integrand of Eq. (\ref{eq:Cell intermediate short}), we may replace the oscillations with their average, $1/2$.  If the spectral index $n_v$ is not too blue ($n_v<4$), the dominant contribution to the integral is given by the value of the integrand at $x\approx \ell$, 
so the angular power spectrum becomes (for $\nu^m\leq 3/2$)
\begin{equation}\label{eq:Cell vector final long}
\begin{split}
	&\ell(\ell+1) C^V_\ell \sim \\ 
	&2\pi  \mathcal{A}_Q^2  
	\left(\frac{\nu^i+\nu^r}{2\nu^r}\right)^2  \!
	\left(\frac{a_\mathrm{dec}}{a_\mathrm{rh}}\right)^{2\nu^r-1}\!
	\left(\frac{a_\mathrm{dec}}{a_\mathrm{eq}}\right)^{\nu^m-2\nu^r-{1\over 2}}
	\! \ell\, {}^{n_v}. 
\end{split}
\end{equation}
This expression is valid for those scales that entered the vector horizon after recombination, 
which corresponds to $\ell\lesssim 50/c_v$ (for a $\Lambda$CDM model with $\Omega_\Lambda=0.7$.) 

If there is a growing mode during matter domination, the integrated term (\ref{eq:integrated}) yields the dominant contribution to the temperature anisotropies. Proceeding along the same lines as above, we find that in this case the angular power spectrum is (for $\nu^m> 3/2$)
\begin{equation}\label{eq:Cell final growing}
\begin{split}
&\ell(\ell+1) C^V_\ell \sim \\
	&2\pi \mathcal{A}_Q^2  
	\left(\frac{\nu^i+\nu^r}{2\nu^r}\right)^2 \!
	\left(\frac{a_\mathrm{eq}}{a_\mathrm{rh}}\right)^{2\nu^r-1}\!
	\left(\frac{a_0}{c_v a_\mathrm{eq}}\right)^{\nu^m-{3\over 2}}
	\ell\, {}^{n_v+3-2\nu^m}.
\end{split}
\end{equation}
Of course, in the crossover case  $\nu^m =3/2$ the two angular power 
spectra (\ref{eq:Cell vector final long}) and (\ref{eq:Cell final growing}) agree.
 
\subsubsection*{Small Angular Scales}

For the scales that enter the vector horizon before decoupling a precise estimate of the integrated term 
in  (\ref{eq:integrated}) becomes more difficult. For these modes the derivative of the transfer function 
is an oscillatory function, whose  amplitude decreases in time. Hence, it is most important at earlier 
times $x\approx x_\mathrm{dec}$ and sharply decays within an interval $\Delta x=k \eta_\mathrm{dec}$. 
Whereas the latter is small for modes that cross after decoupling, for those scales that enter the horizon well 
before that time $\Delta x=k\eta_\mathrm{dec}$ is large, and  the approximation 
of a constant Bessel function that led to (\ref{eq:approximation decaying}) breaks down. Nonetheless,  
if we are interested in the order of magnitude of the Bessel function, and not in
the oscillations, we can still 
use Eq. (\ref{eq:approximation decaying}), since the ``amplitude" of $j_\ell(x)$ only changes 
significantly within  $\Delta x=x_\mathrm{dec}\approx k\eta_0\gg k \eta_\mathrm{dec}$.  
In that case, the integrated term is at most of the same order of 
magnitude as the non-integrated one, and Eqs. (\ref{eq:Cell vector}) imply that the 
temperature anisotropies at any given angular scale will depend on the 
vector anisotropies on the appropriate comoving distance at the time of decoupling.  

Under the assumption that $\mathcal{N}$ and $\mathcal{I}$ are of the same order, 
the angular power spectrum on small angular scales  can be now calculated as before.  
For simplicity, let us concentrate on relatively small scales, which cross the sound horizon
before equality of matter and radiation densities. The relevant expression for the transfer function 
is given by the third line of Eq. (\ref{eq:vector transfer}). Substituting Eqs.  
(\ref{eq:power Q}) and (\ref{eq:vector transfer}) into (\ref{eq:Cell vector})  we obtain
\begin{multline}\label{eq:Cell intermediate long}
	C^V_\ell\approx 4\pi\,  \mathcal{A}_Q^2
	\mathcal{T}^2_{\eta^{-1}_0}(\eta_\mathrm{dec}) \times \\
		{}\times\int \frac{dx_0}{x_0}
		\frac{\ell(\ell+1)}{x_0^2} x_0^{n_v-2\nu^r-1} \cos^2\left(\frac{c_v \eta_\mathrm{dec}}{\eta_0} x_0\right)
		j_\ell^2(x_0).
\end{multline}
As before, the power is proportional to the primordial contribution $x^{n_v}$ times an additional factor $x^{-2\nu^r-1}$, 
which  just reflects that modes enter the horizon at different times, and thus evolve differently. 
The cosine represents a snapshot of the ``acoustic oscillations" of the vector perturbations at decoupling. 

Before we proceed, we should mention an additional effect that 
influences the anisotropies on very small scales. So far, we have been assuming that the decoupling 
of the photons from the baryons is instantaneous.   This is an accurate approximation for scales in 
which the argument of the cosine in Eq. (\ref{eq:vector transfer}) does not change much during 
the duration of decoupling. On scales in which the cosine does change significantly, the spread in time 
at which a photon last scatters dampens the fluctuations by an exponential 
factor $\exp\left(-x_0^2/2 \sigma^2\right)$ \cite{Mukhanov:2005sc}. A similar suppression is also due 
to  Silk-damping, which originates from the breakdown of the tight-coupling approximation at scales of 
the order of the mean free path of photons in the plasma. For the observed values of the cosmological parameters, 
both effects yield an overall value of the suppression scale $\sigma\approx 500$.

\begin{figure}
  \begin{center}
 \includegraphics[width=9cm]{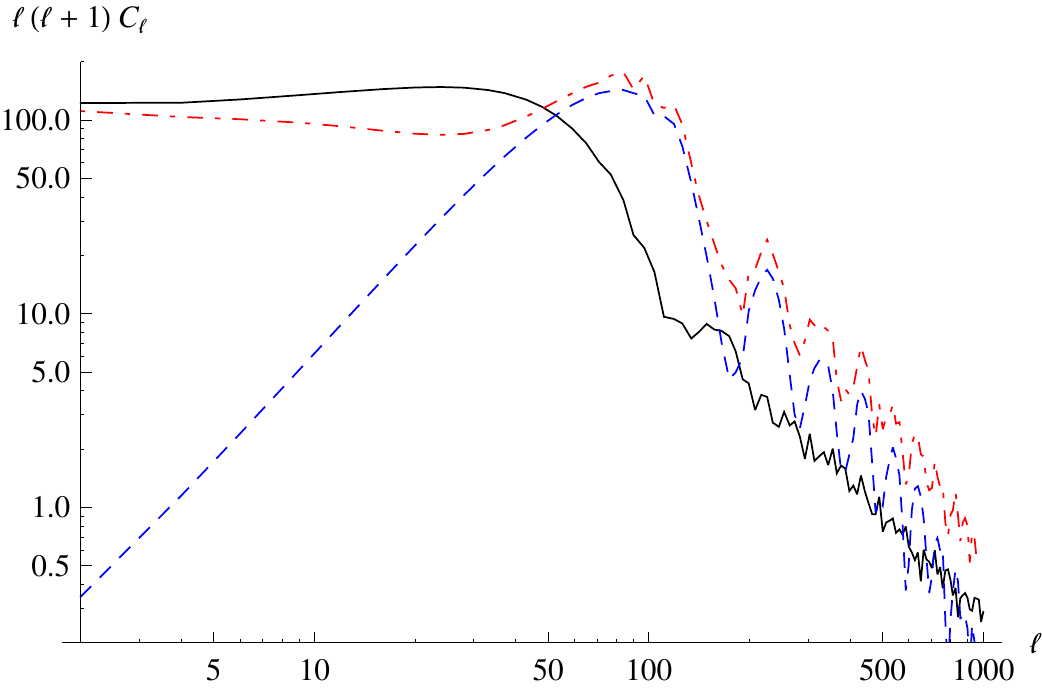}
  \end{center}
  \caption{The contribution of vector modes to the temperature anisotropy power spectrum for models with $c_{14}=-\alpha$ and $c_v=1$ (the normalization is arbitrary.) The black (continuous) curve is the added contribution of the integrated and non-integrated terms, 
Eq. (\ref{eq:Cell vector}). The contribution of the integrated term alone is shown in red (dashed-dotted), while the contribution of the non-integrated term alone is shown in blue (dashed).  On large angular scales, the spectrum is well approximated by 
Eq. (\ref{eq:Cell final growing}). On small scales, Eq. (\ref{eq:Cell vector short final}) gives a qualitatively correct approximation.}
   \label{fig:Cell}
\end{figure}

Due to the exponential damping, the integral over modes converges for any power-law spectrum.  As before,  
if the effective spectral index is not too blue, the dominant contribution to the integral is given by the value 
of the integrand at $x_0\approx \ell$,  so the angular power spectrum becomes 
\begin{equation}\label{eq:Cell vector short final}
\begin{split}
	&\ell(\ell+1) C^V_\ell \sim \\
	& 2c_{13}^2(\nu_r+1)  \mathcal{A}_Q^2 
	\left(\frac{\nu^i+\nu^r}{2\nu^r}\right)^2
	\left(\frac{a_\mathrm{rh}}{a_\mathrm{dec}}\right)^2 
		\left(\frac{2\eta_0}{c_v\eta_\mathrm{rh}}\right)^{2 \nu^r-1} \times\\
		&\times \ell\, {}^{n_v-2\nu^r-1}\,
		\cos^2\left(\frac{c_v \eta_\mathrm{dec}}{\eta_0} \ell \right)
		e^{-\ell^2/2\sigma^2}.
		\end{split}
\end{equation}
This equation is qualitatively valid at small angular scales, those corresponding to modes that crossed the horizon before equality, $\ell\gtrsim 120/c_v$. The acoustic oscillations subtend an angle  $c_v \eta_\mathrm{dec}/\eta_0$ on the sky, the ratio of the comoving size of the sound horizon at decoupling to the comoving distance to the last scattering surface. A plot of the  angular power spectrum for vector modes for a specific set of parameters is shown in Figure \ref{fig:Cell}.

\subsubsection*{Comparison with Tensor Modes}
\label{sec:tensor modes}

It is also illustrative to compare the contribution of the vector modes to the angular power spectrum (\ref{eq:Cell vector}) to  that of the tensor modes \cite{Turner:1993vb,Mukhanov:2005sc}, which, in the limit of  instantaneous decoupling, is given by 
\begin{equation}\label{eq:Cell tensors}
C_\ell^h=\pi\frac{(\ell+2)!}{(\ell-2)!} 
	\int \frac{dk}{k} \, \mathcal{P}_t^\mathrm{rh} \,
	\Bigg|
	\int\limits_{0}^{x_\mathrm{dec}} dx \, \frac{d\mathcal{T}_k}{dx}
		\frac{j_\ell(x)}{x^2}\Bigg|^2,
\end{equation}
where this time $\mathcal{T}_k$ is the transfer function of the tensor modes. 
Up to a factor $\sim (\ell/x)^2$ this is just what the vector modes would contribute 
if the non-integrated term in (\ref{eq:Cell vector}) were negligible. In fact,  
because each power of $x$ in the integral over momenta typically yields a factor $\ell$, 
this scales with $\ell$ in the same way as the contribution from the integrated term of the vector modes. 

The amplitude of the tensor modes remains constant on superhorizon scales and decays also with $1/a$ inside the horizon. Thus, for modes that cross the horizon after decoupling but well before the present, the analogue of  the large-scale approximation (\ref{eq:approximation decaying}) is
\begin{equation}\label{eq:tensor approximation}
\int\limits_{0}^{x_\mathrm{dec}} dx \, \frac{d\mathcal{T}_k}{dx}
	\frac{j_\ell(x)}{x^2}\approx \mathcal{T}_k(\eta_\mathrm{dec})
	\frac{j_\ell(x_\mathrm{dec})}{x^2_\mathrm{dec}}.
\end{equation}
Substituting then Eq. (\ref{eq:tensor approximation}) into (\ref{eq:Cell tensors}) and following the same steps as before we obtain 
\begin{equation}\label{eq:Cell tensor long final}
	\ell(\ell+1)C^h_\ell\approx \frac{\pi}{2}\mathcal{A}_t^2 \ell\,^{n_t},
\end{equation}
which, again, holds for $\ell\lesssim 50$. Thus, on large scales the shape of the  spectrum for tensor and vector modes roughly agree if the spectral indices are the same, and there is no growing mode in the vector sector.  On small scales, the situation is different though.  Along the same lines as before, the contribution of the integrated term can be estimated qualitatively by 
Eq. (\ref{eq:tensor approximation}). Carrying out the same approximations as for the vector modes, the angular power spectrum from the tensor modes then becomes
\begin{multline}
	\ell(\ell+1) C^h_\ell\approx \frac{\pi}{2}\mathcal{A}_t^2 
	\left(\frac{a_\mathrm{eq}}{a_\mathrm{dec}}\right)^2
	\left(\frac{\eta_0}{\eta_\mathrm{eq}}\right)^2\times \\	
	 \times\ell\, ^{n_t-2} \cos^2\left(\frac{\eta_\mathrm{dec}}{\eta_0} \ell\right)
	  \exp\left(-\frac{1}{2}\frac{\ell^2}{\sigma^2}\right).
\end{multline}
This result agrees qualitatively well with numerical simulations \cite{Turner:1993vb,Mukhanov:2005sc}, 
and lends further support to the approximation (\ref{eq:integrated approximation}) for the vector modes 
on small scales. 
It shows that, at these scales, the angular power spectrum of vector modes with spectral index $n_v$ is 
essentially the same  as the angular power spectrum of tensor modes with spectral index
\begin{equation}\label{eq:correspondence}
	n_t=n_v+1-2\nu^r.
\end{equation}

A particularly relevant example of the equivalence arises for $\alpha+c_{14}=0$,
which leads to $\nu^r=3/2$, $\nu^m=5/2$. In this case, 
the spectral index of the vector modes is $n_v\approx 2$, which is just the spectral index of a massless  
field in flat spacetime. Even though the amplitude of this spectrum is negligible on large scales 
at the end of inflation, because there is a growing mode during radiation domination, 
the amplitude of the spectrum at decoupling may be sizable. In any case, during radiation domination
the spectrum at time is $\eta$ is proportional to $(k\eta)^2$, so all modes enter the sound horizon 
$c_v k\eta=1$ with the \emph{same} amplitude. After horizon crossing the amplitude decays as $1/a$, 
as for tensor modes. Hence, up to the frequency of the acoustic oscillations, this case is almost equivalent 
to that of tensor modes with a nearly scale-invariant primordial spectrum,  as stated by (\ref{eq:correspondence}). 
The equivalence also extends to large angular scales. With $\nu^m=5/2$, Eq.~(\ref{eq:Cell final growing}) yields a flat plateau in the contribution of vector modes to the temperature 
anisotropies, just as for a scale-invariant spectrum of gravitational waves.

\subsubsection*{Comparison with Observations}

Present measurements of the CMB temperature anisotropies seem to be well-fit by a nearly scale-invariant 
spectrum of  \emph{scalar} perturbations \cite{Komatsu:2008hk}. Therefore, if vector modes do contribute 
to the temperature  anisotropies at  observable scales, their contribution must be subdominant. This requirement 
places constraints on the parameters of aether theories, which follow from demanding
\begin{equation}\label{eq:subdominant vectors}
	C^V_\ell\lesssim C^s_\ell.
\end{equation}

Let us obtain a very rough estimate of the contribution of vector modes to the temperature anisotropies 
on large scales.  It follows from Eqs. (\ref{eq:vector power}), (\ref{eq:Cell vector final long}) 
and (\ref{eq:Cell final growing}) that the bulk part of the contribution stems from the four large factors
\begin{multline}
	\ell (\ell+1) C^V_\ell\sim {c_{13}^2\over c_{14}}\frac{H^2_\mathrm{rh}}{M_P^2} \,  
\exp\left(\frac{N n_v}{q}\right) 
\left(\frac{a_\mathrm{eq}}{a_\mathrm{rh}}\right)^{2\nu^r-1}\\
	{}\sim {c_{13}^2\over c_{14}}\frac{H^2_\mathrm{rh}}{M_P^2}
	\left(\frac{a_\mathrm{eq}}{a_\mathrm{rh}}\right)^{2\nu^r-n_v-1}.
\end{multline}
Here, we have ignored the (recent) stage of cosmic acceleration, the difference between 
equality and decoupling, and the redshift to the time of equality of matter
and radiation densities. Using Eqs. (\ref{eq:nu}) and (\ref{eq:vector spectral})  we find
\begin{equation}
	2\nu^r-n_v-1\approx-4+\sqrt{1+4\kappa_i}+3\sqrt{1+(4/9)\kappa_r}\approx {2\over 3} \kappa_r
\end{equation}
where $\kappa$ is defined in (\ref{kappaeq}), and the indices $i$ and $r$ refer to inflation
and the radiation era respectively. We have also expanded for small $\kappa$ in the
last step, and neglected $\kappa_i$ in front of $\kappa_r$, 
because of a relative slow roll suppression factor.
The sign of $\kappa$ is determined by the sign of $(1+\alpha/c_{14})$. 
Hence, if $(1+\alpha/c_{14})>0$, vector modes are primordially suppressed, and the subsequent growth during 
radiation domination cannot compensate for this suppression.  
On the other hand if $(1+\alpha/c_{14})<0$, the 
growth during radiation domination may bring the signal well above what is observationally allowed. 
Hence, it seems that the range which is the most interesting from the point of view of observation
is when $|1+\alpha/c_{14}|\ll 1$, which corresponds to $\kappa \ll 1$. 
Therefore, it is not excluded that the present amplitude of these modes is quite sizable, producing
detectable signals in the CMB, or dipole contributions to the gravitational potentials of 
massive bodies through the effect of the velocity field of the vector modes of the aether 
with respect to matter, as discussed at the end of the previous section.

An interesting question is whether,  in the range where $|1+\alpha/c_{14}|\ll 1$, the contribution of the 
scalar isocurvature mode to observables will be larger or smaller than that of the vector modes.
As we noted around Eq. (\ref{comparison}), the relative amplitude of the longitudinal to the transverse velocity
field power spectra is of order $c_{14}/(1+w)$, which is likely to be quite small if the scale of
Lorentz symmetry breaking is low.
Nonetheless, depending on parameters, both situations seem possible. 
Furthermore, in a theory such as BPSH, the vector mode is completely absent, and we only have the scalar 
contribution. A full analysis of the CMB signatures for the scalar mode is left for further research.

\section{Summary and Conclusions}\label{sumcon}

In this article we have studied cosmological perturbations in Einstein-aether theory, where 
the scalar and transverse vector sectors of general relativity are enlarged by an additional dynamical 
field each. We find that inflation can induce sizable perturbations in both of these new massless fields on 
observable scales. Our analysis also applies to the low energy limit of BPSH gravity, 
where the transverse vector is missing by construction \cite{jbpsh}.

For the purposes of summarizing our results, we shall assume that the aether parameters $c_i$\, ($i=1,\ldots,4$) are small.
This is natural, since they can be thought of as proportional to the square of the ratio of the symmetry breaking 
scale $M$ to the Planck scale $c_i \sim (M/M_P)^2\ll 1$. 

To motivate the choice of the range of parameters we shall use below, let us recall that in the 
Einstein-aether theory, 
the effective gravitational constant on small scales $G_N$ can be different from the effective gravitational
constant which appears in the Friedmann equation $G_\mathrm{cos}$. We shall call $\alpha$ and $c_{14}$ the parameters which 
relate these two constants to the bare Newton's constant $G$. They are given in Eqs. (\ref{abbreviations}) as 
linear combinations of the standard $c_i$. In terms of $\alpha$ and $c_{14}$ the effective gravitational constants 
are given by
\begin{equation}
G  = \left(1-{\alpha\over 2}\right) G_\mathrm{cos} = \left(1+{c_{14}\over 2}\right) G_N.
\end{equation}
Note that for $\alpha+c_{14}=0$ we have $G_\mathrm{cos} = G_{N}$. The difference $G_\mathrm{cos}-G_{N}$ 
is constrained by nucleosynthesis to be less than 10 \%, so it seems natural to consider the range
\begin{equation}\label{deltarange}
 |\tilde{\kappa}| \ll 1, \quad {\rm where}\quad  \tilde{\kappa} \equiv -\left(1+{\alpha \over c_{14}}\right). 
\end{equation}
This range guarantees the similarity of $G_\mathrm{cos}$ and $G_{N}$, but it is typically more restrictive 
than required by the nucleosynthesis bound, since $c_{14}\sim (M/M_P)^2$ is naturally small. 
If the parameters are such that we are outside of the range (\ref{deltarange}), 
the effects we are investigating would be either too small to be of phenomenological interest, or too large to be 
compatible with observations.

The main results of the paper are the following. First, we find that in the scalar sector, aside from the
standard adiabatic mode $\zeta$ (which corresponds to the 
curvature of surfaces of constant matter density), there is an additional isocurvature mode which can be important for phenomenology. 
Geometrically, the isocurvature mode can be described as the differential e-folding number $\delta N$ which 
separates the surfaces of constant
matter density from the surfaces orthogonal to the aether. This plays the role 
of a velocity potential $v$ for the aether with respect to matter. At the time of horizon exit during inflation, 
the amplitudes
of $\delta N$ and $v$ are comparable to that of the standard adiabatic mode $\zeta$:
\begin{equation}
	v \sim \delta N \sim \zeta \sim {H \over M_P} \epsilon^{-1/2}  \quad ({\rm horizon\ exit}).
\end{equation}
Here $H$ is the Hubble rate and $\epsilon \ll 1$ is the slow roll parameter during inflation, which is 
independent of the aether
parameters. 

After horizon crossing, the curvature perturbation $\zeta$ stays constant, while
the behaviour of $\delta N$ depends on the parameter $\tilde{\kappa}$ defined above. For $\tilde{\kappa} < 0$, the isocurvature
perturbation slowly decays on large scales, while for $\tilde{\kappa} > 0$ it grows. On the other hand, the velocity
perturbation is given by $v \sim (k/\dot a) \delta N$, where $k$ is the co-moving wave number and $\dot a$ is
the derivative of the scale factor with respect to proper time. Hence, during inflation, when $\dot a$ grows, the long 
wavelength velocity field decays, roughly in proportion to the inverse of the scale factor. After inflaton, the universe
decelerates and the velocity field grows again. At the time of horizon reentry, on cosmologically
relevant scales, we have
\begin{equation}
	v \sim \delta N \sim e^{N\tilde{\kappa}} \zeta \sim 
	e^{N\tilde{\kappa}}\ 10^{-5} \lesssim 1,\quad ({\rm horizon\ reentry})
\end{equation}
where $N\sim 60$ is the number of e-foldings of inflation since the time when the cosmological scale first crossed the 
horizon. The last inequality 
indicates the 
limit of validity of the linear approximation. Note that for $\tilde{\kappa}=0$, the isocurvature perturbation and the 
velocity field of the aether are comparable to $\zeta \sim 10^{-5}$ at horizon reentry. However, 
with $\tilde{\kappa} \lesssim 10/N$, we can have $\delta N \lesssim 1$.
If $\tilde{\kappa}$ is large enough to saturate the inequality, 
this still allows for mildly relativistic speeds for the aether field
$v\sim 1$ within the observable universe. 

Similar results hold for the vector sector. Denoting by $V$ the transverse component of the aether velocity field
with respect to matter, we find that on superhorizon scales
$$
V \sim \left({\epsilon \over c_{14}}\right)^{1/2} v.
$$
Hence, if $c_{14} < \epsilon$ (which seems quite natural if the scale of Lorentz symmetry breaking is low), 
the vector contribution to the velocity field will be
dominant with respect to that of the longitudinal component. On the other hand, in a theory such as BPSH, the 
transverse component $V$ is missing, and the scalar part $v$ is the dominant one. 

We also find that the longitudinal gauge gravitational potentials $\phi$ and $\psi$ can be different even for the adiabatic
mode. On superhorizon scales, we find that this effect (which can be attributed to anisotropic stress of the
aether energy momentum tensor) is of order:
\begin{equation}
(\phi-\psi)_\mathrm{adiab} \sim \phi_\mathrm{adiab}\ c_{13} \sim  \zeta\ c_{13} \sim 10^{-5}c_{13}.
\end{equation}
where $c_{13} \sim (M/M_P)^2$ is another combination of the aether parameters $c_i$, given in Eqs. (\ref{abbreviations}). Physically, this parameter can be expressed in terms of the propagation speed of tensor modes 
$c_{13}=c_t^{-2}-1$.
The isocurvature mode contributes maximally to the anisotropic stress, but the potential due
to the isocurvature mode is suppressed by $c_{13}$:
\begin{equation}
(\phi-\psi)_\mathrm{isoc} \sim \phi_\mathrm{isoc} \sim c_{13}\ \delta N.
\end{equation}
Since $\delta N$ can be larger than $\zeta$, the anisotropic stress can be dominated by the isocurvature mode.
The anisotropic stress on observable scales is suppressed from its value at horizon crossing, due to the
dynamics of the aether on subhorizon scales. For $\tilde{\kappa}=0$, the effect scales like $k^{-2}$ for modes
that crossed the horizon during the matter era. For modes that crossed the horizon during
the radiation era, the behaviour changes to $k^{-1}$. Current constraints on $\phi-\psi$ on cosmological scales
are not very restrictive, and $|c_{13}|\lesssim 1$ seems to be allowed by observations.

The aether manifests itself in PPN parameters through frame dependent effects, which cause anisotropies in the
gravitational field of bodies which move with respect to the aether. In this way, the velocity
field generated during inflation might be detectable. It should be noted, however, that it seems difficult with 
present technology to observe the statistical properties of the random field from this particular type of observations. 
Even if the velocity field were relativistic on cosmological scales 
$v\sim 1$, it falls with scale as $k^{-2}$. In particular, 
the component which varies on scales of the order of 100 Mpc 
would then be below the virial velocity $v_\mathrm{vir} \sim 10^{-3}$ of objects bound in galaxies, 
and it seems unlikely that we can directly sample frame dependent effects in objects which
are located at distances larger than that. On the other hand, at the relatively small distances where the 
observation of frame dependent effects is accessible, we may still detect a large but fairly homogeneous velocity 
field, even one that is much larger than the virial velocity of bound objects. 

Finally, we have computed the 
contribution of transverse vector fields $V$ to the angular power spectrum of CMB anisotropies. We find that for
$\tilde{\kappa}=0$, the spectrum of multipole coefficients $C^{V}_\ell$ has the same shape as that of tensor modes. 
The amplitude, on the other hand, is related to the spectrum $C^{h}_\ell$ for tensor modes and $C^{\zeta}_\ell$ for
the adiabatic scalar mode as
\begin{equation}\label{comparativa}
C^{V}_\ell \sim\ {c_{13}^2 \over c_{14}}e^{2 N \tilde{\kappa}} C^{h}_\ell\  
\sim\  {\epsilon\ c_{13}^2\over c_{14}}e^{2 N \tilde{\kappa}} C^{\zeta}_\ell.
\end{equation}
This means that the vector modes in Einstein-aether theory can easily dominate the signal from tensor modes.
The analysis of polarization induced by the vector modes is therefore of phenomenological interest, and is left 
for further research. Moreover, we know that the CMB is well-fit with a primordial 
spectrum of \emph{scalar} adiabatic perturbations. This imposes additional phenomenological restriction 
amongst the parameters $c_{13}$ and $\tilde{\kappa}$ of Einstein-aether theories, of the form 
\begin{equation}
\tilde{\kappa} \lesssim {1\over 2 N}\ln\left|{c_{14}\over\epsilon\ c_{13}^2}\right|.
\end{equation} 

So far, we have not included the constraints which follow from the frame-dependent effects on the PPN parameters.
These are summarized in Appendix \ref{sec:other constraints}, and take the form
\begin{equation}\label{ppnconst}
\omega \, \alpha_1\lesssim 10^{-7},\quad \omega^2\alpha_2\lesssim 10^{-13}.
\end{equation}
Here, $\omega= \max \{V,v,v_\mathrm{vir}\}$, is the velocity of the aether with respect to the object whose gravitational 
field is being tested at post-Newtonian order, and $v_\mathrm{vir} \sim 10^{-3}$ is the typical virial velocity for 
bound objects with respect to the CMB frame. The post-Newtonian parameters $\alpha_1$ and $\alpha_2$ 
are combinations of the four 
aether parameters $(\alpha,c_{14},c_+, c_-)$. Here, following \cite{Jacobson:2008aj}, 
we have introduced $c_+\equiv c_{13}=c_1+c_3$ and 
$c_- \equiv c_1-c_3$. Phenomenologically, it is possible to set $\alpha_1=\alpha_2=0$, which determines $\alpha$ 
and $c_{14}$ as functions of the other two parameters in the model,
\begin{equation}
\alpha=-c_{14}= - 2 {c_+ c_- \over (c_+ +c_-)}.
\label{restriction}
\end{equation}
The parameters $c_+$ and $c_-$ remain rather unconstrained by 
observations. Stability requirements and superluminality (or Cherenkov) constraints are satisfied provided that 
$-1\leq c_{+} \leq 0$, $c_{+} /3(1+c_{+}) \leq c_{-} \leq 0$. Constraints from radiation damping in binary 
systems determine further constraints on the $(c_+,c_-)$ plane, but a sizable coefficient
\begin{equation}
|c_{13}|\lesssim 1 
\end{equation}
still seems 
to be allowed by all observations \cite{Jacobson:2008aj}. This is important, since the gravitational effects
of the aether are suppressed by this coefficient. For instance the contribution of vectors to the CMB anisotropies
is of order 
\begin{equation}
C_\ell^V \sim c_{13}^2 V^2, 
\end{equation}
where $V\lesssim 1$ is the aether velocity field. Hence, the observability of the effect depends crucially on 
$c_{13}$ being sufficiently large.

This brings us to the question of fine tuning amongst the parameters of the model. In a low energy theory, 
one might have expected all dimensionless parameters to be of the same order,
$$
c_i \sim (M/M_P)^2.
$$ 
Observability of $C_\ell^V$ requires an inequality
of the form $c_{13} V \gtrsim 10^{-6}$, which would be natural provided that 
\begin{equation}\label{osbs}
(M/M_P)_\mathrm{obs}^2 \gtrsim 10^{-6} V^{-1}.
\end{equation}
On the other hand, in Eq. (\ref{restriction}) we have adjusted the parameters 
so that $\alpha_1=\alpha_2=0$, but the actual restriction (\ref{ppnconst}) is of the
form $\alpha_2 \lesssim 10^{-13}\omega^{-2}$. Hence,  $\alpha_2$ must be well below
the natural scale (\ref{osbs}) by a considerable suppression factor
\begin{equation}
\alpha_2 \lesssim 10^{-7} \omega^{-1} (V/\omega) (M/M_P)_\mathrm{obs}^2, 
\end{equation}
with $10^{-3}<\omega<1$. In the classical theory, the parameter $\alpha_2$ can always 
be chosen by hand to have any particular value. However, in an effective field theory (EFT) a 
parameter is considered to be finely tuned or technically unnatural if quantum 
corrections to it are larger than the desired renormalized value of the parameter. 
The question, therefore is whether the very small values of $\alpha_2 \ll (M/M_P)^2$ are 
stable or not under quantum corrections. Withers \cite{Withers} has recently analyzed 
the Einstein-Aether theory as an EFT, with the conclusion that the parameters $c_i$ receive only negligible 
logarithmic corrections. A similar result may hold in BPSH theory \cite{bps2}. This subject is left 
for further study.

To conclude, the results presented here show that the preferred frame singled out by the aether field $A^{\mu}$, or by
the preferred foliation of the BPSH theory, may have picked up a large random velocity field seeded by
quantum fluctuations during inflation. Depending on the parameters, this may even be mildly 
relativistic on cosmological scales. The effects of this velocity field may be detectable in observations
of frame dependent PPN effects, or in specific features in the CMB spectrum such as a sizable contribution from 
vector modes. These issues deserve further investigation.

\begin{acknowledgments}
We would like to thank Diego Blas, Eugene Lim, Oriol Pujol\`as and Yuko Urakawa for useful conversations.
The work of CAP is supported in part by the National Science Foundation under Grant No. PHY-0855523. 
The work of JG is supported by grants FPA2007-66665C02-02, DURSI 2009-SGR-168,
and CPAN CSD2007-00042.
The work of NFS is supported by the Comissionat per a Universitats i Recerca del 
Departament d'Innovaci\'o, Universitats i Empresa de la Generalitat de Catalunya and the European Social Fund.
\end{acknowledgments}

\appendix

\section{Other Constraints}
\label{sec:other constraints} 

In this article we have derived the constraints on the parameters of Einstein-aether theories that follow from classical and quantum stability, and from phenomenological considerations related to the primordial perturbations. In addition to these constraints, there are further conditions that the $c_i$'s have to satisfy, which arise from the Post-Newtonian limit of the theory,  Big-Bang nucleosynthesis, and from the arrival of high-energy cosmic rays to earth. An extensive summary of these constraints can be found in \cite{Jacobson:2008aj}.

\subsection*{Post-Newtonian Limits}

In any metric theory, the gravitational field  created by non-relativistic bodies 
can be characterized beyond the Newtonian limit by a set of post-Newtonian PPN parameters, 
whose values are tightly constrained by solar 
system tests of gravity \cite{Will:1981cz}. The parameters $\beta$ and $c_{13}$ in  aether theories agree with 
those of general relativity, and also agree with the measured ones \cite{Eling:2003rd}. But because the aether defines a preferred frame,  it also introduces additional departures from general relativity, 
which manifest themselves as gravitational potentials that depend on the velocity of the interacting bodies with 
respect to the aether. These preferred-frame effects are  encoded in  the PPN parameters $\alpha_1$ and $\alpha_2$. 
One of the most stringent limits on the value of $\alpha_1$ comes from measurements of the eccentricity of the binary 
pulsar J2317+1439 (which would change if $\alpha_1$ were non-zero) \cite{Bell:1995jz}, while one of the most stringent 
limits on $\alpha_2$ stems from the alignment of the sun spin with the solar system angular momentum 
(a non-zero $\alpha_2$ would lead to a misalignment) \cite{Nordtvedt}. These limits lead to the conditions
\begin{subequations}\label{eq:PPN constraints}
\begin{align}
 	\alpha_{1}&=\frac{-8(c_{1}c_{4}+c_{3}^{2})}{2c_{1}-c_{1}^{2}+c_{3}^{2}}\leq1 .7 \times 10^{-4},\\
	\alpha_{2}&=\frac{\alpha_{1}}{2}
		-\frac{(2c_{13}-c_{14})(\alpha+c_{14})}{\beta (2-c_{14})}\leq 1.2\times10^{-7}.
\end{align}
\end{subequations}
It is important to stress that both  limits assume that the velocity of the sun with respect to the aether $\omega$ is the velocity with respect to the frame in which  CMB dipole vanishes, $\omega\sim 10^{-3}$. Roughly speaking, the limit on $\alpha_1$ is actually a limit on $\alpha_1\omega$, while the limit on $\alpha_2$ actually constraints $\alpha_2\omega^2$. Thus, if $\omega$ were larger that assumed, as the results of our work seem to allow, the limits on $\alpha_1$ and $\alpha_2$ would be actually  tighter. In other words, the constraints on the PPN parameters $\alpha_1$ and $\alpha_2$ actually are
\begin{subequations}
\begin{align}
 	\alpha_{1}&\lesssim 10^{-4}\times \left(\frac{10^{-3}}{\omega}\right),\\
	\alpha_{2}&\lesssim 10^{-7} \times\left(\frac{10^{-3}}{\omega}\right)^2,
\end{align}
\end{subequations}
where $\omega$ is again the velocity of the sun with respect to the preferred frame.  The constraints (\ref{eq:PPN constraints}) are typically satisfied if the norm of the $\tilde{A}$, defined in Section \ref{sec:Theory} is of order ${M\sim 10^{-4}M_P}$. 

\subsection*{Big-Bang Nucleosynthesis}

The agreement between the predicted light element abundances, 
and those actually observed (or indirectly measured using CMB observations \cite{Komatsu:2008hk})  
constrains the value of the Hubble constant at the time the light elements formed, 
at temperatures of about $T\approx 10^9\,\mathrm{K}$. Because the expansion rate depends on the value of 
the renormalized Newton constant $G_\mathrm{cos}=2G/(2-\alpha)$ through Eq. (\ref{eq:Friedman}), 
and because the latter is related to the ``Newtonian" gravitational constant by Eq. (\ref{eq:Newtonian}), 
given the measured value of $G_N$ on small scales and the number of relativistic species during nucleosynthesis,
one can determine how light element abundances depend on the parameters $\alpha$ and $c_{14}$. Agreement of such a prediction with observations then implies
\begin{equation}\label{eq:BBN}
	\frac{c_{14}+\alpha}{2-\alpha}\lesssim 10\%.
\end{equation}

\subsection*{Cherenkov Radiation}

If any of the propagation speeds  of tensor, vector or scalar modes we have discussed  were sufficiently smaller than the speed of light, highly relativistic particles traveling close to the speed of light would loose energy into these modes by a process analogous to Cherenkov radiation (this is kinematically possible only if the dispersion relation of the emitted  quanta is subluminal.) Although the emission amplitude is inversely proportional to the Planck mass, the fact that we detect these cosmic rays, and that they must originate at astrophysical distances, allows one to place quite stringent limits on the parameters of the aether \cite{Elliott:2005va},
\begin{subequations}\label{eq:Cherenkov constraints}
\begin{eqnarray}
	c_{13}&<&1\times 10^{-15} \\
	\frac{c_{13}^2(c_{13}^2+2c_4)}{c_1^2}&<&1.4\times 10^{-31} \\
	\frac{(c_3-c_4)^2}{|c_{14}|}&<&1\times 10^{-30} \\
	\frac{c_4-c_2-c_3}{c_1}&<&3\times 10^{-19}.
\end{eqnarray}
\end{subequations}
It is important to realize though that these constraints only apply if the different aether 
modes propagate subluminally. Under this assumption conditions (\ref{eq:Cherenkov constraints}) 
can also be taken to imply the bound  $M\leq 10^{-7}M_P$ on the norm of the aether field 
$\tilde{A}_\mu$ defined in Section \ref{sec:Theory}.

\subsection*{Propagation Speed}

Some authors impose further conditions on the parameters of the aether,  namely, that the propagation speed of the perturbations be subluminal. The origin of this requirement goes back to the violations of causality that appear in \emph{Lorentz-invariant theories} with superluminal signals. As far as we know, there is no link however between superluminality and violations of causality in backgrounds like the ones we are considering. The cosmic aether breaks Lorentz invariance and  defines a preferred reference frame. Signals always travel forward in time in this frame, so no closed timelike curves can arise. Even the construction of \cite{Adams:2006sv}, in which due to the nature of the background closed timelike curves may appear seems difficult to realize here, because the aether satisfies a fixed-norm constraint.  Hence, we shall not require subluminal propagation, though because this is a somewhat controversial issue, we collect the appropriate conditions here for completeness. They easily follow from 
Eqs. (\ref{eq:tensor c}), (\ref{eq:scalar cs}) and (\ref{eq:vector c}). In the limit of small coefficients, $c_i\ll 1$ they read
\begin{subequations}
\begin{eqnarray}
	c_{13}&\geq &0\\
	c_1-c_4-c_{13}^2&\geq& 0\\
	\beta-c_{14}&\geq&0.
\end{eqnarray}
\end{subequations}
Note that because we have taken metric perturbations into account, these conditions differ from those derived in the limit of a decoupled aether \cite{Lim:2004js}.  For alternative views on  superluminal propagation we refer the reader to the references \cite{Adams:2006sv,Bruneton:2006gf,Babichev:2007dw}.

\section{Scalar equations in the longitudinal gauge}

In this appendix we discuss the scalar sector in the longitudinal gauge. This is useful and complementary 
to the next appendix, which deals with the gauge-invariant formulation.

In the longitudinal gauge, there are five fields in the scalar sector: $\phi, \psi$, $C$, $\delta\lambda$ and the 
inflaton perturbation $\delta \varphi$. Therefore, we need five independent equations to uniquely determine their values. 
Contracting Eq. (\ref{eq:vector field motion}) with $A^\beta $ and using the constraint ${A^\beta A_\beta=-1}$ 
yields an equation that expresses the Lagrange multiplier in terms of the aether and the metric, 
\begin{multline}\label{eq:delta lambda}
	\delta\lambda=\frac{1}{a^2}\Bigg[
	-6(\alpha-c_2)\mathcal{H}^2\phi+6c_2\frac{a''}{a}\phi+3c_2\mathcal{H}\phi'+\\
	{}+c_3k^2 \phi-3(2\beta-c_2)\mathcal{H}\psi'+3c_2\psi''-  \\
	{}-(\beta+c_1)\mathcal{H}k^2 C+(\beta-c_1)k^2 C'\Bigg],
\end{multline}
which shows explicitly how the Lagrange multiplier can be expressed in terms of the remaining fields.  
The time component of the linearized aether field Eq. (\ref{eq:vector field motion}) 
is identically satisfied. The linearized spatial components give 
\begin{multline}\label{eq:scalar vector}
	C''+2\mathcal{H}C'
	+\left[\frac{\alpha}{c_{14}}\left(2\mathcal{H}^{2}-\frac{a''}{a}\right)+\frac{a''}{a}\right]C+\\
	{}+\frac{\beta}{c_{14}}k^{2}C
	+\left(1+\frac{\alpha}{c_{14}}\right)\mathcal{H}\phi+\phi'
	+\frac{\alpha}{c_{14}}\psi'=0,
\end{multline} 
which combined with the ${}^0{}_i$ Einstein equation results in
\begin{equation}\label{eq:delta inflaton}
	\mathcal{H}\phi+\psi'-\frac{\beta}{2-\alpha} k^2 C=4\pi G  \frac{2}{2-\alpha}\varphi' \delta\varphi.
\end{equation}
Eq. (\ref{eq:delta inflaton}) expresses $\delta\varphi$ in terms of the remaining scalars, and allows us to eliminate 
$\delta\varphi$ from our system of equations. On large scales, this equation has the same form it would have in the absence of the aether, 
with the difference that the effective Newton's constant has the renormalized value implied by Eqs. (\ref{eq:Einstein}).
The part of the ${}^i{}_j$ Einstein equations which is not proportional to $\delta^i{}_j$ is
\begin{equation}\label{eq:scalar stress}
	\phi=\psi+c_{13} (C'+2\mathcal{H}C),
\end{equation}
which immediately reveals that the Einstein-aether is a source of anisotropic stress in the scalar sector. 
This equation allows us to express $\phi$ in terms of $\psi$ and $C$, and thus eliminate yet another variable from the equations. 
Note that scalar fields and perfect fluids cannot source anisotropic stress, which is why a value of $\psi-\phi$ different from 
zero is sometimes attributed to modified  gravity.   Finally, the sum of the ${}^0{}_0$ and the ${}^i{}_j$ Einstein equations 
proportional to $\delta^i{}_j$ is
\begin{multline}\label{eq:scalar diagonal}
	\psi''+5\mathcal{H}\psi'+\mathcal{H}\phi'+
	2\left(\frac{a''}{a}+\mathcal{H}^2\right)\phi
	+\frac{c_{14}-1}{2-\alpha}k^2\phi+ \\
	{}+\frac{3}{2-\alpha}k^2\psi+\frac{c_{14}-c_2}{2-\alpha}k^2C'
	+\frac{c_{14}-\alpha-2c_2}{2-\alpha}\mathcal{H}k^2 C\\
	{}=\frac{8\pi G}{2-\alpha}
	3\mathcal{H}(1-w)\varphi'\delta\varphi,
\end{multline}
where we have used Eq. (\ref{eq:delta lambda}) to eliminate $\delta\lambda$, and that during power-law expansion 
the equation of state parameter $w$ is constant. 

Eqs. (\ref{eq:delta lambda}), (\ref{eq:scalar vector}),  (\ref{eq:delta inflaton}), (\ref{eq:scalar stress}) 
and (\ref{eq:scalar diagonal}) form a set of five differential equations for the five unknowns. 
We can use the constraints (\ref{eq:delta inflaton}) and  (\ref{eq:scalar stress}) to eliminate $\phi$ and $\delta\varphi$ 
from Eqs. (\ref{eq:scalar vector}) and (\ref{eq:scalar diagonal}), arriving at 
\begin{equation}\label{eq:final C}
\begin{split}
	C''&+\left(2+\frac{c_{13}(c_{14}+\alpha)}{c_{14}(1+c_{13})}\right)\mathcal{H}C'+\\
	&+\left(\frac{c_{14}-\alpha+2c_{13} c_{14}}{c_{14}(1+c_{13})}\frac{a''}{a}+	\frac{2\alpha}{c_{14}}\mathcal{H}^{2}\right)C+\\
	&+\frac{\beta}{c_{14}(1+c_{13})}k^2 C+\frac{c_{14}+\alpha}{c_{14}(1+c_{13})}(\psi'+\mathcal{H}\psi)=0, 
\end{split}
\end{equation}
and
\begin{equation}\label{eq:final psi}
\begin{split}
	\psi''&+3(1+w)\mathcal{H}\psi'+2\left(\frac{a''}{a}-2\mathcal{H}^{2}\right)\psi+3(1+w)	\mathcal{H}^{2}\psi+\\
	&+c_{13}\mathcal{H}C''+2c_{13}\left(\frac{a''}{a}-\mathcal{H}^{2}\right)C'+3(1+w)c_{13}\mathcal{H}^{2}C'+\\
	&+c_{13}\left(6\frac{a''}{a}-10\mathcal{H}^{2}\right)\mathcal{H}C+6(1+w)c_{13}\mathcal{H}^{3}C+ \\
	&+ \frac{2+c_{14}}{2-\alpha}k^{2}\psi+\frac{c_{14}(1+c_{13})-\beta}{2-\alpha} k^{2}C'+\\
	&+\frac{c_{14}(1+2c_{13})+4\beta-\alpha-3(1+w)\beta}{2-\alpha}\mathcal{H} k^{2}C=0.
\end{split}
\end{equation}
Because this is a system of two second order linear differential equations, 
we need to specify four independent initial conditions, so there must exist four 
linearly independent solutions. This is also what we expect by simply counting matter fields. 
In the limit of weak gravitational couplings, we may neglect metric perturbations, so  we just 
have one degree of freedom in the inflaton perturbations and one degree of freedom in the aether field perturbations, 
for a total of four initial conditions to determine uniquely the evolution of the system. As we deviate from the limit of weak coupling, 
neglecting metric perturbations ceases to be a good approximation, but the number of degrees of freedom in the theory remains unchanged.

\subsection{Short-wavelength Solutions} \label{sec:short-wavelength}

In the short-wavelength regime, $k|\eta|\gg 1$, the solutions of the equations of motion (\ref{eq:final C}) and  
(\ref{eq:final psi})  behave approximately like in flat space. The notion of an approximate solution can be formalized 
by introducing $k\eta$ as an expansion parameter. In the limit $k|\eta|\gg 1$ the solution of Eqs.  
(\ref{eq:final C}) and (\ref{eq:final psi}) can be cast in the form
\begin{equation}\label{eq:WKB}
	\psi=\widetilde{\psi}(k\eta)\,\exp(-i c_s k \eta), \quad 
	C=\frac{1}{k}\widetilde{C}(k\eta) \, \psi,
\end{equation}
where,  $\widetilde{\psi}$ and $\widetilde{C}$ are functions whose power series expansion 
starts at a finite positive power of $k\eta$,   and $c_s$ is a ``sound speed" to be determined. 
Substituting the ansatz (\ref{eq:WKB}) into Eqs. (\ref{eq:final C})  and (\ref{eq:final psi}), and keeping 
the leading powers of $k\eta$  yields a set of  algebraic equations with two positive frequency and two negative 
frequency solutions, for a total of four solutions, as expected. At leading order, $\widetilde{\psi}$ remains 
unconstrained and can be taken to be constant.  The positive frequency solutions are given by 
\begin{subequations}
\begin{align}
	&(c_s)_a= c_a,  &\widetilde{C}_a&=i  \frac{\alpha-2}{\beta}c_a,
\label{eq:A1}
	\\
	&(c_s)_\varphi=1, &\widetilde{C}_\varphi&=i\frac{c_{14}+\alpha}{\beta-c_{14}(1+c_{13})},
\label{eq:A2}	
\end{align}
\end{subequations}
where $c_a$ is the sound speed of Eq. (\ref{eq:scalar cs}). 

These two modes correspond 
to the two independent short wavelength solutions  (\ref{eq:mode phi}) and (\ref{eq:mode a}) that we found in 
Subsection \ref{sec:scalar stability}.
To see that this is the case, we may use the
expression of $\delta N$ and $\zeta_a$ in the longitudinal gauge
\begin{subequations}
\label{eq:relation}
\begin{eqnarray}
	\delta N&=& \frac{\mathcal{H}\delta\varphi}{\varphi'}+\mathcal{H} C, \\
	\zeta_a&=&\psi- \mathcal{H} C.
\end{eqnarray}
\end{subequations}
In the first equation, $\delta\varphi$ should be expressed in terms of $\psi$ and $C$ through the relation
\begin{equation}
\delta\varphi =\frac{M_P^{2}}{\varphi'}\left[(2 - \alpha)\left(\psi' + 	\mathcal{H}\psi + c_{13}\mathcal{H}C'+ 
2 c_{13}\mathcal{H}^2 C\right) - 
\beta k^2  C\right],\label{llaa}
\end{equation}
which follows from (\ref{eq:delta inflaton}) and (\ref{eq:scalar stress}).
By comparison with (\ref{eq:mode phi}) and (\ref{eq:mode a})
we also obtain the overall normalization factor $\widetilde{\psi}$. For the first mode, we have
\begin{equation}\label{eq:short C}
	\psi_a\to {Z_a^{1/2}\over a}\, \frac{e^{-ic_a k \eta}}{\sqrt{2c_a k}},
		\quad 
	C_a\to  \frac{1}{k}\widetilde{C}_a \,\psi_a,
\end{equation}
where $Z_a$ is given in Eq. (\ref{zsn}), and  $\widetilde{C}_a$ in Eq. (\ref{eq:A1}).  
For the second mode, we have
\begin{equation}\label{eq:short phi}
	\psi_\varphi \to {Z_\varphi^{1/2}\over a}\frac{e^{-i k \eta}}{\sqrt{2 k}}, 
	\quad 
	C_\varphi\to \frac{1}{k}\widetilde{C}_\varphi\, \psi_\varphi,
\end{equation}
where $\widetilde{C}_\varphi$ is given in Eq. (\ref{eq:A2}) and
\begin{equation}
 Z_\varphi^{1/2}\equiv i 
        \frac{\varphi'}{k\,M_P^2}\left(2+\frac{c_{14}(\beta+\alpha(1+c_{13}))}{\beta-c_{14}(1+c_{13})}\right)^{-1}.
\end{equation}
Substitution of (\ref{eq:short C}) 
into (\ref{eq:relation}) reproduces Eq. (\ref{eq:mode a}), while 
substitution of (\ref{eq:short phi}) into (\ref{eq:relation}), together with the background 
Eqs. (\ref{eq:Einstein}),  reproduces Eq. (\ref{eq:mode phi}). 
The  vacuum is thus characterized by the two independent solutions of Eqs. (\ref{eq:final C}) and
 (\ref{eq:final psi}) that approach (\ref{eq:short C}) and (\ref{eq:short phi}) in the limit $k|\eta|\to \infty$.

\subsection{Long-wavelength Solutions}
\label{sec:long wavelength}

In the limit of long wavelengths, $k|\eta|\ll 1$, we may neglect terms proportional to $k^2$ in Eqs. (\ref{eq:final C}) and (\ref{eq:final psi}). 
In this limit, the power-law ansatz
\begin{equation}\label{eq:long}
	\psi=(-\eta)^t, \quad C=\mathcal{C}\cdot(-\eta) \cdot \psi
\end{equation}
reduces the two coupled differential equations (\ref{eq:final C}) and (\ref{eq:final psi}) to an algebraic system for the two constants $t$ 
and $\mathcal{C}$,
\begin{subequations}\label{eq:long wavelength system}
\begin{multline}  \label{eq:t} 
	\Big[t^{2}+\frac{c_{14}(1 + c_{13})(5 + 3 w)
	+2c_{13}(c_{14}+\alpha)}{c_{14}(1+c_{13})(1+3w)}t+  \\
	+\frac{2(c_{14}+\alpha)(3(1+w)+c_{13}(5+3w))}{c_{14}(1+c_{13})(1+3 w)^{2}}\Big]	\mathcal{C}=\\
	=\frac{c_{14}+\alpha}{c_{14}(1+c_{13})}\left(t+\frac{2}{1+3w}\right), 
\end{multline}
\begin{multline} \label{eq:A}	
	t(5+t+3w+3wt)(1+3w-2\mathcal{C}c_{13})=0.
\end{multline}
\end{subequations}
Because Eqs. (\ref{eq:long wavelength system}) are linear in $\mathcal{C}$, they may be reduced to a single quartic equation 
for $t$, with four different solutions, as it should be. 

\subsubsection{Adiabatic Modes ($\delta N=0$)}
Two solutions of the coupled equations (\ref{eq:long wavelength system}) follow directly from Eq. (\ref{eq:A}),
\begin{subequations}
\begin{align}
	t_1&=0, &\mathcal{C}_1&=\frac{1+3w}{3(1+w)+c_{13}(5+3w)}, \label{eq:adiabatic 1}\\
	t_2&=-\frac{5+3w}{1+3w},  &\mathcal{C}_2&=-\frac{1+3w}{2} \label{eq:adiabatic 2}.
\end{align}
\end{subequations}
The corresponding perturbations are the two ``adiabatic" modes that always exist at long wavelengths, regardless of the matter content of the universe \cite{Weinberg:2003sw}. 
Along these two modes,  the (spatial) curvature perturbation on comoving slices,\footnote{Recall from equation (\ref{comovz})  that we mean comoving with respect to all forms of matter, excluding the aether. The $^0{}_i$ Einstein equation (\ref{eq:delta inflaton}) however reveals that the contribution of the aether to the total velocity perturbation is negligible on large scales. Hence, on large scales,  hypersurfaces comoving with matter and comoving with matter plus aether are actually the same.}
\begin{equation}
	\mathcal{\zeta}\equiv\psi+\frac{2}{3}\frac{\mathcal{H}\phi+\psi'}{\mathcal{H}(1+w)} ,
\end{equation}
and the difference of the two metric potentials (which is proportional to the anisotropic stress)  are given by
\begin{subequations}
\begin{align}
	\mathcal{\zeta}_1&=\frac{(5+3w)(1+c_{13})}{3(1+w)+c_{13}(5+3w)}\psi_1,
	&\phi_1-\psi_1&=
	-\frac{c_{13}}{1+c_{13}}\mathcal{\zeta}_1,
	\quad  \label{eq:R1}\\
	\mathcal{\zeta}_2&=0, &\phi_2-\psi_2&=0.
\end{align}
\end{subequations}
It can be readily checked that for these modes $\delta N=0$, so that matter is at rest in the aether frame.
Though  these  adiabatic modes have the properties described in \cite{Weinberg:2003sw}, 
they do not share the properties postulated in \cite{Weinberg:2004kr, Weinberg:2004kf, Weinberg:2008zzc}. 
In particular, for the first adiabatic mode, the anisotropic stress is non-zero. The form of the two adiabatic 
modes for an arbitrary expansion history and matter content is derived in Appendix \ref{sec:adiabatic modes}.

\subsubsection{Isocurvature Modes ($\zeta=0, \delta N\neq 0$)}
\label{sec:isocurvature}

The two remaining solutions of Eqs. (\ref{eq:long wavelength system}) require  $\mathcal{C}=({1+3w})/{2c_{13}}$,
which gives
\begin{equation}
\psi = -c_{13} \mathcal{H} C \propto (-\eta)^{t_{\pm}}.
\end{equation}
From (\ref{eq:t}), the exponents are given by
\begin{eqnarray}\label{exponents2}	
2 t_{\pm} &=& -\left({5+3w\over 1+3w}\right) \pm \sqrt{\left({5+3w\over 1+3w}\right)^2 + 4 \kappa},
\end{eqnarray}
where $\kappa$ is given by Eq. (\ref{kappaeq}).
It is straightforward to check that for these modes we have
\begin{subequations}\label{isocmodes}
\begin{eqnarray}
	\mathcal{\zeta}_{\pm}&=&0,  \\
\phi_{\pm}-\psi_{\pm} &=& \left({1+3w \over 2}\right) t_{(\mp)}\ \psi_{\pm} \\
\delta N_{\pm} &=& -\left({1+c_{13}\over c_{13}}\right) \psi_{\pm}\ \propto (-\eta)^{t_{(\pm)}}.\label{isocmodes3}
\end{eqnarray}
\end{subequations}
These are two isocurvature modes, in the sense that the curvature perturbation on comoving slices $\zeta$
vanishes for any value of $w$,

From Eq. (\ref{exponents2}) it is straightforward to check that, for any value of $w$, 
one of the two modes is a decaying one.
Whether the second mode is growing or decaying depends on the sign of $\kappa$, which is in 
turn determined by the sign of 
$1+(\alpha/c_{14})$. For $\kappa>0$ the second solution is also a decaying one, but for $\kappa<0$ there 
is a growing mode.
In the special case $\kappa=0$, there is a constant non-decaying long wavelength solution.

The existence of a growing non-adiabatic isocurvature mode in Einstein-aether theories for 
$(\alpha/c_{14}) < -1$ can have 
important phenomenological consequences, as we discuss in the main text. 

\section{Canonical reduction of the scalar sector.}
\label{sec:gauge invariant}

The normalization of the spectrum of scalar perturbations follows from the normalization of the action 
for the corresponding physical degrees of freedom. Here, we find the reduced set 
of gauge-invariant dynamical variables, and express the second order Lagrangian in terms of these. 
This Lagrangian can also be used, of course, to rederive the scalar equations 
of motion (\ref{eq:final C}) and (\ref{eq:final psi}).

The starting point is the Lagrangian for scalar perturbations in an arbitrary gauge,
which is obtained by substituting 
the metric (\ref{eq:metric perturbations}) into the action (\ref{eq:action}), and expanding 
to second order in the scalar perturbations. Using the constraint (\ref{eq:fixed norm}) 
which is obtained from variation with respect to $\delta\lambda$, we arrive at
\begin{align}\label{eq:gauge free}
\begin{split}
	\mathcal{L}^{(2)}_{s}&=\frac{M_{P}^{2}}{2} a^{2}\Bigg[ 2 k^{2}\psi^{2}-3(2-\alpha)\psi'^{2}-4 k^{2}\psi \phi+4k^{2}\psi' B-\\
	{}&-(2-\alpha)k^{2}\psi' E'+2\alpha k^{2}\psi' C+\beta k^{4}(C+\frac{1}{2}E')^{2}-\\
	{}&-c_{14}k^{2}(\phi+C'+B')^{2}-6(2-\alpha)H\phi  \psi'-\\
	{}&-2(c_{14}-2)\mathcal{H}k^{2}\phi  B-(2-\alpha)\mathcal{H}k^{2}\phi E'+\\
	&+2(\alpha-c_{14}) \mathcal{H}k^{2}\phi C-(2-\alpha)(2\mathcal{H}^{2}+\mathcal{H}')\phi^{2}+\\\
	{}&+(\alpha(\mathcal{H}^{2}-\mathcal{H}')+c_{14}(\mathcal{H}^{2}+\mathcal{H}'))k^{2}(C+B)^{2}+\\
{}&+M_P^{-2}\left(\delta\varphi'^{2}-k^{2}\delta\varphi^{2}+2\varphi'\delta\varphi(3\psi'-k^{2}B+\frac{k^{2}E'}{2}\right)-\\
 {}&-2 \varphi' \delta\varphi'\phi-a^2 V_{,\varphi \varphi}\delta\varphi^{2}-2a^{2}V_{,\varphi}\delta\varphi\phi)\Bigg].
\end{split}
\end{align}
Not all variables in this Lagrangian are dynamical. 
Some linear combinations are gauge modes, while others are constrained. 
We would like to find a Lagrangian that contains dynamical gauge-invariant variables only. 

The identification of constraints and the reduction of phase space is best performed in the
canonical formalism, where the equations of motion are at most of first order in time.
Constraints are equations of motion without any time derivatives, and can be substituted back 
into the first order Lagrangian. Here, we closely follow Fadeev and Jackiw's method 
for dealing with constrained systems \cite{tears}. For a discussion of cosmological perturbation
theory in this framework, see \cite{gmst}. 

We begin by introducing new variables $U$ and $W$ through
\begin{subequations}
\begin{align}
	2W = B + C,\\ 
	2U = B - C.
\end{align}
\end{subequations}
The conjugate momenta of the system are given by
\begin{subequations}
\begin{align}
	\Pi_{\psi}\equiv\frac{\mathcal{L}^{(2)}_{s}}{\partial \psi'}
	={}&M_P^{2}a^{2}\Big[\alpha k^{2}(W-U)+2k^{2}(U+W)- \nonumber\\
	&-\frac{(2-\alpha)}{2}k^{2}E'-3(2-\alpha)(\psi'+\mathcal{H}\phi)+ \nonumber\\
	&+3M_P^{-2}\varphi'\delta\varphi\Big],\\
	\Pi_{E}\equiv\frac{\mathcal{L}^{(2)}_{s}}{\partial E'}
	={}&\frac{1}{2} M_P^{2}a^{2}k^{2} \Big[\beta k^{2}(W-U+\frac{1}{2} E')-\nonumber \\
	&-(2-\alpha)(\psi'+\mathcal{H}\phi) +M_P^{-2}\varphi'\delta\varphi \Big],\\
	\label{eq:pi delta phi}
	\Pi_{\delta\varphi}\equiv\frac{\mathcal{L}^{(2)}_{s}}{\partial \delta\varphi'}
	={}& a^{2} (\delta\varphi'-\phi\varphi'),\\
	\Pi_{W}\equiv\frac{\mathcal{L}^{(2)}_{s}}{\partial W'}={}&-2 c_{14} M_P^{2}a^{2}k^{2}(\phi + 2 W'),
 \end{align}
 \end{subequations}
and we can write the first order Lagrangian
\begin{equation}\label{eq:first order}
\begin{split}
	\mathcal{L}^{(1)}_{s}&=\Pi_{E}E'+\Pi_{W}W'+\Pi_{\delta\varphi}\delta\varphi'+\Pi_{\psi}\psi'-\\
	&-\frac{3 \Pi_{E}^2}{M_P^{2}a^{2}k^{4} (1+c_{13}) } + \frac{\Pi_{W}^2}{8 c_{14} k^2 M_P^{2} a^2}-\frac{\Pi_{\delta\varphi}^2}{2 a^{2}}+\\
	&+ \frac{ \beta\Pi_{\psi}^2}{4 M_{PL}^{2}a^{2} (2 -\alpha) (1+c_{13})} - \frac{2 k^{2} \beta W \Pi_{\psi}}{(2 - \alpha) (1+ c_{13})}+\\
	&+\frac{ \Pi_{E}\Pi_{\psi}}{M_P^{2}a^{2} k^{2} (1+ c_{13})}- 2 U\Pi_{E} -\frac{2 (1 -c_{13}) W\Pi_{E}}{(1+ c_{13})} -\\
	&- \frac{\varphi'\delta\varphi\Pi_{\psi}}{M_P^{2} (2 -\alpha)} +M_P^{2}a^{2} k^{2} \psi^{2} +
	2 M_P^{2}a^{2}k^{2} W^{2}\times\\
	&\times\Bigg(\frac{2k^{2} \beta}{(2 -\alpha) (1+c_{13})}+(c_{14} + \alpha) \mathcal{H}^2 + (c_{14} - \alpha) \mathcal{H}'\Bigg)-\\
	&- \frac{1}{2}a^{2} \delta\varphi^{2}\left(k^{2} +a^{2} V_{, \varphi\varphi}- \frac{3\varphi'^{2}}{M_P^{2}(2 -\alpha)}\right)+\\
	&+\frac{2 \alpha a^{2}k^{2}  \varphi'  \delta\varphi W}{(2 -\alpha)}+ \phi \big[-2 c_{14}M_P^{2}a^{2} k^{2}\mathcal{H}W -\\
	&-2 M_P^{2}a^{2}k^{2}\psi +\frac{1}{2}\Pi_{W} +\mathcal{H}\Pi_{\psi} -\varphi'\Pi_{\delta\varphi}-\\
	 &- a^{2} \delta\varphi(a^{2} V_{, \varphi }+ 3\mathcal{H}\varphi')\big].
\end{split}
\end{equation}
Variation with respect to the \emph{independent} variables $\psi$, $\Pi_\psi$, $E$, $\Pi_E$, 
$\delta\varphi$, $\Pi_{\delta\varphi}$, $W$, $\Pi_W$, $U$ and $\phi$ leads to the same equations 
of motion as those derived from the variation of (\ref{eq:gauge free}) with respect to $\psi$, $E$, 
$\delta\varphi$, $B$, $C$ and $\phi$.

Note that the time derivatives of $U$ and $\phi$ do not appear in Eq. (\ref{eq:first order}), 
so variation with respect to these variables leads to the two constraints
\begin{subequations}
\begin{align}
	\Pi_{E}={}&0,\\
	\label{eq:pi constraint}
	\Pi_{\delta\varphi}={}&\frac{-4 c_{14}M_P^{2}a^{2}k^{2} \mathcal{H}W - 2 a^{4} V_{, \varphi}\delta\varphi - 4M_P^{2}a^{2} k^{2}\psi}{2\varphi'} + \nonumber\\
	&+\frac{\Pi_{W} + 2\mathcal{H}\Pi_\psi-6 a^{2}\mathcal{H}\varphi'\delta\varphi}{2\varphi'}.
\end{align} 
\end{subequations} 
Substitution of these constraints also causes $E, \phi$ and $U$ to drop from the Lagrangian,
which therefore depends only on the five independent canonical variables $\psi$, $\Pi_\psi$, 
$\delta\varphi$, $W$, $\Pi_W$. Five is one too many, since we expect two canonical pairs only.
Indeed, one of the variables is redundant, and it corresponds to the residual gauge invariance 
of the Lagrangian. Let us introduce the gauge-invariant combinations
\begin{subequations}
\begin{align}
	\zeta\equiv {}&\psi +\frac{\mathcal{H}}{\varphi'}\delta\varphi,
	\label{eq:Psi}\\
	\delta N\equiv 2\mathcal{H}\Omega\equiv{}& 2\mathcal{H} W +\frac{\mathcal{H}}{\varphi'}\delta\varphi.
	\label{eq:Omega}
\end{align}
\end{subequations}
Geometrically, these can be interpreted as follows. The variable $\zeta$ is the curvature 
perturbation on surfaces of constant inflaton field $\varphi$. The variable $\delta N$ is the
same as the one introduced in (\ref{deltaN}), and can be interpreted as the differential e-folding 
number between hypersurfaces of constant inflaton field and surfaces orthogonal to the aether field.

The momenta conjugate to the gauge-invariant variables $\zeta$ and $\Omega$ are given by
\begin{subequations}
\begin{align}
	\Pi_{\zeta}\equiv{}&\Pi_{\psi} +\frac{2M_P^{2}a^{2} k^{2}}{\varphi'}
	\delta\varphi,\\
	\Pi_{\Omega}\equiv{}& \Pi_{W}+\frac{2 c_{14}M_P^{2}a^{2}k^{2}\mathcal{H}}{\varphi'}\delta\varphi.
\end{align}
\end{subequations}
In terms of the new variables, the field perturbation $\delta\varphi$ disappears from the Lagrangian (\ref{eq:first order}) 
and we have
\begin{equation}\label{eq:first order invariant}
\begin{split}
	&\mathcal{L}^{(1)GI}_{s}
	=M_P^{2}a^{2}k^{2}\Bigg[\zeta^{2}+\\
	&+2\left(\frac{2\beta k^{2}}{(2-\alpha)(1+ c_{13})}
	+ (c_{14}+\alpha)\mathcal{H}^{2} +(c14 - \alpha)\mathcal{H}'\right)\Omega^{2}-\\
	&- \frac{2M_P^{2} k^{2}}{\varphi'^{2}}(\zeta + c_{14}\mathcal{H}	\Omega)^2)\Bigg]
	+\left(\frac{1}{8c_{14}M_P^{2}a^{2}k^{2}}
	-\frac{1}{8a^{2}\varphi'^{2}}\right)\Pi_{\Omega}^{2}+ \\
	&+\left(\frac{\beta}{4M_P^{2}a^{2}(2 -\alpha) (1 + c_{13})}
	- \frac{\mathcal{H}^{2}}{2a^{2}\varphi'^{2}}\right)\Pi_{\zeta}^{2}
	- \frac{\mathcal{H}}{2a^{2}\varphi'^{2}}\Pi_{\zeta}\Pi_{\Omega}+\\
	&+ \frac{c_{14}M_P^{2} \mathcal{H}k^{2}}{\varphi'^{2}}(2\mathcal{H}\Pi_{\zeta}
	+\Pi_{\Omega}) \Omega-\frac{2 \beta k^{2}}{(2 -\alpha) (1 + c_{13})}\Pi_{\zeta}\Omega+\\
	&+\frac{2M_P^{2} \mathcal{H}k^{2}}{\varphi'^{2}}\zeta\Pi_{\zeta}
	+\frac{M_P^{2}k^{2}}{\varphi'^{2}}\zeta\Pi_{\Omega}+\Pi_{\zeta}\zeta'+\Pi_{\Omega}\Omega'.
\end{split}
\end{equation}
Expression (\ref{eq:first order invariant}) gives the first order Lagrangian we have been looking for, since it 
is a function of two canonical pairs, corresponding to two field degrees of freedom.

To see this more explicitly, we may vary with respect to $\Pi_\Omega$ and $\Pi_\zeta$, and plug the resulting 
equations back into Eq. (\ref{eq:first order invariant}) to obtain the second order Lagrangian. 
For reference we just reproduce the leading terms in the limit 
$k|\eta|\gg 1$ (the full expression is cumbersome and not very illuminating). In terms of $\zeta$ and $\delta N$, this
is given by
\begin{equation}\label{eq:Lred}
\begin{split}
	\mathcal{L}^{(2)GI}_{s}&=\frac{M_P^{2} a^2}{2}
	\Big[\frac{-4(2 + c_{14}) k^2}{c_{14}} \zeta (\delta N) +\\
	&+ \frac{k^2 (2(4 +  c_{14} \alpha) + c_{14} (\alpha - 2) (1 + 3 w ))}{2 c_{14}}(\delta N)^{2}+ \\
	&+ \frac{2(2 + c_{14}) k^2}{c_{14}} \zeta^2 + \frac{4  (2 -\alpha) (1 + c_{13})}{\beta}(\delta N)'\zeta'-\\
	&-\frac{2 (2-\alpha) (1 + c_{13})}{\beta}\zeta'^2- \\
	&- \frac{(2 -\alpha) (4(1 + c_{13}) - 3\beta(1 + w))}{2\beta}(\delta N)'^2+\cdots \Big],
\end{split}
\end{equation}
where the ellipsis denote terms which are subleading in the momentum expansion.

Variation of (\ref{eq:Lred}) with respect to $\zeta$ and $\delta N$ 
(including the terms that we do not explicitly write down) 
yields two second order differential equations for $\zeta$ and $\delta N$.  
These equations of motion are valid in any gauge. To find their form in the longitudinal gauge, we may use
Eq. (\ref{llaa}) to express the inflaton perturbations in terms of metric and aether perturbations.
Substituting in (\ref{eq:Psi}) and (\ref{eq:Omega}),
 we can cast the equations of motion for $\zeta$ and $\Omega$ as two third order differential equations 
for the longitudinal gauge variables $\psi$ and $C$. The latter happen to be  precisely linear combinations of 
Eqs.  (\ref{eq:final C}), (\ref{eq:final psi}) and the time derivative of (\ref{eq:final psi}).\footnote{Note in particular that Eqs.  (\ref{eq:final C}) and (\ref{eq:final psi}) cannot follow from a variational principle from a 
reduced Lagrangian depending quadratically on $C$ and $\psi$. If $\alpha=-c_{14}$, the evolution of $C$ decouples from that of $\psi$,  
while the evolution of the latter does depend on the evolution of the former.}

In Eq. (\ref{eq:Lred}), the curvature 
perturbation $\zeta$ on surfaces of constant inflaton field is coupled to the variable $\delta N$. However,
if we replace $\zeta$ by the curvature perturbation $\zeta_a=\zeta-\delta N$ on hypersurfaces orthogonal to the aether, 
this leads to a Lagrangian for two decoupled 
variables, $\zeta_a$ and $\delta N$:
\begin{equation}
	\mathcal{L}_{k\eta\gg1}=\frac{1}{2Z_{N}}\left[(\delta N)'^2 - k^2 (\delta N)^2\right] + 
\frac{1}{2Z_{a}}(\zeta_a'^2 - c_{a}^{2} k^2 \zeta_a^2) +\cdots,\label{lswa}
\end{equation}
where the ellipsis denote terms which are subleading in the momentum expansion, and we have introduced
\begin{equation}
Z_N^{-1}={3(1 + w) (2 - \alpha)\over 2}M_P^{2} a^2,
\end{equation}
and 
\begin{equation}
	Z_a^{-1}=\frac{2(1+c_{13})(\alpha-2)}{\beta}M_{P}^{2}\, a^2, \,
	c_a^2=-\frac{(2+c_{14})\beta}{c_{14}(\alpha-2)(1+c_{13})}.
\end{equation}
This form of the Lagrangian will be used in order to normalize the positive frequency modes associated with the
initial vacuum fluctuations.

\section{Long Wavelength Adiabatic and Isocurvature Modes}
\label{sec:adiabatic modes}

The properties of the two long wavelength adiabatic and isocurvature modes for arbitrary 
expansion history and  fairly general matter content can be also obtained by following a procedure outlined by Weinberg in \cite{Weinberg:2003sw}.

\subsection{Adiabatic Modes}
Consider the gauge transformations generated by 
\begin{equation}\label{eq:gauge transformation}
	\eta\to \eta+\epsilon(\eta)\quad  \text{and}\quad
	 x^i\to x^i+\omega\,  x^i,
\end{equation}
where $\omega$ is a constant. Using the transformation properties of the metric  one finds that  these transformations preserve the structure of longitudinal gauge. In particular, they induce the following transformations on the metric and aether perturbations,
\begin{equation}
	\phi \to \phi-\epsilon{}'-\mathcal{H}\epsilon,\quad
	\psi\to \psi+\omega+\mathcal{H}\epsilon, \quad
	C\to C+\epsilon.
\end{equation}
Because the equations of motion are invariant under gauge transformations, the difference of two sets of perturbations that differ by a gauge transformation is a solution of the linearized equations,
\begin{equation}\label{eq:adiabatic metric}
	\phi=-\epsilon{}'-\mathcal{H}\epsilon, \quad
	\psi=\omega+\mathcal{H}\epsilon, \quad C=\epsilon.
\end{equation}
The corresponding values of the remaining perturbation variables can be also determined by their transformation properties under (\ref{eq:gauge transformation}). For instance, for any scalar perturbation $\delta\varphi$ or any velocity perturbation  $\delta u_i\equiv \partial_i\delta u$ the solutions have 
\begin{equation}\label{eq:adiabatic matter}
	\delta\varphi=-\epsilon \varphi',\quad \delta u=a\, \epsilon.
\end{equation}

Of course, these space-independent solutions are just gauge modes, physically equivalent to no perturbation at all. But they can be extended to actual space-dependent perturbations if the linearized $^0{}_i$ and $^i{}_k$ Einstein equations are satisfied for these putative solutions. The $^0{}_i$ equation is automatically satisfied for the ansatz (\ref{eq:adiabatic metric}) and (\ref{eq:adiabatic matter}). On the other hand, in the presence of the aether the $^i{}_j$ Einstein equation (\ref{eq:scalar stress}) imposes the constraint
\begin{equation}\label{eq:two modes}
	\epsilon'+2\mathcal{H}\epsilon+\frac{1}{1+c_{13}}\omega=0,
\end{equation} 
where we have assumed that the remaining matter does not contribute to the scalar anisotropic stress. The general solution of Eq. (\ref{eq:two modes}) is the superposition of two solutions, with
\begin{subequations}
\begin{align}\label{eq:two solutions}
	\epsilon_1&=-\frac{1}{a^2}\frac{\omega}{1+c_{13}}
	\int^\eta d\tilde{\eta} \, a^2(\tilde{\eta}),&
		\omega_1&=\omega,
		\\
	\epsilon_2&=\frac{C_0}{a^2},&  \omega_2&=0,
\end{align}
\end{subequations}
where $C_0$ is an integration constant. The first solution yields the non-decaying mode, which in the ``gravity" sector reads
\begin{subequations}
\begin{eqnarray}
	\phi_1&=&\frac{\omega}{1+c_{13}}\left(1-\frac{\mathcal{H}}{a^2}\int d\tilde{\eta}\,
	a^2(\tilde{\eta})\right), \\
	\psi_1&=&\omega\left(1-\frac{\mathcal{H}}{a^2}\frac{1}{1+c_{13}}\int d\tilde{\eta}\,
	a^2(\tilde{\eta})\right), \\
	C_1&=&-\frac{\omega}{1+c_{13}}\frac{1}{a^2}\int d\tilde{\eta}\, 
	a^2(\tilde{\eta}).
\end{eqnarray}
\end{subequations}
This reduces to the adiabatic mode (\ref{eq:adiabatic 1}) for a constant equation of state. 
For this mode the curvature perturbation is constant, $\mathcal{\zeta}=\omega$, 
and the anisotropic stress is non-zero (if $c_{13}\neq 0$).  
The second solution in Eq. (\ref{eq:two solutions}) corresponds to a decaying mode, which, 
for a constant equation of state, agrees with the adiabatic mode in Eq. (\ref{eq:adiabatic 2}),
\begin{equation}
	\phi_2=C_0\frac{\mathcal{H}}{a^2},\quad
	\psi_2=C_0\frac{\mathcal{H}}{a^2},\quad
	C=\frac{C_0}{a^2}.\label{2ad}
\end{equation}
For this second adiabatic mode, the curvature perturbation vanishes, $\mathcal{\zeta}=0$, and so does the anisotropic stress. 

\subsection{Isocurvature Modes}

An extension of the previous method also unveils the two  isocurvature modes, under the assumption that the aether does not couple to matter.  Consider the ansatz 
\begin{equation}\label{eq:iso potentials}
	\phi=c_{13}(C'+\mathcal{H}C), \quad
	\psi=-c_{13}\mathcal{H}C, \\
\end{equation}
which arises from the gauge transformation (\ref{eq:gauge transformation}) with $\omega=0$ and $\epsilon=-c_{13} C$.   Acting on any velocity $u_\mu$ and any scalar $\varphi$ (not necessarily the inflaton),  the same gauge transformation leads to the matter perturbations 
\begin{equation}\label{eq:iso matter}
	\delta\varphi=c_{13} \varphi' C,\quad 
	\delta u=-c_{13} a C.
\end{equation}

Since by assumption the aether does not couple to matter, and for the same reasons as in the adiabatic case, we expect Eqs. (\ref{eq:iso potentials})  and (\ref{eq:iso matter}) then to be a solution of the matter equations of motion, no matter what the aether perturbation $C$ actually is.  Of course, for arbitrary values of $C$, we cannot expect the ansatz to satisfy Einstein's equations, since the aether does couple to gravity. Inspection of the latter however reveals that  the $^0{}_0$, $^0{}_i$ and diagonal $^i{}_j$  equations  only contain spatial gradients of the aether field,   which can be neglected in the long-wavelength limit.  The only equation in which the aether perturbation  is not negligible at long wavelengths is (\ref{eq:scalar stress}),   which is actually satisfied by the ansatz (\ref{eq:iso potentials}). Hence, it only remains to find out what the aether perturbation $C$ is. Substituting Eq. (\ref{eq:iso potentials}) into the aether field equation (\ref{eq:scalar vector}) results in a differential equation for the yet undetermined aether perturbation,
\begin{equation}\label{efc}
	C''+2\mathcal{H}C'+
	\left[\left(1+\frac{\alpha}{c_{14}}\right)\mathcal{H}^2+\left(1-\frac{\alpha}{c_{14}}\right)\mathcal{H}'\right]C=0.
\end{equation}
This equation has two independent solutions, which when plugged into (\ref{eq:iso potentials}) and (\ref{eq:iso matter})  give the to two independent isocurvature modes, for which $\zeta=\omega=0$. None of these solutions can be adiabatic, as the adiabatic mode has $\epsilon=C$, while along these solutions $\epsilon=-c_{13} C$ (recall that $c_{13}=-1$ is a singular case.)  A measure of the non-adiabaticity of these modes is the difference  in the e-folding number  between surfaces comoving with aether, and those  comoving with  matter, which equals
\begin{equation}
	\delta N=(1+c_{13})\mathcal{H} C=-\frac{1+c_{13}}{c_{13}}\psi,
\end{equation}
and thus differs from zero if $c_{13}\neq -1$. Since along these solutions all matter components 
(aside from the aether) share the same velocity, the two modes describe a matter-aether 
isocurvature perturbation, which is the only kind of isocurvature perturbation that  can be generated if the aether does not couple to matter.  For a constant equation of state, these two isocurvature modes  
reproduce those found in Subsection \ref{sec:isocurvature}. 

Note that this method of generating  solutions would break down if the anisotropic stress  
of matter on large scales were not negligible, as would happen for instance if the matter sector contained a second aether field.

\section{CMB anisotropies in the vector sector}
\label{sec:vector anisotropies}

The effect of vector perturbations on the amplitude of CMB anisotropies is easily estimated in the 
approximation of a
sharp transition between thermal equilibrium and complete transparency at the moment of decoupling. 
Before the transition, photons and baryons are approximated as a perfect fluid, whereas after the 
transition the radiation will be described in terms of a distribution of free photons. 

The number of 
photons in a phase space cell can be written as
\begin{equation}
dn = n({\bf x},{\bf p})\ \prod_k dx^k \ \prod_i dp_i,
\end{equation}
where $x^k$ are space coordinates and $p_k$ are the spatial components of the momentum.
For a gas of free photons, the number density in phase space obeys the collisionless Boltzmann equation
\begin{equation}
	\frac{\partial n}{\partial \eta}
	+\frac{\partial n}{\partial x^k}\frac{d x^k}{d \eta}
	+ \frac{\partial n}{\partial p_k}\frac{d p_k}{d \eta} =0. \label{bolt}
\end{equation}
Further, we assume that the distribution of photons 
traveling in a given direction ${\bf l}$ at any given point has the Planckian spectrum,
\begin{equation}
n=n(E/T). \label{y}
\end{equation}
Here, 
\begin{equation}
E= -p_{\mu}u^\mu = - a^{-1} p_0 \label{E}
\end{equation}
is the energy of a photon as measured by an observer at rest in the coordinates ${\bf x}$. 
The four-velocity of this observer
is given by $u^{\mu}=(-g_{00})^{-1/2} \delta_{0}^{\mu}$, and in the last equality we have used that 
$-g_{00}=a^2$ is unperturbed in the linearized vector sector.
The local temperature 
\begin{equation}
T = T_0(\eta) + \delta T(\eta,{\bf x},{\bf l})
\end{equation}
depends not only on position, but also on the direction of arrival of the photons,
\begin{equation}
l_i \equiv {p_i/ p}, \quad \text{where}\quad 
p\equiv (\delta^{ij} p_ip_j)^{1/2}.
\end{equation}
Before decoupling, when the system is in thermal equilibrium, the temperature anisotropy is just a dipole, corresponding to the 
local motion of the photon fluid. This is characterized by the four-velocity $\delta u^{\mu}$. Note that $n$ is a scalar, and so $T$ is defined
in such a way that the ratio $y \equiv E/T$ transforms as a scalar. In the co-moving frame, where the fluid is at rest, we have
\begin{equation}
	y=\frac{E_c}{T_c} = \frac{-(u^{\mu}+\delta u^{\mu}) p_\mu}{T_c} 
	= \frac{E-\delta u^i p_i}{T_c}, \label{yscalar}
\end{equation}
where the co-moving temperature $T_c = T_0 + \delta_0(\eta,{\bf x})$ is isotropic, and we have used $\delta u^0=0$ 
(to linear order in $\delta u^i$). 
Since $y=E/T = E_c/T_c$, it follows from (\ref{yscalar}) that at the time of decoupling
\begin{equation}
	\frac{\delta T}{T_0}(\eta_\mathrm{dec}, {\bf x},{\bf l})
	 = \delta_0 + a\ \delta u^i\ l_i. \label{mondi}
\end{equation}
Later, after decoupling, the photons arriving from different directions at a 
given spacetime point have originated at different locations on the surface of last scattering, which leads to anisotropies also in 
the higher multipoles.

The monopole and dipole components in (\ref{mondi}) are related to the perturbations in $T^{0}_{0}$ and $T^{0}_i$, which can 
be obtained from the expression
\begin{equation}
	T^{\mu}_{\nu} = 
		\frac{1}{\sqrt{-g}} \int n(y) \frac{p^{\mu} p_{\nu}}{p^0} d^3 \bf p.
\end{equation}
Here $\bf p$ stands for the spatial components of the momentum, with lower indices. Let us consider the perturbation in the energy 
density. This will be related to the monopole component in the temperature anisotropy.
For vector perturbations, 
\begin{eqnarray}
a^{-2}\delta g_{0i} &=& a^2 \delta g^{0i} = S^i\\
a^{-2} \delta g_{ij} &=& - a^2 \delta g^{ij}=  (F^{i,j} + F^{j,i}), 
\end{eqnarray}
the linearized metric determinant is $\sqrt{-g}=a^4$, and the condition $p_{\mu}p^\mu=0$ leads to
\begin{equation}
p_0 = -p (1 - S^i l_i - F^{i,j}\ {l_il_j}),\label{en}
\end{equation}
The energy density of photons is given by
\begin{equation}
	\rho_{c_{13}} = 
		-T^0_0 = -\frac{1}{a^4} \int n(y) p_0 p^2 dp d^2{\bf l}.
\end{equation}
We can now use that $p_0 = -a T_0 y (1+{\delta T/T})$ and $p= a T_0 y (1+S^i l_i + F^{i,j}\ {l_il_j}+{\delta T/T})$  to eliminate $p$ and $p_0$ in favour of $y$. After simple manipulations,
one obtains
\begin{equation}
	\rho_{c_{13}} = \rho_{c_{13}}^{(0)} 
	\left[ 1+ 4 \int \frac{d^2{\bf l}}{4\pi} 
	\frac{\delta T}{T}\right]. \label{delrho}
\end{equation}
Since vector perturbations do not change the energy density, we have $\delta\rho_{c_{13}}=0$. Therefore, 
using (\ref{mondi}) in (\ref{delrho}) we find $\delta_0=0$. Hence, the temperature anisotropy (\ref{mondi})
for vector perturbations in the perfect fluid is purely dipolar:
\begin{equation}
	\frac{\delta T}{T_0}(\eta_\mathrm{dec}, {\bf x},{\bf l}) 
	= - S^i l_i +\frac{\delta u_i}{a}\ l_i. \label{mondi2}
\end{equation}
For later convenience, here we have expressed the result in terms of the velocity perturbation with lower indices,
which is gauge-invariant.

The evolution of the temperature anisotropy after decoupling can be inferred from the Boltzmann equation. Defining 
\begin{equation}
E_0 = p/a,
\end{equation}
we have $\partial_\eta (E_0/T_0) =\partial_{x^k}(E_0/T_0)=0$, and $\partial_{p_k}(E_0/T_0)= (l_k/p)(E_0/T_0)$.
Substituting (\ref{y}) in (\ref{bolt}), and linearizing in perturbations, it is straightforward to show that
\begin{equation}
	\left(\frac{\partial}{\partial \eta}+l_k \frac{\partial}{\partial x^k}\right)
	\left(\frac{\delta E}{E_0}-\frac{\delta T}{T_0}\right) 
	+ \frac{l_k}{p} \frac{d p_k}{d \eta}=0,\label{bolt2}
\end{equation}
where $\delta E=E-E_0= -(p_0+p)/a$. The geodesic equation reads
\begin{equation}
	\frac{dp_k}{d\eta} = \frac{1}{2 p^0}
	\frac{\partial g_{\mu\nu}}{\partial x^k} p^{\mu}p^{\nu} = S^i,_k p_i 
	+ F^{i,j},_{k} \frac{p_i p_j}{p}.\label{geo}
\end{equation}
Using (\ref{en}) and (\ref{geo}) in (\ref{bolt2}) we have
\begin{equation}
	\frac{d}{d\eta}
	\left(\frac{\delta T}{T_0}+{\bf F}'\cdot {\bf l}\right)
	 = {\bf Q}'\cdot {\bf l}. \label{bolt3}
\end{equation}
Here, $d/d\eta=\partial_\eta +l_i \partial_{x^i}$ is the total derivative along the line of sight, and primes indicate partial derivatives with 
respect to $\eta$. The result is expressed in terms of the gauge-invariant combinations 
 $(\delta T/T_0)+{\bf F}'\cdot {\bf l}$ and $Q^i \equiv {F^i}'-S^i$. Eq. (\ref{bolt3}) can be integrated along the trajectory of the
photons ${\bf x}(\eta)=(\eta-\eta_0){\bf l}$, 
from the time of decoupling $\eta_\mathrm{dec}$ to the present time $\eta_0$, to obtain the temperature 
anisotropy which is observed at present:
\begin{equation}
	\left(\frac{\delta T}{T_0}+{\bf F}'\cdot {\bf l}\right)_0 = \left(Q^i l_i
	+  \frac{\delta u_i}{ a}\ l_i\right)_\mathrm{dec}
	 + \int_{\eta_\mathrm{dec}}^{\eta_0} d\eta\ {\bf Q}'\cdot {\bf l}. \label{full}
\end{equation}
Here we have used the initial condition determined by (\ref{mondi2}).  
As we mention in Subsection \ref{sec:vectors deceleration}, a non-vanishing velocity perturbation $\delta u_i$ cannot be generated as long as the perfect fluid description is valid, so we shall ignore $\delta u_i$  in Eq. (\ref{full}), and simply write
\begin{equation}
	\left(\frac{\delta T}{T_0}\right)_0 = 
	\left({\bf Q}\cdot {\bf l}\right)_\mathrm{dec} +
	\int_{\eta_\mathrm{dec}}^{\eta_0} d\eta\ {\bf Q}'\cdot {\bf l}. \label{full2}
\end{equation}
Here we have also dropped the dipole term at the time of observation, since this is always subtracted.


\begin{thebibliography}{99}

\bibitem{DGP} G.~R.~Dvali, G.~Gabadadze and M.~Porrati,
  ``4D gravity on a brane in 5D Minkowski space,''
  Phys.\ Lett.\  B {\bf 485}, 208 (2000)
  [arXiv:hep-th/0005016].

\bibitem{galileon}
A.~Nicolis, R.~Rattazzi and E.~Trincherini,
  ``The galileon as a local modification of gravity,''
  Phys.\ Rev.\  D {\bf 79}, 064036 (2009)
  [arXiv:0811.2197 [hep-th]].

\bibitem{galileoncov}
C.~Deffayet, G.~Esposito-Farese and A.~Vikman,
  Phys.\ Rev.\  D {\bf 79}, 084003 (2009)
  [arXiv:0901.1314 [hep-th]].

\bibitem{Peebles:1987ek}
  P.~J.~E.~Peebles and B.~Ratra,
  ``Cosmology with a Time Variable Cosmological Constant,''
  Astrophys.\ J.\  {\bf 325}, L17 (1988).

\bibitem{mod}
This includes the case where we ``modify" the Einstein-Hilbert action to an arbitrary function of the 
Ricci scalar \cite{Capozziello:2002rd,Carroll:2003wy}, since the resulting theory can
be reformulated as a standard scalar-tensor theory \cite{Will:1981cz}.

\bibitem{Capozziello:2002rd}
  S.~Capozziello,
  ``Curvature Quintessence,''
  Int.\ J.\ Mod.\ Phys.\  D {\bf 11}, 483 (2002)
  [arXiv:gr-qc/0201033].

\bibitem{Carroll:2003wy}
  S.~M.~Carroll, V.~Duvvuri, M.~Trodden and M.~S.~Turner,
  ``Is cosmic speed-up due to new gravitational physics?,''
  Phys.\ Rev.\  D {\bf 70}, 043528 (2004)
  [arXiv:astro-ph/0306438].

\bibitem{Will:1981cz}
  C.~M.~Will,
  ``Theory And Experiment In Gravitational Physics,''
{\it  Cambridge, Uk: Univ. Pr. ( 1981) 342p}

\bibitem{bopo}
R.~Bousso and J.~Polchinski,
  ``Quantization of four-form fluxes and dynamical neutralization of the
  cosmological constant,''
  JHEP {\bf 0006}, 006 (2000)
  [arXiv:hep-th/0004134].
  
  \bibitem{kessence}
  C.~Armendariz-Picon, T.~Damour and V.~F.~Mukhanov,
  ``k-Inflation,''
  Phys.\ Lett.\  B {\bf 458}, 209 (1999)
  [arXiv:hep-th/9904075];
 J.~Garriga and V.~F.~Mukhanov,
  ``Perturbations in k-inflation,''
  Phys.\ Lett.\  B {\bf 458}, 219 (1999)
  [arXiv:hep-th/9904176];
C.~Armendariz-Picon, V.~F.~Mukhanov and P.~J.~Steinhardt,
  ``A dynamical solution to the problem of a small cosmological constant  and
  Phys.\ Rev.\ Lett.\  {\bf 85}, 4438 (2000)
  [arXiv:astro-ph/0004134];
 C.~Armendariz-Picon, V.~F.~Mukhanov and P.~J.~Steinhardt,
  ``Essentials of k-essence,''
  Phys.\ Rev.\  D {\bf 63}, 103510 (2001)
  [arXiv:astro-ph/0006373].

\bibitem{ghostcondensation}
N.~Arkani-Hamed, H.~C.~Cheng, M.~A.~Luty and S.~Mukohyama,
  ``Ghost condensation and a consistent infrared modification of gravity,''
  JHEP {\bf 0405}, 074 (2004)
  [arXiv:hep-th/0312099].

\bibitem{massive}
N.~Arkani-Hamed, H.~Georgi and M.~D.~Schwartz,
  ``Effective field theory for massive gravitons and gravity in theory space,''
  Annals Phys.\  {\bf 305}, 96 (2003)
  [arXiv:hep-th/0210184].


\bibitem{phases}
S.~L.~Dubovsky,
  ``Phases of massive gravity,''
  JHEP {\bf 0410}, 076 (2004)
  [arXiv:hep-th/0409124];
  D.~Blas, D.~Comelli, F.~Nesti and L.~Pilo,
  ``Lorentz Breaking Massive Gravity in Curved Space,''
  Phys.\ Rev.\  D {\bf 80}, 044025 (2009)
  [arXiv:0905.1699 [hep-th]].

 \bibitem{slavaali}
  A.~H.~Chamseddine and V.~Mukhanov,
 ``Higgs for Graviton: Simple and Elegant Solution,''
 arXiv:1002.3877 [hep-th].

\bibitem{lbmass}
 V.~A.~Rubakov,
  ``Lorentz-violating graviton masses: Getting around ghosts, low strong
  arXiv:hep-th/0407104.

\bibitem{pheno}
S.~L.~Dubovsky, P.~G.~Tinyakov and I.~I.~Tkachev,
  ``Massive graviton as a testable cold dark matter candidate,''
  Phys.\ Rev.\ Lett.\  {\bf 94}, 181102 (2005)
  [arXiv:hep-th/0411158];
 S.~Dubovsky, R.~Flauger, A.~Starobinsky and I.~Tkachev,
  ``Signatures of a Graviton Mass in the Cosmic Microwave Background,''
  arXiv:0907.1658 [astro-ph.CO],
 D.~Bessada and O.~D.~Miranda,
 ``Cmb Anisotropies Induced By Tensor Modes In Massive Gravity,''
 JCAP {\bf 0908} (2009) 033
 [arXiv:0908.1360 [astro-ph.CO]].

  
\bibitem{Damour:2002ws}
  T.~Damour and I.~I.~Kogan,
  ``Effective Lagrangians and universality classes of nonlinear bigravity,''
  Phys.\ Rev.\  D {\bf 66}, 104024 (2002)
  [arXiv:hep-th/0206042].

\bibitem{lbbig}
Z.~Berezhiani, D.~Comelli, F.~Nesti and L.~Pilo,
  ``Spontaneous Lorentz breaking and massive gravity,''
  Phys.\ Rev.\ Lett.\  {\bf 99}, 131101 (2007)
  [arXiv:hep-th/0703264];
D.~Blas, C.~Deffayet and J.~Garriga,
  ``Bigravity and Lorentz-violating Massive Gravity,''
  Phys.\ Rev.\  D {\bf 76}, 104036 (2007)
  [arXiv:0705.1982 [hep-th]];
  Z.~Berezhiani, D.~Comelli, F.~Nesti and L.~Pilo,
  ``Exact Spherically Symmetric Solutions in Massive Gravity,''
  JHEP {\bf 0807}, 130 (2008)
  [arXiv:0803.1687 [hep-th]].
  
  \bibitem{ArmendarizPicon:2009ai}
 C.~Armendariz-Picon and A.~Diez-Tejedor,
  ``Aether Unleashed,''
  JCAP {\bf 0912}, 018 (2009)
  [arXiv:0904.0809 [astro-ph.CO]].

\bibitem{Jacobson:2000xp}
  T.~Jacobson and D.~Mattingly,
  ``Gravity with a dynamical preferred frame,''
  Phys.\ Rev.\  D {\bf 64}, 024028 (2001)
  [arXiv:gr-qc/0007031].
 
\bibitem{mond}
A fixed norm vector field determining a preferred frame has also been used in relativistic versions of MOND 
\cite{Milgrom:1983ca}, such as TeVeS \cite{Bekenstein:2004ne}, which attempt to explain the rotation curves of
galaxies without introducing cold dark matter.

\bibitem{Milgrom:1983ca}
  M. Milgrom,
  ``A Modification Of The Newtonian Dynamics As A Possible Alternative To The Hidden Mass Hypothesis,''
  Astrophys.\ J.\  {\bf 270}, 365 (1983).
  
\bibitem{Bekenstein:2004ne}
  J.~D.~Bekenstein,
 ``Relativistic gravitation theory for the MOND paradigm,''
  Phys.\ Rev.\  D {\bf 70}, 083509 (2004)
  [Erratum-ibid.\  D {\bf 71}, 069901 (2005)]
  [arXiv:astro-ph/0403694].
 
 \bibitem{Low:2001bw}
  I.~Low and A.~V.~Manohar,
  ``Spontaneously broken spacetime symmetries and Goldstone's theorem,''
  Phys.\ Rev.\ Lett.\  {\bf 88}, 101602 (2002)
  [arXiv:hep-th/0110285].

  
\bibitem{hg}
P.~Horava,
  ``Quantum Gravity at a Lifshitz Point,''
  Phys.\ Rev.\  D {\bf 79}, 084008 (2009)
  [arXiv:0901.3775 [hep-th]];
  ``Spectral Dimension of the Universe in Quantum Gravity at a Lifshitz
  Phys.\ Rev.\ Lett.\  {\bf 102}, 161301 (2009)
  [arXiv:0902.3657 [hep-th]].

\bibitem{bps1}
  D.~Blas, O.~Pujolas and S.~Sibiryakov,
  ``On the Extra Mode and Inconsistency of Horava Gravity,''
  JHEP {\bf 0910}, 029 (2009)
  [arXiv:0906.3046 [hep-th]].
  
\bibitem{bps2}
  D.~Blas, O.~Pujolas and S.~Sibiryakov,
  ``A healthy extension of Horava gravity,''
  arXiv:0909.3525 [hep-th];
  ``Comment on `Strong coupling in extended Horava-Lifshitz gravity',''
  arXiv:0912.0550 [hep-th].
  
\bibitem{jbpsh}
 T.~Jacobson,
  ``Extended Horava gravity and Einstein-aether theory,''
  arXiv:1001.4823 [hep-th].

\bibitem{Lim:2004js}
  E.~A.~Lim,
  ``Can we see Lorentz-violating vector fields in the CMB?,''
  Phys.\ Rev.\  D {\bf 71}, 063504 (2005)
  [arXiv:astro-ph/0407437].
  
\bibitem{Li:2007vz}
  B.~Li, D.~Fonseca Mota and J.~D.~Barrow,
  ``Detecting a Lorentz-Violating Field in Cosmology,''
  Phys.\ Rev.\  D {\bf 77}, 024032 (2008)
  [arXiv:0709.4581 [astro-ph]].
  
\bibitem{Zuntz:2008zz}
  J.~A.~Zuntz, P.~G.~Ferreira and T.~G.~Zlosnik,
  ``Constraining Lorentz violation with cosmology,''
  Phys.\ Rev.\ Lett.\  {\bf 101}, 261102 (2008)
  [arXiv:0808.1824 [gr-qc]].
  
\bibitem{Zuntz:2010jp}
  J.~Zuntz, T.~G.~Zlosnik, F.~Bourliot, P.~G.~Ferreira and G.~D.~Starkman,
  ``Vector field models of modified gravity and the dark sector,''
  arXiv:1002.0849 [astro-ph.CO].

\bibitem{yuko}
  T.~Kobayashi, Y.~Urakawa and M.~Yamaguchi,
  ``Cosmological perturbations in a healthy extension of Horava gravity,''
  arXiv:1002.3101 [hep-th].

\bibitem{Carroll:2004ai}
  S.~M.~Carroll and E.~A.~Lim,
  ``Lorentz-violating vector fields slow the universe down,''
  Phys.\ Rev.\  D {\bf 70}, 123525 (2004)
  [arXiv:hep-th/0407149].
  
\bibitem{Lucchin:1984yf}
  F.~Lucchin and S.~Matarrese,
  ``Power Law Inflation,''
  Phys.\ Rev.\  D {\bf 32}, 1316 (1985).
        
  \bibitem{Jacobson:2004ts}
  T.~Jacobson and D.~Mattingly,
  ``Einstein-aether waves,''
  Phys.\ Rev.\  D {\bf 70}, 024003 (2004)
  [arXiv:gr-qc/0402005].
  
  \bibitem{Cline:2003gs}
  J.~M.~Cline, S.~Jeon and G.~D.~Moore,
  ``The phantom menaced: Constraints on low-energy effective ghosts,''
  Phys.\ Rev.\  D {\bf 70}, 043543 (2004)
  [arXiv:hep-ph/0311312].
  
   \bibitem{ArmendarizPicon:2003gd}
  C.~Armendariz-Picon and E.~A.~Lim,
  ``Vacuum choices and the predictions of inflation,''
  JCAP {\bf 0312}, 006 (2003)
  [arXiv:hep-th/0303103].
    
  \bibitem{Weinberg:2003sw}
  S.~Weinberg,
  ``Adiabatic modes in cosmology,''
  Phys.\ Rev.\  D {\bf 67}, 123504 (2003)
  [arXiv:astro-ph/0302326].
  
  \bibitem{Weinberg:2004kr}
  S.~Weinberg,
  ``Can non-adiabatic perturbations arise after single-field inflation?,''
  Phys.\ Rev.\  D {\bf 70}, 043541 (2004)
  [arXiv:astro-ph/0401313].
  
  \bibitem{Weinberg:2004kf}
  S.~Weinberg,
  ``Must cosmological perturbations remain non-adiabatic after multi-field
  inflation?,''
  Phys.\ Rev.\  D {\bf 70}, 083522 (2004)
  [arXiv:astro-ph/0405397].

\bibitem{Weinberg:2008zzc}
  S.~Weinberg,
  ``Cosmology,''
{\it  Oxford, UK: Oxford Univ. Pr. (2008) 593 p}.  
  
 \bibitem{Eling:2003rd}
  C.~Eling and T.~Jacobson,
  ``Static post-Newtonian equivalence of GR and gravity with a dynamical
  preferred frame,''
  Phys.\ Rev.\  D {\bf 69}, 064005 (2004)
  [arXiv:gr-qc/0310044]. 
  
  \bibitem{Koivisto:2005mm}
  T.~Koivisto and D.~F.~Mota,
  ``Dark Energy Anisotropic Stress and Large Scale Structure Formation,''
  Phys.\ Rev.\  D {\bf 73}, 083502 (2006)
  [arXiv:astro-ph/0512135].
  
   \bibitem{Daniel:2009kr}
  S.~F.~Daniel {\it et al.},
  ``A Multi-Parameter Investigation of Gravitational Slip,''
  Phys.\ Rev.\  D {\bf 80}, 023532 (2009)
  [arXiv:0901.0919 [astro-ph.CO]].
  
 \bibitem{Giannantonio:2009gi}
  T.~Giannantonio, M.~Martinelli, A.~Silvestri and A.~Melchiorri,
  ``New constraints on parametrised modified gravity from correlations of the
  CMB with large scale structure,''
  arXiv:0909.2045 [astro-ph.CO].
  
  \bibitem{Bean:2009wj}
  R.~Bean,
  ``A weak lensing detection of a deviation from General Relativity on cosmic
  scales,''
  arXiv:0909.3853 [astro-ph.CO].

  
    \bibitem{Damour:1993ev}
  T.~Damour and G.~Esposito-Farese,
  ``Testing for preferred frame effects in gravity with artificial earth
  satellites,''
  Phys.\ Rev.\  D {\bf 49}, 1693 (1994)
  [arXiv:gr-qc/9311034].
  
  \bibitem{Graesser:2005bg}
  M.~L.~Graesser, A.~Jenkins and M.~B.~Wise,
  ``Spontaneous Lorentz violation and the long-range gravitational
  preferred-frame effect,''
  Phys.\ Lett.\  B {\bf 613}, 5 (2005)
  [arXiv:hep-th/0501223].
  
    \bibitem{Dodelson:2003vq}
  S.~Dodelson and L.~Hui,
  ``A horizon ratio bound for inflationary fluctuations,''
  Phys.\ Rev.\ Lett.\  {\bf 91}, 131301 (2003)
  [arXiv:astro-ph/0305113].
  
   \bibitem{Hu:1997hp}
  W.~Hu and M.~J.~White,
  ``CMB Anisotropies: Total Angular Momentum Method,''
  Phys.\ Rev.\  D {\bf 56}, 596 (1997)
  [arXiv:astro-ph/9702170].
  
  \bibitem{Mukhanov:2005sc}
  V.~Mukhanov,
  ``Physical foundations of cosmology,''
{\it  Cambridge, UK: Univ. Pr. (2005) 421 p}

\bibitem{bessel}
I.~S. Gradshteyn and I~.M~. Ryzhik, ``Table of Integrals, Series and Products,"  {\it  New York, US: Academic Press (1980)}.

\bibitem{Turner:1993vb}
  M.~S.~Turner, M.~J.~White and J.~E.~Lidsey,
  ``Tensor perturbations in inflationary models as a probe of cosmology,''
  Phys.\ Rev.\  D {\bf 48}, 4613 (1993)
  [arXiv:astro-ph/9306029].
      
\bibitem{Withers}
B.~Withers,
  ``Einstein-aether as a quantum effective field theory,''
  Class.\ Quant.\ Grav.\  {\bf 26}, 225009 (2009)
  [arXiv:0905.2446 [gr-qc]].

  \bibitem{Komatsu:2008hk}
  E.~Komatsu {\it et al.}  [WMAP Collaboration],
  ``Five-Year Wilkinson Microwave Anisotropy Probe (WMAP)
  Observations:Cosmological Interpretation,''
  Astrophys.\ J.\ Suppl.\  {\bf 180}, 330 (2009)
  [arXiv:0803.0547 [astro-ph]].
    
    
  \bibitem{Jacobson:2008aj}
  T.~Jacobson,
  ``Einstein-aether gravity: a status report,''
  PoS {\bf QG-PH}, 020 (2007)
  [arXiv:0801.1547 [gr-qc]].

\bibitem{Bell:1995jz} J.~F.~Bell, F.~Camilo and T.~Damour, ``A Tighter Test of Local Lorentz Invariance using PSR J2317+1439,'' Astrophys.\ J.\ {\bf 464}, 857 (1996) [arXiv:astro-ph/9512100].
  
\bibitem{Nordtvedt} K.~Nordvedt, ``Probing Gravity to the Second Post-Newtonian Order and to One Part in $10^7$ Using the Sping Axis of the Sun," Astrophys.\ J.\ {\bf 320}, 871 (1987).


\bibitem{Elliott:2005va}
  J.~W.~Elliott, G.~D.~Moore and H.~Stoica,
  ``Constraining the new aether: Gravitational Cherenkov radiation,''
  JHEP {\bf 0508}, 066 (2005)
  [arXiv:hep-ph/0505211].

\bibitem{Adams:2006sv}
  A.~Adams, N.~Arkani-Hamed, S.~Dubovsky, A.~Nicolis and R.~Rattazzi,
  ``Causality, analyticity and an IR obstruction to UV completion,''
  JHEP {\bf 0610}, 014 (2006)
  [arXiv:hep-th/0602178].
  
\bibitem{Bruneton:2006gf}
  J.~P.~Bruneton,
 ``On causality and superluminal behavior in classical field theories. Applications to k-essence theories and MOND-like theories of gravity,''
  Phys.\ Rev.\  D {\bf 75}, 085013 (2007)
  [arXiv:gr-qc/0607055].
  
  \bibitem{Babichev:2007dw}
  E.~Babichev, V.~Mukhanov and A.~Vikman,
  ``k-Essence, superluminal propagation, causality and emergent geometry,''
  JHEP {\bf 0802}, 101 (2008)
  [arXiv:0708.0561 [hep-th]].

\bibitem{tears}
  R.~Jackiw,
  ``(Constrained) quantization without tears,''
  arXiv:hep-th/9306075.



\bibitem{gmst}
  J.~Garriga, X.~Montes, M.~Sasaki and T.~Tanaka,
  ``Canonical quantization of cosmological perturbations in the one-bubble
  open universe,''
  Nucl.\ Phys.\  B {\bf 513}, 343 (1998)
  [arXiv:astro-ph/9706229].


  
\end{thebibliography}
\end{document}